\documentclass[twocolumn]{aastex62}
\usepackage[utf8]{inputenc}
\usepackage[T1]{fontenc}
\usepackage{lmodern}

\bibstyle{aasjournal}

\newcommand{\naiw}{\mbox{\ion{Na}{1} $\lambda$8183,8195}}

\newcommand{\teff}{$T_{\rm eff}$}
\newcommand{\logg}{$\log g$}

\shorttitle{SDSS-IV MaStar}
\shortauthors{Yan et al.}

\begin{document}
\title{SDSS-IV MaStar -- A Large and Comprehensive Empirical Stellar Spectral Library: First Release}

\correspondingauthor{Renbin Yan}
\email{rya225@g.uky.edu, yanrenbin@gmail.com}

\author[0000-0003-1025-1711]{Renbin Yan}
\affil{Department of Physics and Astronomy, University of Kentucky, 505 Rose St., Lexington, KY 40506-0057, USA}

\author[0000-0001-8821-0309]{Yanping Chen}
\affil{New York University Abu Dhabi, Abu Dhabi, P.O. Box 129188, United Arab Emirates}

\author{Daniel Lazarz}
\affil{Department of Physics and Astronomy, University of Kentucky, 505 Rose St., Lexington, KY 40506-0057, USA}

\author[0000-0002-3601-133X]{Dmitry Bizyaev}
\affil{Apache Point Observatory and New Mexico State University, P.O. Box 59, Sunspot, NM 88349, USA}
\affil{Sternberg Astronomical Institute, Moscow State University, Universitetskij pr. 13, Moscow, Russia}

\author{Claudia Maraston}
\affil{Institute of Cosmology \& Gravitation, University of Portsmouth, Dennis Sciama Building, Portsmouth, PO1 3FX, UK}

\author[0000-0003-1479-3059]{Guy S. Stringfellow}
\affil{Center for Astrophysics and Space Astronomy, Department of Astrophysical and Planetary Sciences, University of Colorado, 389 UCB, Boulder, CO 80309-0389, USA}

\author{Kyle McCarthy}
\affil{Department of Physics and Astronomy, University of Kentucky, 505 Rose St., Lexington, KY 40506-0057, USA}

\author[0000-0001-6744-1204]{Sofia Meneses-Goytia}
\affil{Institute of Cosmology \& Gravitation, University of Portsmouth, Dennis Sciama Building, Portsmouth, PO1 3FX, UK}
\affil{Department of Physics, University of Surrey, Guildford, GU2 7XH, UK}

\author[0000-0002-9402-186X]{David R. Law}
\affil{Space Telescope Science Institute, 3700 San Martin Drive, Baltimore, MD 21218, USA}

\author[0000-0002-6325-5671]{Daniel Thomas}
\affil{Institute of Cosmology \& Gravitation, University of Portsmouth, Dennis Sciama Building, Portsmouth, PO1 3FX, UK}

\author[0000-0002-0608-9574]{Jesus Falcon Barroso}
\affil{Instituto de Astrof\'isica de Canarias, V\'ia L\'actea s/n, E-38205 La Laguna, Tenerife, Spain}
\affil{Departamento de Astrof\'isica, Universidad de La Laguna, E-38200 La Laguna, Tenerife, Spain}

\author{José R. Sánchez-Gallego}
\affil{Department of Astronomy, Box 351580, University of Washington, Seattle, WA 98195, USA}

\author{Edward Schlafly}
\affil{Lawrence Berkeley National Laboratory, 1 Cyclotron Road, Berkeley, CA 94720, USA}

\author{Zheng Zheng}
\affil{National Astronomical Observatories, Chinese Academy of Sciences, 20A Datun Road, Chaoyang District, Beijing 100012, China}

\author{Maria Argudo-Fern{\'a}ndez}
\affil{Centro de Astronom\'ia (CITEVA), Universidad de Antofagasta, Avenida Angamos 601 Antofagasta, Chile}

\author[0000-0002-1691-8217]{Rachael L. Beaton}
\altaffiliation{Hubble Fellow}
\altaffiliation{Carnegie-Princeton Fellow}
\affiliation{Department of Astrophysical Sciences, Princeton University, 4 Ivy Lane, Princeton, NJ~08544}
\affiliation{The Observatories of the Carnegie Institution for Science, 813 Santa Barbara St., Pasadena, CA~91101}

\author{Timothy C. Beers}
\affil{Department of Physics and JINA Center for the Evolution of the Elements, University of Notre Dame, Notre Dame, IN 46530  USA}

\author{Matthew Bershady}
\affiliation{Department of Astronomy, University of Wisconsin-Madison, 475 N. Charter St., Madison, WI 53726, USA}
\affiliation{South African Astronomical Observatory, Cape Town, South Africa}

\author{Michael R. Blanton}
\affiliation{Center for Cosmology and Particle Physics, Department of Physics, New York University, 726 Broadway, Room 1005, New York, NY 10003, USA}

\author{Joel Brownstein}
\affil{Department of Physics and Astronomy, University of Utah, 115 S. 1400 E., Salt Lake City, UT 84112, USA}

\author{Kevin Bundy}
\affil{University of California Observatories, University of California, Santa Cruz, CA 95064, USA}

\author{Kenneth C. Chambers}
\affil{Institute for Astronomy, University of Hawaii, Honolulu, HI}

\author{Brian Cherinka}
\affil{Department of Physics and Astronomy, Johns Hopkins University, 3400 N. Charles St., Baltimore, MD 21218, USA}

\author[0000-0002-3657-0705]{Nathan De Lee}
\affil{Department of Physics, Geology, and Engineering Technology, Northern 
Kentucky University, Highland Heights, KY 41099, USA}
\affil{Department of Physics \& Astronomy, Vanderbilt University, Nashville, TN 37235, USA}

\author{Niv Drory}
\affil{McDonald Observatory, The University of Texas at Austin, 1 University Station, Austin, TX 78712, USA}

\author{Lluís Galbany}
\affil{PITT PACC, Department of Physics and Astronomy, University of Pittsburgh, Pittsburgh, PA 15260, USA}

\author{Jon Holtzman}
\affil{Department of Astronomy, New Mexico State University, Box 30001, MSC 4500, Las Cruces NM 88003, USA}

\author{Julie Imig}
\affil{Department of Astronomy, New Mexico State University, Box 30001, MSC 4500, Las Cruces NM 88003, USA}

\author{Nick Kaiser}
\affil{Institute for Astronomy, University of Hawaii, Honolulu, HI}

\author{Karen Kinemuchi}
\affil{Apache Point Observatory and New Mexico State University, P.O. Box 59, Sunspot, NM 88349, USA}

\author{Chao Liu}
\affil{Key Laboratory of Optical Astronomy, National Astronomical 
Observatories, Chinese Academy of Sciences, Beijing 100012, China}

\author{A-Li Luo}
\affil{Key Laboratory of Optical Astronomy, National Astronomical 
Observatories, Chinese Academy of Sciences, Beijing 100012, China}

\author[0000-0002-7965-2815]{Eugene Magnier}
\affil{Institute for Astronomy, University of Hawaii, Honolulu, HI}

\author{Steven Majewski}
\affil{Department of Astronomy, University of Virginia, P.O. Box 400325, Charlottesville, VA 22904-4325}

\author{Preethi Nair}
\affil{Department of Physics and Astronomy, University of Alabama, Tuscaloosa, AL 35487-0324, USA}

\author{Audrey Oravetz}
\affil{Apache Point Observatory and New Mexico State University, P.O. Box 59, Sunspot, NM 88349, USA}

\author{Daniel Oravetz}
\affil{Apache Point Observatory and New Mexico State University, P.O. Box 59, Sunspot, NM 88349, USA}

\author{Kaike Pan}
\affil{Apache Point Observatory and New Mexico State University, P.O. Box 59, Sunspot, NM 88349, USA}

\author{Jennifer Sobeck}
\affil{Department of Astronomy, Box 351580, University of Washington, Seattle, WA 98195, USA}

\author{Keivan Stassun}
\affil{Department of Physics \& Astronomy, Vanderbilt University, Nashville, TN 37235, USA}

\author{Michael Talbot}
\affil{Department of Physics and Astronomy, University of Utah, 115 S. 1400 E., Salt Lake City, UT 84112, USA}

\author{Christy Tremonti}
\affiliation{Department of Astronomy, University of Wisconsin-Madison, 475 N. Charter St., Madison, WI 53726, USA}

\author[0000-0003-1989-4879]{Christopher Waters}
\affil{Institute for Astronomy, University of Hawaii, Honolulu, HI}

\author[0000-0002-5908-6852]{Anne-Marie Weijmans}
\affiliation{School of Physics and Astronomy, University of St Andrews, North Haugh, St. Andrews KY16 9SS, UK}

\author{Ronald Wilhelm}
\affil{Department of Physics and Astronomy, University of Kentucky, 505 Rose St., Lexington, KY 40506-0057, USA}

\author{Gail Zasowski}
\affil{Department of Physics and Astronomy, University of Utah, 115 S. 1400 E., Salt Lake City, UT 84112, USA}

\author{Gang Zhao}
\affil{Key Laboratory of Optical Astronomy, National Astronomical 
Observatories, Chinese Academy of Sciences, Beijing 100012, China}

\author{Yong-Heng Zhao}
\affil{Key Laboratory of Optical Astronomy, National Astronomical 
Observatories, Chinese Academy of Sciences, Beijing 100012, China}

\begin{abstract}
We present the first release of the MaNGA Stellar Library (MaStar), which is a large, well-calibrated, high-quality empirical library covering the wavelength range of 3622-10,354\AA\ at a resolving power of  $R\sim1800$. The spectra were obtained using the same instrument as used by the Mapping Nearby Galaxies at Apache Point Observatory (MaNGA) project, by piggybacking on the SDSS-IV/APOGEE-2N observations. Compared to previous empirical libraries, the MaStar library will have a higher number of stars and a more comprehensive stellar-parameter coverage, especially of cool dwarfs, low-metallicity stars, and stars with different [${\rm \alpha}$/Fe], achieved by a sophisticated target selection strategy that takes advantage of stellar-parameter catalogs from the literature. This empirical library will provide a new basis for stellar population synthesis, and is particularly well-suited for stellar-population analysis of MaNGA galaxies. The first version of the library contains 8646 high-quality per-visit spectra for 3321 unique stars. Compared to photometry, the relative flux calibration of the library is accurate to 3.9\% in $g-r$, $2.7\%$ in $r-i$, and 2.2\% in $i-z$. The data are released as part of Sloan Digital Sky Survey Data Release 15. We expect the final release of the library to contain more than 10,000 stars. 
\end{abstract}

\section{Introduction and Motivation}

A stellar library is a collection of spectra of individual stars, empirical or theoretical, with a given wavelength range and intrinsic resolution covering a certain parameter space of atmospheric properties. These stellar spectral libraries play an essential role in a wide range of astrophysics applications. In extragalactic astronomy, they are essential ingredients in stellar population synthesis, which has been widely used to derive properties such as stellar population age, stellar mass, stellar metallicity, initial mass function, and to model the broadband spectral energy distribution in order to measure redshifts \citep[e.g.,][]{Leitherer99, BC03, Maraston05, MarastonS11, Vazdekis10, Vazdekis12, Conroy09, Conroy13, Rock16}. Stellar libraries are also used directly in modeling the stellar continuum in integrated spectra, in order to remove the continuum for emission-line studies of star formation and active galactic nuclei, or to model the stellar kinematics to infer the baryonic and dark matter mass distributions. For stellar and Galactic astronomy, they are often used to model the continuum spectra of stars, in the absence of spectroscopy, and to estimate stellar parameters (temperature, surface gravity, metallicity) and other properties, such as foreground dust and distances.

Theoretical libraries are produced by calculations of stellar atmosphere and radiative-transfer processes. Empirical libraries are obtained through observations of real stars. Both have strengths and shortcomings. Theoretical libraries can cover a wide range of stellar parameters and chemical-abundance pattern variations, including even those kinds of stars that are not available in the Milky Way. Theoretical spectra can cover wavelength ranges inaccessible to observations, without noise and have nearly unlimited spectral resolution \citep[e.g.,][]{Kurucz79, Lejeune97, Lejeune98, Westera02, Barbuy03, MurphyM04, Zwitter04, Martins05, Munari05, Rodriguez-Merino05, Fremaux06, Coelho05, Coelho07, Coelho14, Leitherer10, Palacios10, Sordo10, Kirby11, deLaverny12, BohlinM17}. However, theoretical libraries are not yet sufficiently realistic. There are many physical effects that are difficult to model across broad spectral range, such as the sphericity, non-local-thermodynamic-equilibrium (non-LTE) effects, line-blanketing, expansion, non-radiative heating, and convection. Furthermore, we do not yet have a complete atomic and molecular line list. Many lines are theoretically computed and do not have empirical lab measurements, and thus they can have incorrect wavelengths and strengths. Current theoretical models are not able to reproduce the observed spectra for some stars.  For example, they cannot yet reproduce all of the observed features in an ultra-high-resolution solar spectrum \citep{Kurucz11}. Therefore, to properly model the observed spectra of external galaxies, we still need to rely on empirical libraries, at least for those stellar types that are not well modeled theoretically. 

On the other hand, current empirical stellar spectral libraries also have serious shortcomings. First, the spectral resolution and wavelength coverage are limited to the capabilities of the instruments used. Secondly, they are more limited in their coverage of stellar-parameter space than theoretical libraries.  

Some empirical libraries target only certain stellar types, others aim to cover a wide range of stellar types. We focus our discussion on the latter as they are more relevant to applications in extragalactic studies. Examples of such libraries that are widely used include \cite{GunnS83}, Pickles \citep{Pickles85, Pickles98}, \cite{Diaz89}, \cite{SilvaC92}, Lick/IDS \citep{Worthey94}, \cite{LanconW00}, STELIB \citep{LeBorgne03}, ELODIE \citep{Soubiran98, PrugnielS01, PrugnielS04, Prugniel07}, INDO-US  \citep{Valdes04}, CaT \citep{Cenarro01}, MILES \citep{Sanchez-Blazquez06,FalconBarroso11}, HST NGSL \citep{Gregg06}, X-Shooter Stellar Library (XSL, \citealt{Chen14}), the NASA Infrared Telescope Facility (IRTF) Library \citep{Rayner09}, and the Extended IRTF library \citep{Villaume17}. 


\begin{deluxetable*}{lrcrc}
\caption{Summary of Current Optical Empirical Libraries (see text for references)}\label{tab:currentlibraries}
\tablehead{
\colhead{Empirical Libraries} & 
\colhead{Number of stars} &
\colhead{Wavelength } &
\colhead{Approx. R} &
\colhead{Number of dwarfs}\\
\colhead{}&\colhead{}&
\colhead{Coverage (\AA)} & 
\colhead{($\lambda/ \Delta \lambda$)} &
\colhead{with $T_{\rm eff}<4200$K}
}
\startdata
MILES & 985 & 3525-7500 & $2100$ & 15  \\
STELIB & 249 & 3200-9500 & $1600$ &3 \\
LICK/IDS & 425 & 4100-6300 & $500$ & 1 \\
INDO-US & 1273 & 3460-9464 & 5000 & 1  \\
ELODIE & 1388 &4100-6800 & 50000 & 4  \\
HST-NGSL & 374 & 1675-10250 & $1000$ & 9  \\
X-shooter Library & 668 (237 in DR1) & 0.3 - 2.5 ${\rm \mu m}$ & 10000 & $25$  \\
IRTF Library & 210 & 0.8 - 2.5 (5) ${\rm \mu m}$ & $2000$ & $\sim27$ \\
Extended IRTF Library & 284\tablenotemark{a} & 0.7-2.5 ${\rm \mu m}$ & $2000$ & 7\tablenotemark{a}  \\
MaStar & $>10,000$ & 3622-10354 & $1800$ & Hundreds \\
\enddata

\tablenotetext{a}{These are in addition to the numbers of the IRTF Library.}
\end{deluxetable*}

The most severe limitation of all these empirical libraries is their lack of adequate coverage of the stellar-parameter space (\teff, \logg, [Fe/H], and [${\rm \alpha}$/Fe]). Naturally, we are limited by the kind of stars and abundance patterns available within the Solar neighborhood, the Milky Way galaxy, and its satellites. However, even within stellar types and abundance patterns available in the Milky Way, the coverage is quite incomplete. There is much room for improvement, particularly for low-metallicity stars, cool dwarfs, and cool giants, in particular C- and O-stars along the thermally-pulsating asymptotic giant branch (TP-AGB) phase. In addition, previous libraries are often limited to relatively bright stars, which means they are relatively close to the Sun and have smaller abundance-pattern variations. By pushing the observations to fainter magnitudes, we could sample a larger portion of the Milky Way and sample a greater variation of abundance patterns, in particular a wide range of [${\rm \alpha}$/Fe] at fixed [Fe/H]. 

In Table~\ref{tab:currentlibraries}, we summarize the specifications of several widely used libraries. We also list the number of stars in one example part of the parameter space to demonstrate the need for a larger and more inclusive library. For studying stellar populations and modeling the stellar continuum in external galaxies, we need an empirical library of stellar spectra that have sufficient resolution, wide wavelength coverage, and adequate coverage of stellar-parameter space. 

%



One of the state-of-the-art spectroscopic surveys of galaxies is the MaNGA (Mapping Nearby Galaxies at Apache Point Observatory) survey \citep{Bundy15, Yan16b}, which is one of the three main surveys in the 4th generation of the Sloan Digital Sky Survey (SDSS-IV, \citealt{Blanton17}). MaNGA provides spatially-resolved spectroscopy for 10,000 nearby galaxies, covering 3622-10,354\AA\ with a resolving power of $R\sim1800$. To model MaNGA spectra, we need a stellar library with similar wavelength coverage, similar or higher spectral resolution, and including all types of stars that are detectable in an integrated spectrum. For the last point, some stars can be detected at certain wavelengths, e.g., M dwarfs at \naiw, even though they do not contribute significantly to the broadband luminosity. However, no stellar library that can satisfy this need existed at the time when MaNGA started. Currently, the Data Analysis Pipeline of MaNGA (Westfall et al. in prep) uses the MILES library, which stops at 7500\AA. Thus, we are not taking full advantage of the important features (TiO bands, \naiw, FeH Wing-Ford, [SIII], etc.) at the red end of the spectra.

Motivated by the need to model MaNGA galaxy spectra, we have carried out a project called MaNGA Stellar Library (MaStar) to build a large, comprehensive stellar library that satisfies the above requirements. A library that can cover more comprehensive stellar-parameter space must have a bigger sample size and have its stars selected over large areas of the sky. It is observationally expensive to observe these stars one by one. An ideal opportunity is provided in SDSS-IV, in which we can piggyback on the Apache Point Observatory Galaxy Evolution Experiment 2 (APOGEE-2) to observe in parallel a large number of stars over many hundreds of fields in the sky. We describe how this is achieved in detail in Section \ref{sec:observations}. 

Target selection is highly critical for a stellar library, as the primary goal of a library is to cover as wide a range of stellar parameters as possible. For this, we can take advantage of the many existing stellar spectroscopic surveys to preselect our targets according to their measured parameters. However, this is still insufficient, and we have to devise multiple ways to complement the parameter coverage. We describe all of these in Section \ref{sec:targeting}. 

The rest of the paper is organized as follows. Section  
\ref{sec:reduction} describes the data reduction procedure;
Section \ref{sec:quality} presents quality evaluations of the spectra;
Section \ref{sec:parametercoverage} presents an evaluation of the stellar-parameter coverage of the current version of the library and compares it to MILES; and Section \ref{sec:summary} gives the summary. 

This paper provides an overview and a technical summary for the first release of the MaStar Library. We also have a few other papers forthcoming in the near future, including a paper discussing flux calibration issues regarding templates choices (Chen et al., in prep), papers presenting stellar-parameter measurements with different methodologies (Imig et al. in prep, Lazarz et al. in prep, and Meneses-Goytia et al. in prep), and a paper presenting stellar population models based on MaStar (Maraston et al., in prep).

\section{Observations}\label{sec:observations}
SDSS-IV has three major survey components: APOGEE-2, MaNGA, and the Extended Baryon Oscillation Spectroscopic Survey (eBOSS). APOGEE-2 is a medium resolution infrared H-band spectroscopy survey of stars in the Milky Way \citep{Majewski16}. It has a northern component executed with the 2.5-meter Sloan Foundation Telescope \citep{Gunn06} at Apache Point Observatory (APO) (APOGEE-2N) , and a southern component executed with the 2.5-meter du Pont Telescope at Las Campanas Observatory (APOGEE-2S). MaNGA is an integral field spectroscopy survey of 10,000 nearby galaxies \citep{Bundy15,Yan16b, Law15, Wake17}. eBOSS is a spectroscopic survey of galaxies and quasars in the more distant Universe \citep{Dawson16}. All these surveys use a fiber-plug-plate system to conduct observations. MaNGA and eBOSS are both using the Baryon Oscillation Spectroscopic Survey (BOSS) spectrographs \citep{Smee13}, although they are fed with different fiber feed systems. eBOSS uses single fibers in the same way as done in previous generations of SDSS, while MaNGA uses fiber bundles \citep{Drory15}. APOGEE-2 uses the APOGEE infrared spectrograph, with its own set of fibers \citep{Wilson12}. This setup allows for both BOSS and APOGEE spectrographs to collect data simultaneously from different targets. The spectrographs are each fed with different, dedicated fibers, which can share the same focal plane. 


All the fiber assemblies and plates are installed in large cylindrical housings called `cartridges'. These cartridges allow efficient switching of fields during the night observations. During the day, technicians put a plate into each cartridge, and plug the fibers in that cartridge into the plate. At night, the observers only need to mount each cartridge to the telescope sequentially to quickly observe multiple fields.

The MaNGA fiber bundles are installed in six of the nine cartridges that have APOGEE fibers. Thus, we can piggyback on APOGEE-2 as long as these 6 cartridges are used by APOGEE-2 observations. In each cartridge, there are 17 MaNGA science fiber bundles with 12 calibration mini-bundles. Thus, we can observe 17 science targets along with 12 standard stars. This makes our survey efficient at building up large samples of stellar spectra. There are also 92 single sky fibers for sky subtraction purpose. The details of the MaNGA fiber feed system are described by \cite{Drory15}. 



Because MaStar piggybacks on APOGEE-2N, it is necessary to briefly describe the observation strategy of APOGEE-2N \citep{Zasowski13}. APOGEE-2N is a program focused on a survey of red giant stars in the Milky Way. They planned a few hundred fields to be observed from Apache Point Observatory, often with multiple designs for each field. Each design has a different set of targets and is assigned to a different plate. Some of these designs have multiple visits, meaning they will be observed multiple nights with a cadence appropriate for their science goals. The same design could also corresponds to multiple physical plates, with different plate numbers. This is to make it possible to observe the same field at multiple hour angles, with slightly different fiber hole positions corresponding to different corrections for the effect of atmosphere refraction. 

At the end of April 2018, we have had about 550 unique APOGEE-2N designs co-designed with MaStar targets. There are 162 more planned to be designed in the future. These are in a total of 370 fields. Given 17 target stars per design, this means we can expect to observe a total of 12,000 stars, including repeated objects. We would also have observations for a total of 8500 standard stars. 

Because there are three APOGEE cartridges without MaNGA fibers, if a co-designed plate is observed on these APOGEE-only cartridges, MaStar will lose the chance to co-observe. In our observing scheduling, we check existing MaStar data on all the plates requested for each night, and  preferentially put those plates with zero or fewer previous visits on those six shared cartridges so that we can maximize the opportunity to collect MaStar data. 

Usually, each visit by APOGEE-2N is 67 minutes. Within this time, we could do 4x15 minute exposures with the BOSS spectrographs, which have a slightly higher overhead.  We define a `visit' of MaStar to be the set of exposures taken for a given plate on a single night. Each `visit' of a MaStar plate typically consists of up to 4 exposures. 

\begin{figure}
\begin{center}
\includegraphics[width=0.45\textwidth]{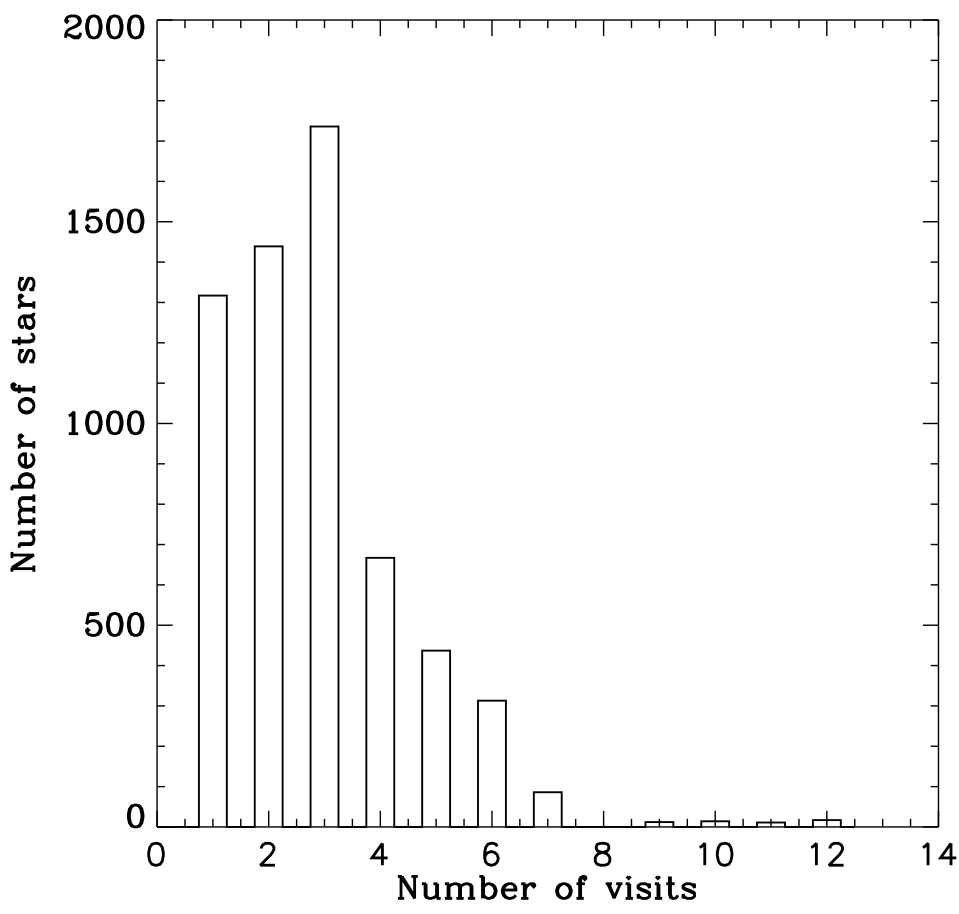}
\caption{Distribution of the number of visits (observations) for all the MaStar targets observed before July 7, 2017.}
\label{fig:visitdistribution}
\end{center}
\end{figure}

Up to Jul 7, 2017 (MJD 57942), which is the cutoff date for SDSS Data Release 15 (DR15), we have obtained 64,309 exposures during 17309 visits for 6042 unique stars. Some stars have a large number of visits. Some stars have a single visit. The distribution of the number of observations for the targets are shown in Fig.~\ref{fig:visitdistribution}.  However, not all of these visits result in high-quality spectra. We report the statistics of the high-quality spectra in Section \ref{sec:summarystat} after introduction of the quality-evaluation procedure. 

\section{Target Selection}\label{sec:targeting}

In this section, we briefly describe the selection of targets for MaStar. We refer the readers to Appendix~\ref{sec:app_target} for more details on this. The goal of the library is to build a sample of stars covering as wide a parameter space as possible. It is not meant to provide unbiased and/or statistical samples. Thus the selection is very different from other stellar spectroscopy surveys. 

\subsection{Magnitude Limits and Isolation Constraints}

In general, we select stars that are brighter than 17.5\ in either the $g$-band or the $i$-band, and fainter than 12.7\ in both the $g$- and $i$-bands to avoid saturation. On a fraction of the plates, we included brighter stars by offsetting the fibers or shortening the exposure time, in order to sample certain parameter space. 

Because we need accurate flux calibration, we select only those stars that are relatively isolated in images. The exact isolation criteria are described in Appendix~\ref{sec:isolationconstraints}. This isolation requirement does not mean we exclude all binary stars. Close binaries that are not photometrically resolved or with large contrast in magnitudes could still be included. And stars in very wide binaries that satisfy our isolation criterion would still be included.

\subsection{Optical Photometry for our Targets}

Our targets are primarily selected from either large stellar-parameter catalogs or large photometry catalogs. To ensure observability and to design plates, we need astrometry and optical photometry information for these stars. The SDSS photometry would be a natural choice. However, the
great majority (about 75\%) of the APOGEE-2N fields are outside the SDSS imaging footprint, and many stars we want to target (especially from the APOGEE stellar-parameter catalogs) are brighter than the saturation limit of SDSS imaging. Therefore, we have to make use of photometry catalogs other than SDSS to obtain astrometry and optical photometry information. 

The Panoramic Survey Telescope and Rapid Response System-1 (Pan-STARRS1 or PS1 \citealt{Chambers16}) provided $grizy$ photometry for the entire sky above Dec$=-30$ deg. We use a customized photometry catalog including the same data as in the public PS1 Data Release 1, but with a slightly different photometric calibration (Edward Schlafly, priv. comm.). The catalog includes all stars with either the $g$- or $i$-band mag brighter than 17.5 mag. The saturation limits are approximately 14.0, 14.4, 14.4, 13.8, and 13.0 mag for $g-,r-,i-,z-$,and $y$-bands, respectively. Therefore, for stars brighter than these limits, we have to resort to other catalogs. 

The American Association of Variable Star Observers (AAVSO) Photometric All-Sky Survey (APASS\footnote{https://www.aavso.org/apass}) provides all-sky BV$gri$ photometry for all stars between 7 and 17 magnitude. We use DR8 of the APASS catalog. 

Combining PS1 and APASS catalogs provides the basis photometry and astrometry system that our targeting is based on. For all stars with known stellar parameters, we first match them to this combined APASS+PS1 catalog, and use those coordinates for targeting. We also make use of this combined catalog to select standard stars, and use PS1 to select empty sky locations to obtain sky spectra for sky subtraction. 

\subsection{General Target Selection Scheme}

Target selection is critical for the success of the stellar library. Our primary goal is to cover as wide a parameter space as possible. There are 4 parameters we consider: $T_{\rm eff}$, $\log g$, [Fe/H], and $[{\rm \alpha}/{\rm Fe}]$. We would like to sample this 4-dimensional parameter space uniformly, and reach to the extreme ends of the distribution. To achieve this goal, we select stars from existing stellar-parameter catalogs, including the APOGEE Stellar Parameter and Chemical Abundance Pipeline (ASPCAP) catalogs \citep{Holtzman15, GarciaPerez16, Holtzman18} from SDSS Data Releases 13 and 14 \citep{SDSSDR13, SDSSDR14}, the Sloan Extension for Galactic Understanding and Exploration (SEGUE, \citealt{Yanny09}) Stellar Parameter Pipeline (SSPP) catalog \citep{Lee08a, Lee08b, AllendePrieto08, AllendePrieto14} from SDSS Data Release 12 \citep{SDSSDR12}, and the Large Sky Area Multi-Object Fiber Spectroscopic Telescope (LAMOST, \citealt{Cui2012, Zhao12, Deng12}) Experiment for Galactic Understanding and Exploration (LEGUE) Data Release 2 AFGK catalog \citep{LAMOSTDR1}. This allows us to efficiently pick the stars we need.

Roughly speaking, our selection scheme is the following. We first applied small systematic shifts to the parameters from these catalogs to put them on a roughly consistent system. We then assigned a selection weight to each star that is inversely proportional to the local density of stars in the parameter space, taking into account the number of chances we could target them. We adjust the weight based on the availability and value of [$\alpha$/Fe] measurements, on our preference among the catalogs, and on the distributon of the already-observed sample. Last, we randomly select the stars with a probability proportional to their adjusted weight. There are a few other practical constraints due to the co-observing arrangement with APOGEE-2N. 

In fields without stars with known stellar parameters, we utilitze photometry to select very hot stars and very cool stars to patch those parameter space. We also have ancillary programs to pick stars for specific regions of the parameter space that are not easily populated by the above selection. 

Our resulting sample covers a very wide range of stellar parameters, with significant oversampling for rare combinations of stellar parameters. 

More details of this selection is described in Appendix~\ref{sec:app_target}. 

\subsection{Selection of Standard Stars}

All of our science targets are observed simultaneously with a set of standard stars. The simultaneous observation is important to correct for short time-scale variations in the atmosphere transparency. This is a critical difference between our library and other stellar libraries. The selection of the standard stars is similar to what we do in the MaNGA survey. More details can be found in Appendix~\ref{sec:app_fluxcal}. 

\section{Data Reduction}\label{sec:reduction}

Data reduction for MaStar is handled by the MaNGA Data Reduction Pipeline (MaNGA DRP; \citealt{Law16}). It has two stages. The first stage, which is referred to as the 2D pipeline, is shared between the MaNGA galaxy program and the MaStar program. It processes the raw data frames to produce a sky-subtracted, flux-calibrated, camera-combined spectrum for every fiber in every exposure. In the second stage of the DRP, the processing of the MaStar data is completely different from that of the galaxy program. For the MaNGA galaxy program, the pipeline turns these individual fiber spectra into data cubes through an image reconstruction process, while for the MaStar program, the pipeline uses these individual fiber spectra to derive the final 1D spectra for the stellar targets.

There have been various updates to the pipeline since SDSS Data Release 14 \citep{SDSSDR14}. Here we first describe some of the relevant updates to the first stage of the pipeline, specifically about flux calibration and line spread function characterization. Then we describe the MaStar-specific reduction in the second stage. 

\subsection{Update to the 2D Pipeline}
\subsubsection{Update to the Flux Calibration}

The flux calibration is done as part of the MANGA 2D pipeline. Thus, it is the same as what is applied to galaxy observations. The general procedure is described in detail by \cite{Yan16}.  There are two major changes to the flux-calibration procedure compared to the version released in DR14. First, we have updated the spectral templates used for standard stars. Secondly, we changed the default calibration curve and the smoothing scale when generating the calibration vector. 

In DR14 and previously, the templates were generated using the SPECTRUM code \citep{GrayO94}, based on the Kurucz model-atmosphere grids \citep{Kurucz79, KuruczA81}. The Kurucz grid used was produced in 2003. The set of templates we used had $T_{\rm eff}$ ranging from 5000K to 7000K with 250K intervals, with log g equal to 4.0 or 4.5, and [Fe/H] ranging from -2.0 to 0.0 with 0.5 dex intervals. These parameters are sufficient to cover late-F stars that are used on galaxy plates. But the lack of $u$-band photometry on most MaStar plates meant some hotter stars are often included as standards. Therefore, we need to add in templates with hotter temperatures. In addition, there have been significant updates to the Solar abundance table, model-atmosphere grids, and atomic and molecular line list, so that an update to the model templates seemed appropriate. 

In DR15 \citep{SDSSDR15}, we have updated the flux calibration templates to use the BOSZ spectral template set made by \cite{BohlinM17}. The BOSZ template is computed using the ATLAS9 model-atmosphere grid \citep{Meszaros12}, which employs the updated Solar abundances from \cite{Asplund05}. This is a significant change relative to previous models. We also adopted a much bigger grid of templates to cover a wider stellar-parameter space. The new grid has $T_{\rm eff}$ ranging from 5000K to 10,000K also with 250K intervals, log g ranging from 3.5 to 5.0 with 0.5 dex intervals, [Fe/H] ranging from -2.5 to 0.5 with 0.25 dex intervals, and with a fixed [${\rm \alpha}$/Fe]=0.0. The inclusion of hotter stars is  necessary for the calibration of MaStar plates, which included hotter stars as calibration standards. The new templates also differ from the previous template in the overall spectral shape and the width of some spectral features. These all have subtle implications for the final flux calibration of the data. In a separate paper, Chen et al. (in prep) will describe these effects and the evidence that the BOSZ templates are better than the original templates. 

Another major change we made is that we have revised the way to correct for high-frequency variation in the spectrophotometry. This helps improve the accuracy of the high-frequency correction so that it does not introduce artificial wiggles, even at the 1-2\% levels. 
The calibration step involves two calibration curves. One is the default calibration curve that is applied to all spectra. The other is a per-exposure calibration curve that makes an exposure-specific correction on top of the default curve. 

In DR14 and previous releases, both the default calibration curves and the per-exposure curves are derived using a bspline fit, with break points spaced every 10 pixels in the blue and every 1.5 pixels in the red. 
This yielded some high-frequency wiggles. The high-frequency information is necessary for the telluric absorption correction, but is unnecessary outside the telluric bands. In addition, the default calibration vector includes some artificial wiggles due to slight template mismatches. It also contained the telluric features. This could make the residual per-exposure correction harder to fit when the observed telluric features do not fully match the telluric feature in the default curve. 

In DR15, we have revised both the derivation of the default calibration curve and the treatment of the per-exposure calibration curve. We first derived a new version of the default calibration curve, using a much larger dataset selected to have the best template-matching, high S/N, and observed under the most typical conditions without significant extinction. The curve is derived using a bspline fit, with break points spaced every 10 pixels in both the blue camera and the red camera outside telluric regions. We bridge the telluric regions with smooth curves, so that the default curve does not contain any telluric features. 
Then we apply this default calibration curve for a first-order correction on the data. On top of this, individual exposures can have different large-scale variations due to atmosphere transparency variations, and high-frequency telluric absorptions. The per-exposure calibration curve is allowed to vary at high frequency only in the telluric regions, but not outside them. Inside the telluric regions, we use a bspline with breakpoint spaced every pixel. Outside the telluric regions, we use a bspline with breakpoints spaced every 160 pixels. 

\subsubsection{Update to the Characterization of the Line Spread Function}\label{sec:pipelineLSF}
The line spread function (LSF) describes the broadening profile produced by the instrument in the dispersion direction, given a delta function as input. Compared to the DR14 pipeline, we have significantly improved the accuracy of the line spread function estimates for the data. Based on the MaStar data, we discovered that the spectral resolution at the very blue end of the wavelength coverage is in fact better than what the DR14 pipeline reports. The change is due to a problematic line list used by the pipeline at the very blue end, which dates back to the early days of SDSS-III. After updating the line list, we achieved a much better fit to the line width of the arc lines as a function of wavelength. The resulting line spread function has been verified empirically. We observed the star HD 37828 with very short exposures. Comparing the MaStar spectrum of this star with the high-resolution spectra available from XSHOOTER Stellar Library \citep{Chen14}, we empirically derived the instrumental broadening as a function of wavelength. This matches very well to the LSF provided by our pipeline. More details of this LSF update will be described by Law et al. (in prep). 

The LSF as a function of wavelength are provided for each spectrum in the final summary file for spectra. We provide two versions of it: DISP and PREDISP. The DISP array contains the $\sigma$ (in Angstroms) for a Gaussian fit to the pixel-convolved LSF. The PREDISP array contains the $\sigma$ (in Angstroms) for a Gaussian fit to the LSF before integrating over the size of each pixel. Which version one should use depends on whether the user's software includes the effect of pixel integration. 

In order to make use of the spectra in spectral fitting, one needs to convolve it with a varying-width kernel with wavelength. Here is the procedure we recommend for dealing with this issue. 
\begin{enumerate}
    \item Compare the resolution vector of the template with the resolution vector of the data to be fitted. For our spectra which have wavelength sampling spaced evenly in logarithmic space, it is most convenient to express the quadratic difference in resolution in velocity units.
    \item Build a sequence of kernels with widths covering the range from the smallest resolution difference to the largest resolution difference. 
    \item Convolve the template spectrum with all the kernels and store the resulting flux in a 2D array with one dimension in wavelength and the other dimension corresponding to the different kernel widths.
    \item Given the resolution difference at each wavelength, interpolate the 2D flux array along the kernel-width dimension. This will yield a convolved spectra with a wavelength-dependent convolution kernel.
\end{enumerate} 

\subsection{2nd Stage: MaStar-specific Reduction}

\subsubsection{Aperture Correction}\label{sec:aperturecorrection}
The MaStar-specific reduction includes the following steps. It uses the camera-combined spectra from the previous stage, then employs the flux ratios between the central fiber and its surrounding fibers to determine the exact position of the star relative to the fiber aperture, to better than 0.1\arcsec\ accuracy. This is needed because the 2\arcsec\ fibers do not fully cover the flux in a point spread function. We use the guider images to measure the in-focus point spread function (PSF) at 5400\AA, then adjust it for other wavelengths and different positions on the plate in the same way as described by \cite{Yan16}. This information is used along with the position of the star relative to the fiber to derive an aperture-covering fraction for the central fiber as a function of wavelength. We then divide the flux in the central fiber by this aperture-covering fraction to obtain the total flux in the PSF. This procedure is very similar to that of the flux calibration, as described by \cite{Yan16}. The only differences here are that we do not make use of a model spectrum in the derivation of the flux ratios, and that the flux ratios are not derived in large wavelength windows. In this case, we derive the flux ratios at each wavelength and then fit them directly with the PSF models to search for the relative positioning. 

This is a key difference between our data and the previous generation of stellar spectroscopy from SDSS, SEGUE, and ancillary stellar programs done as part of SDSS and BOSS. Previously, all those observations were done with single fibers. Because those single fibers do not cover the PSF completely, especially when the seeing is poor, they lose a fraction of the flux as a function of wavelength. The pipeline attempts to correct for those fluxes, but the corrections are not perfect due to the unknown positioning of the stars relative to the fiber aperture. Due to mechanical drilling error, fiber centering error within the metal ferrule, the designed mechanical tolerance between the fiber ferrule and the plate hole, the guiding error, and the distortion due to the atmosphere refraction, there can be a significant amount of offset between the center of star PSF and the center of the fiber, which is also different for each fiber. Part of the offset is also unknown. The SDSS and BOSS pipeline derives corrections as a smooth function of position on the plate and as a low-order function of wavelength, but these corrections cannot be perfect, as the fiber misalignment is not a smooth function of the positions on the plate. The final calibration is better in SDSS-I and -II than in SDSS-III/BOSS, as the former used 3\arcsec\ fibers while BOSS used 2\arcsec\ fibers. 
As shown by \cite{SDSSDR6}, for SDSS-I and -II, post-DR6, the difference between the synthetic magnitudes from the spectra and the PSF magnitudes from photometry shows an RMS of 0.05 mag in the $r$-band, with 14.3\% of the stars having errors larger than $3\sigma$. For the $g-r$ color, the RMS is 0.05 mag, with 8.0\% distributed beyond 3$\sigma$. For the $r-i$ color, the RMS is 0.03 mag, with 8.2\% distributed beyond $3\sigma$. For SDSS-III/BOSS, the RMS difference is 0.058 mag in $r$-band, 0.063 mag in $g-r$ color, and 0.035 mag in $r-i$ color \citep{Dawson13}, with no statistics given for the fraction of stars in the tail of the distribution. As we demonstrate below, our method can achieve a better calibration accuracy than both of these, and reduce the fraction of stars in the tails of the distribution. 

The software is included in the {\it pro/spec3d} folder of the {\it mangadrp} package, and is released as part of DR15. 

\subsubsection{Radial Velocity Correction}

We fit each spectrum with BOSZ templates to determine the best radial velocity. Before fitting, we convolve each of the templates to the spectral resolution of the spectrum to be fitted. The resolution is a function of wavelength for each spectrum. In order to have a varying convolution kernel with wavelength, and to save computation time, we convolve each template with a number of kernels with different widths, and store the results in a 2d array with each row, giving the convolved template at a different resolution. Then for each spectrum, we only need to do an interpolation in the 2d array of each template to obtain the appropriate template convolved with a varying LSF kernel. 

The radial velocity determination is done in four steps. First, without knowing the velocity, we smooth the spectra heavily to find a subset of best-fit templates (10\% of all templates). Secondly, we use the best-fit template to get a first guess of velocity. Thirdly, we use this new velocity to redo the search for the best-fit template among the subset identified above. Finally, we use the new best-fit template to refine the velocity measurement. In all these steps, the fitting is performed on the continuum-normalized spectrum, obtained by dividing the original spectrum by a smoothed version using a sliding window of 599 pixels in width. The same continuum normalization is applied to all of the theoretical templates. In the first step above, in order to minimize the impact of radial velocity mismatches, we further smooth both the continuum-normalized spectrum and the continuum-normalized models using a sliding window of 31 pixels in width (about 34\AA\ in the blue and 43\AA\ in the red). To derive the velocity, we shift the logarithmically-sampled spectra one pixel at a time, and search for the optimal shift, which can be a non-integer, that would minimize $\chi^2$. 

The templates are the same as employed for the flux calibration, which are limited to $T_{\rm eff}$ of 5000-10,000K, and $\log g$ of 3.5-5.0. Therefore, for some stellar types, such as very cool stars and white dwarfs, these templates cannot provide a good fit. These stars are flagged during the radial velocity search; they can be identified because their $\chi^2$ variations with velocity do not have nice, Gaussian-like troughs far away from the boundary values. These stars are flagged with a nonzero and negative error code (``V\_ERRCODE'').

After obtaining the radial velocity from multiple exposures, we check the consistency among them. For each plate-IFU combination, we select only those good exposures that do not have the LOWCOV bit (bit 4) set in the exposure quality flag (``EXPQUAL'', see Section~\ref{sec:qualityflagging}) and do not have a non-zero error code resulting from the velocity fitting. If there are more than one good exposure, we compute the sigma-clipped mean velocity and the standard deviation among all velocities, and set these as the final velocity and the velocity error. If there is only one good exposure, the only measurement is taken as the final velocity, and the velocity error is set to 999.0. If there is no good exposure for a given plate-IFU, then we set the final velocity to 0.0 and the error to 999.0. Since this procedure is followed for each plate-IFU, all visits associated with a single plate-IFU have the same velocity and error. 

If any of the exposures have a non-zero error code, or if the standard deviation among the velocities from those good exposures are larger than 10 km/s, we set the final error code (``V\_ERRCODE'') to 1 and flip the BADHELIORV bit (bit 6) in the MJDQUAL bitmask (see Section~\ref{sec:qualityflagging}) for all visits of the plate-IFU. These spectra should be excluded when building stellar population synthesis models with this library, unless one re-derives the velocities for them separately.

For those plate-IFUs that have only one good exposures and do not have a non-zero error code in fitting other exposures, we do not set their final error code or the BADHELIORV bit, as we do not know whether the velocity is reliable or not. The users may want to exclude these cases as well to be conservative. They can be identified by having {VERR=999.0}. 

This velocity is derived from the camera-combined spectra that are sampled on a wavelength grid evenly spaced in $\log \lambda$. These spectra are already the result of an interpolation from the native wavelength sampling. Correcting the radial velocity to put spectra into the restframe would involve another interpolation. To reduce the number of times we interpolate, we go back to the sky-subtracted and flux-calibrated spectra with the original native wavelength sampling, shift the native wavelength grid by the derived radial velocity, then resample them to the logarithmically-spaced wavelength grid.  The spectrum will now be in the restframe, and we only interpolate once to maintain the high fidelity of the data. 

All spectra have been corrected to restframe according to the reported heliocentric velocity, unless the latter is zero. This includes those cases where we deemed the velocity to be unreliable, as having either larger than 10~km/s errors or having a non-zero error code.

The version released in DR15 used a limited set of spectral templates when deriving the radial velocity. The set is limited to $5,000{\rm K} \leq T_{\rm eff} \leq 10,000 {\rm K}$ and $3.5 \leq \log g \leq 5$ . This results in a poor match between the templates and some stellar types, such as very cool and very hot stars. The radial velocity for these stars can be inaccurate. These spectra can be identified with the BADHELIOV bit (bit 6) in the MJDQUAL bitmask. 

\subsubsection{Stacking of Spectra}

We usually take 3-4 exposures on each night for a plate. A plate can also get several nights of observations, which we refer to as `visits'. We combine the spectra from multiple exposures together to form these stacked per-visit spectra. We first take out the low-order broadband shape difference between the spectra from multiple exposures during the same visit. The difference between each spectrum and the inverse-variance-weighted average spectrum is fitted with a 4th-order polynomial. We divide each spectrum by its corresponding 4th-order polynomial so that they all agree with each other on the broadband shape. We then do a bspline fitting among all of the flux points, taking into account the inverse variance. This is the combined nightly spectrum. All exposures on the same night for a plate usually have very similar line spread functions. We average the line spread functions among the spectra and assign that to the per-visit spectrum. The per-visit spectrum of all stars are collected together in the master file called ``MaStar-goodspec-(versions).fits.gz''. There can be multiple entries for each star in this file corresponding to the multiple visits during which it has been observed. 

\subsubsection{Quality Flagging of Spectra}\label{sec:qualityflagging}

The final spectra files contain a quality flag to provide information on issues encountered in data reduction and post-reduction quality assessment. For the per-exposure spectrum, this flag is called ``EXPQUAL''; for the per-visit stacked spectrum, this flag is called ``MJDQUAL''; for each plate-IFU combination, we also have ``MSTRQUA'' in the header of the file. These quality flags all follow the MASTAR\_QUAL bitmask scheme, which is listed in detail in the Appendix. The MJDQUAL is a bitwise AND product of all the EXPQUAL bitmasks for this star from all the exposures taken during the given visit. The MSTRQUAL is a bitwise OR product of all the EXPQUAL bitmasks for this star from all the exposures taken for this plate. 

Here we describe the meanings of the different bits and how they are set.

EXPQUAL provides the quality of a spectrum derived from individual exposures. Bits NODATA (bit 0), SKYSUBBAD (bit 1), HIGHSCATT (bit 2), BADFLUX (bit 3), and LOWCOV (bit 4) are all set individually for each exposure and each IFU. 
Here SKYSUBBAD and HIGHSCATT are both set by lower levels of the pipeline and are set on a per-exposure, per-camera basis. LOWCOV is set in the aperture-correction step for MaStar-specific reduction. We set this bit if the median aperture covering fraction for the brightest, unsaturated fiber is less than 10\%. Such cases usually happen only when the center fiber is saturated, in which case a peripheral fiber with very low aperture covering fraction becomes the brightest unsaturated fiber. For example, if its covering fraction is less than 10\%, then this bit would be set. Bit NODATA (bit 0) and BADFLUX (bit 3) are never set by the pipeline.

The POORCAL bit (bit 5) is designed to capture those cases where the flux-calibration procedure yields a grossly-incorrect calibration vector due to an issue related with how we deal with dust extinction for standard stars. In our flux-calibration step, we assume the standards are behind all of the dust measured by the SFD dust map. We apply the SFD extinction on the models and compare the observed spectra with the extincted models to derive the calibration vector. This assumption holds for most high Galactic latitude fields, but they fail at low Galactic latitudes or regions with a significant amount of dust distributed far away from us, behind the stars. In those regions, the standard stars could have less extinction in front of them than specified by the SFD dust map. In this case, we would over-redden the models and derive a throughput curve that is too blue. To identify these problems, we compare the shape of the derived throughput curve with our expectations, based on regions at high Galactic latitudes, to identify those data that are affected. For each exposure, we derive a normalized blue throughput at 4354.1\AA\ by normalizing the blue-camera throughput curve at 6200.1\AA\ (in the dichroic region) to 0.08, and a normalized red throughput at 8687.6\AA\ by normalizing the red-camera throughput curve at 6200.1\AA\ to 0.18. We average the normalized blue and red throughputs among all exposures on the spectrograph that the IFU belongs to. We then take the ratio between the average blue ratio and the average red ratio. If the final ratio is above 0.94 for Spectrograph 1,  or 0.80 for spectrograph 2, we flag this bit in the bitmask for all exposures (EXPQUAL) and all mjds (MJDQUAL) for that IFU. Since flux-calibration vectors are derived per spectrograph, this flag is set together for all IFUs on the same spectrograph and on the same plate. If this bit is flagged, then the spectra are excluded from the primary set released in DR15 of the MaStar library. We will correct for this problem in a future version of the reduction. 

BADHELIORV (bit 6) is set on a per-IFU basis. We derive radial velocities for each exposure as mentioned above. (Exposures with LOWCOV bit set are not run through the radial velocity determination). If the radial velocities from all the exposures have a standard deviation greater than 10~km/s, or if any good (LOWCOV not set) exposures failed in the derivation of the radial velocity (errcode $< 0$), then we set the BADHELIORV flag for all exposures and all MJDs of this IFU. 

The MANUAL bit (bit 7) and the EMLINE bit (bit 8) are set based on visual inspection. We describe these inspections separately in the next section. 

The LOWSN bit (bit 9) is flagged if the median S/N per pixel in the per-visit spectrum is less than 15. This bit is set at the per-MJD level, but we propagate it down to EXPQUAL to ensure consistency between EXPQUAL and MJDQUAL. 

The bits NODATA (bit 0), BADFLUX (bit 3), and CRITICAL (bit 30) are originally designed for MaNGA galaxy plates. They are currently not used for MaStar reductions,  thus they can be safely ignored. 

The MJDQUAL flag is the AND product of all EXPQUAL that contributes to the stacked spectrum. Each bit in MJDQUAL will be set only if all the exposures on that MJD has that bit set. This is an optimistic approach, assuming the best-quality exposures would dominate in the stacked spectrum. In the header of the file, we also include MSTRQUAL which is the OR product. In other words, each bit of MSTRQUAL will be set if at least one exposure on that MJD has that bit set. This is a conservative approach. 

We select the subset of visits with good-quality spectra as the primary product for the current release of the library. This 
excludes those spectra affected by extinction issues in their flux calibration, those that are marked as bad in visual inspection, or those with median S/N per pixel less than 15. This includes objects that contain emission lines, objects that have unreliable radial velocity measurements, and objects that may be affected by scattered light. We refer to this subset of visits as `Good Visits' and their associated spectra as `Good Spectra'. We refer to the subset of stars that has at least one good visit as `Good Stars'.

We include a full catalog for all the stars observed and all of the visits. But for all science purposes, we strongly advise the user to use the good-quality subsets.  

\subsubsection{Visual Inspection}
In order to ensure the quality of the library spectra released, we have visually inspected all the per-visit spectra that do not have either POORCAL or LOWSN bit set. Using the Zooniverse interface, we organized a campaign within the collaboration to visually inspect all the spectra to check for quality issues. With 28 volunteers, we inspected all 10,797 per-visit spectra, each by at least 3 volunteers. 
The inspectors are first asked whether the spectrum is free of problems. If they answered no, they were also asked to identify the specific issue. The options include large flux calibration discrepancies between individual exposures, red upturn, poor sky subtraction, poor telluric correction, and other catastrophic reduction issues. The median decision among the 3 or more inspectors is adopted as the decision for whether a per-visit spectrum is considered bad and has the MANUAL bit set in the quality bitmasks (EXPQUAL, MJDQUAL, and MSTRQUAL). If it is set, that means the spectrum has one of the above problems. This bit is also propagated down to EXPQUAL for consistency. 

The visual inspection also identified objects containing emission lines as a byproduct. These could include  flaring M dwarfs or stars embedded in HII regions and planetary nebula. Objects having emission lines are masked as 'EMLINE' in their per-visit spectrum. This bit is propagated down to EXPQUAL for consistency. Since this is done as a byproduct of the visual inspection rather than by an automated code, the identification of spectra containing emission lines may be incomplete.

The visual-inspection step is performed after an initial run of the data reduction pipeline. The resulting spectra are then staged for inspection. The results from visual inspection are collated into two Yanny-style par files called ``bogey\_mastar.par'' and ``emline\_mastar.par''. We then rerun the whole pipeline, which incorporates the info from these files and flag the MANUAL and EMLINE bits in the EXPQUAL and MJDQUAL bitmasks.

\subsection{Summary Files}\label{sec:summarystat}

After the above flagging procedure to ensure good spectral quality, we arrive at a sample of 3321 unique stars with 8646 per-visit spectra. 

We make a summary file called mastarall-[DRPVER]-[MPROCVER].fits to summarize the various metadata associated with each unique MaNGAID object and each per-visit spectrum. This file contains multiple extensions. The first extension (named GOODSTARS) contains all the Good Stars with unique MaNGAIDs that have at least one good per-visit spectrum. In this table, we have included the version numbers of the reduction pipelines, the version of the post-processing pipeline that collates the information, the MaNGAID, the range of observing period, the number of visits and the number of plates each target is on, the astrometry, the info from the input photometry catalog, and the input stellar parameters if available. We also included the MNGTARG2 bitmask to help identify the source of the astrometry and photometry. Five of the stars among the 3321 unique stars have had two MaNGAIDs to be associated with them. Thus, our summary file for the stars have 3326 entries, corresponding to 3326 unique MaNGAIDs. The GOODSTARS table and the GOODVISITS table are also available in their entirety in the electronic edition of the {\it Astrophysical Journal}. We show the format of these tables in Table~\ref{tab:goodstars} and Table~\ref{tab:goodvisits} in Appendix~\ref{apdx:mastarall}.


The second extension (named GOODVISITS) contains the information associated with all the Good Visits for these Good Stars. If a star has more than one good visit, it will have multiple entries in the GOODVISITS table. The order of entries in this table corresponds exactly to the order of entries in the summary spectra file -- mastar-goodspec-(mangadrp\_version)-(mastarproc\_version).fits.gz . This table includes the versions of the pipelines, the MaNGAID, the plate number, IFU number, MJD of the observation, coordinates of the IFU, coordinates of the star (for the epoch of the expected observation date), the PSF magnitudes, the targeting bitmask, number of exposures taken during the visit, heliocentric velocity and uncertainty, and the quality bitmask. 

The third extension (named ALLSTARS) contains the information for all stars, regardless of whether there are any good-quality visits for them. It follows the same format as the GOODSTARS extension. 

The fourth extension (named ALLVISITS) contains the information for all the visits. It follows the same format as the GOODVISITS extension. 

We also provide a summary spectra file that contains all the high-quality per-visit spectra. For each of the Good Visits, it contains the vacuum wavelength, flux, inverse variance, instrumental dispersion ($\sigma$ of the LSF), and mask in arrays, in addition to all the information given in the 'GOODVISITS' table. The wavelength array for all spectra is logarithmically spaced with $\Delta \log \lambda = 1.e-4$. Note that the spectra are {\it not} corrected for foreground extinction. 


\section{Quality of MaStar spectra}\label{sec:quality}

\subsection{Signal-to-Noise Distribution of the Spectra}

\begin{figure*}
\begin{center}
\includegraphics[width=0.99\textwidth]{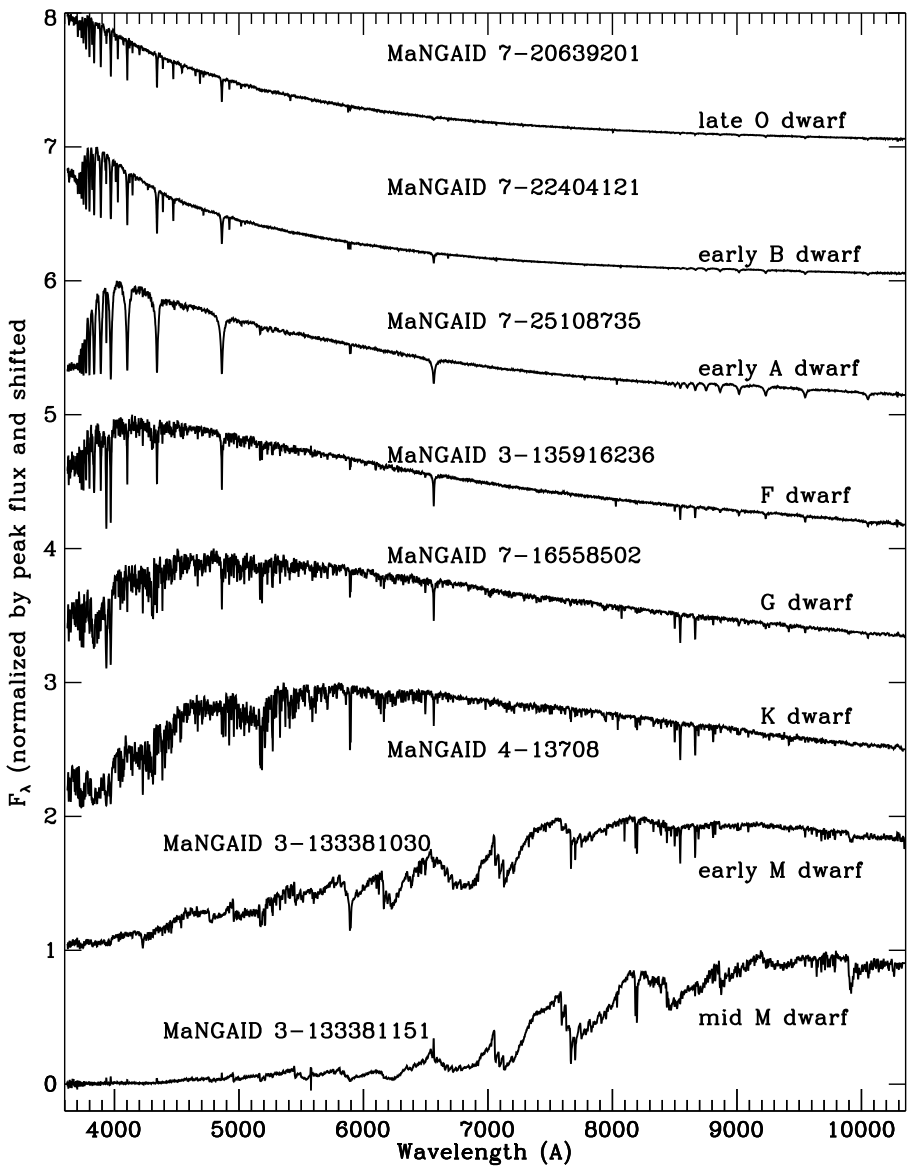}
\caption{Example per-visit spectra for some main sequence stars in the MaStar Library.}
\label{fig:sample1}
\end{center}
\end{figure*}

\begin{figure*}
\begin{center}
\includegraphics[width=0.99\textwidth]{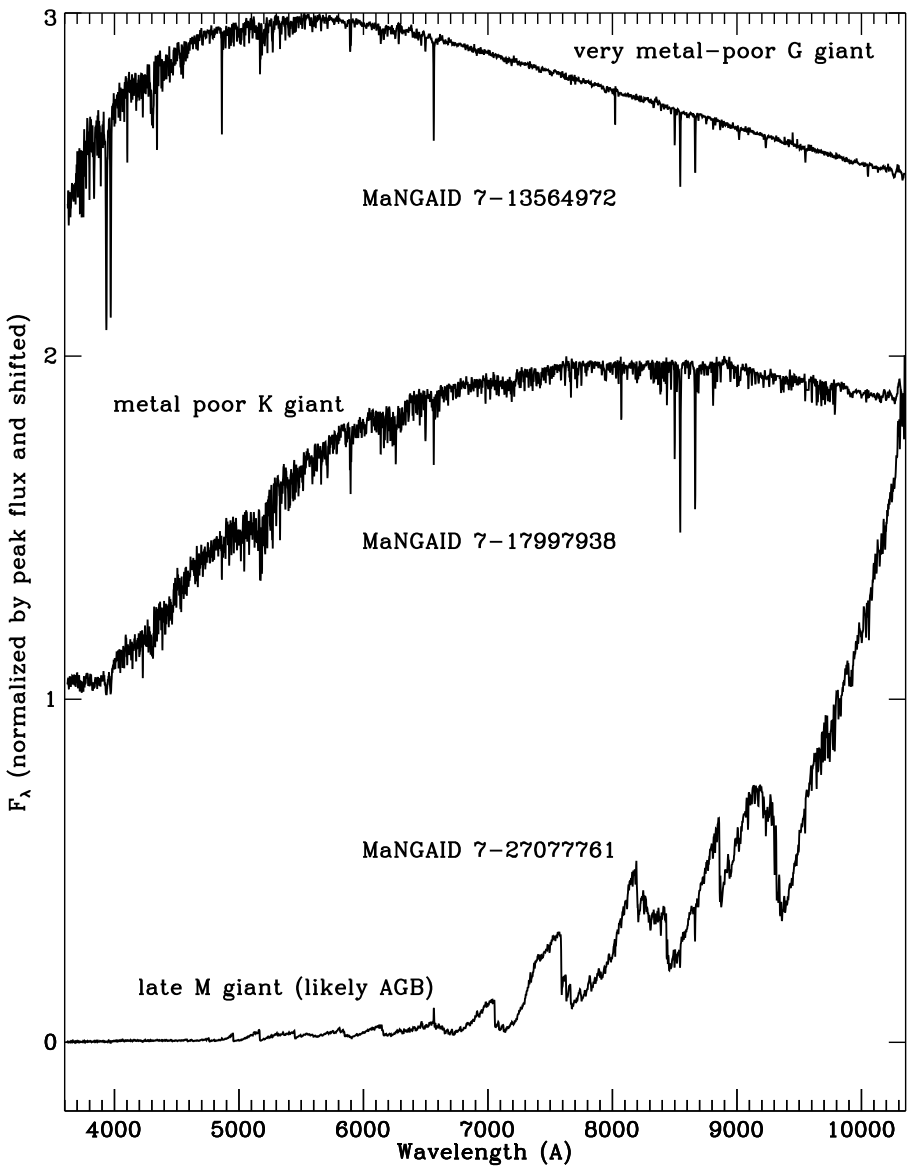}
\caption{Example per-visit spectra for some giant stars in the MaStar Library.}
\label{fig:sample2}
\end{center}
\end{figure*}

We show some example spectra in two figures. Figure~\ref{fig:sample1} shows spectra for a few stars on the main sequence. Figure~\ref{fig:sample2} shows spectra for a few giant stars with different colors. 

\begin{figure}
\begin{center}
\includegraphics[width=0.49\textwidth]{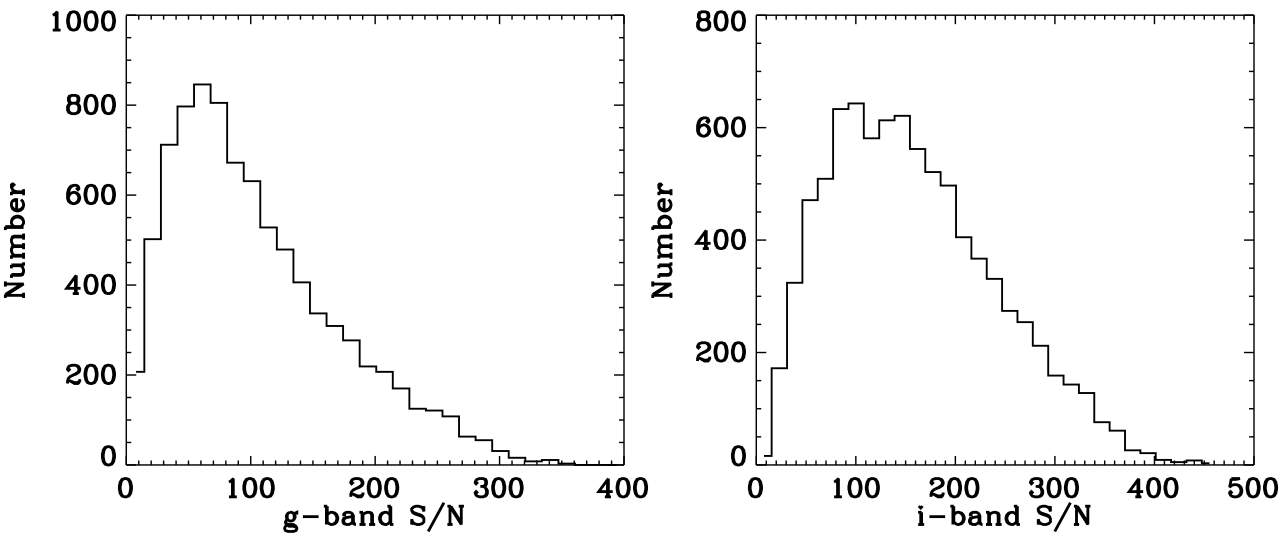}
\caption{Distributions of median S/N per pixel for all the good per-visit spectra in the $g$-band (left) and the $i$-band (right).}
\label{fig:sn_distribution}
\end{center}
\end{figure}

Figure~\ref{fig:sn_distribution} shows the distributions of median S/N per pixel in the $g$-band and $i$-band in each spectrum for all of the good per-visit spectra. The median S/N among all per-visit spectra is 89.7 for the the $g$-band and 148.3 for the the $i$-band. Taking the median S/N per pixel over the entire wavelength window, the median of the median S/N among all per-visit spectra is 113.5, with 87.6\% of the spectra having median S/N per pixel greater than 50. 

\subsection{Flux-Calibration Accuracy}

We evaluate the flux-calibration accuracy in multiple ways. 

First, we evaluate it by measuring the synthetic magnitudes on the observed per-visit spectra and compare them with photometry. Among the 3321 good stars, 1918 have PS1 photometry. However, there are a small fraction of objects for which the PS1 photometry are in error in some of the bands, which can be identified by them being outside the nominal stellar locus on a color-color diagram. We use $g-r$ vs. $r-i$ and $r-i$ vs. $i-z$ color-color diagrams to reject those stars with problematic PS1 photometry. This results in 1592 unique MaNGAIDs with 4242 visits. This selection should not affect our evaluation of flux calibration as the selection is blind to the spectral data. 

\begin{figure*}
\begin{center}
\includegraphics[width=0.9\textwidth]{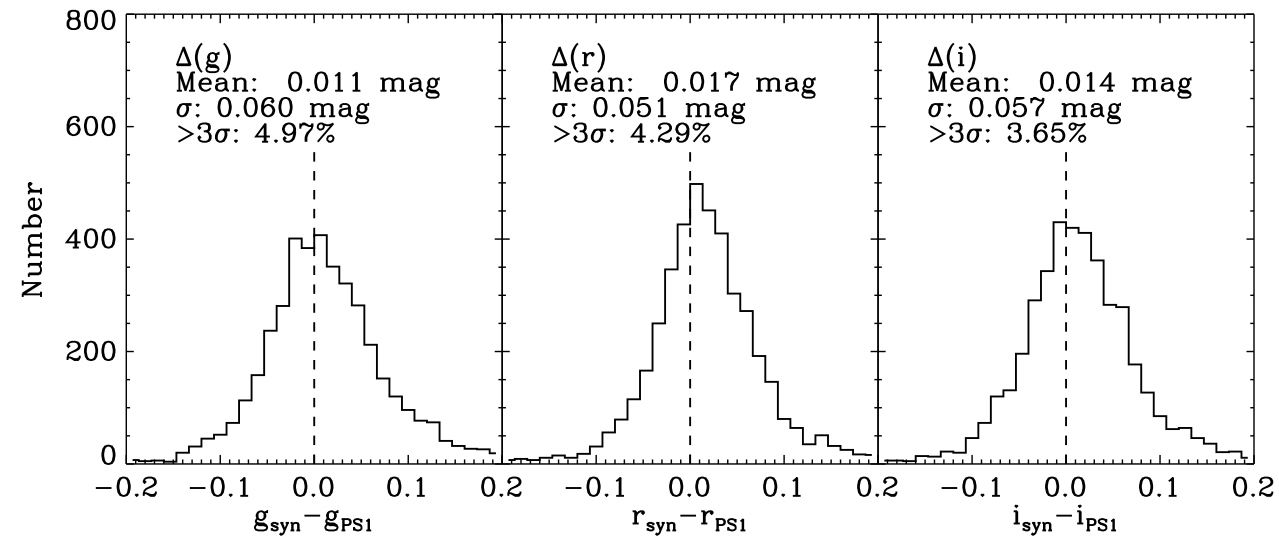}
\includegraphics[width=0.9\textwidth]{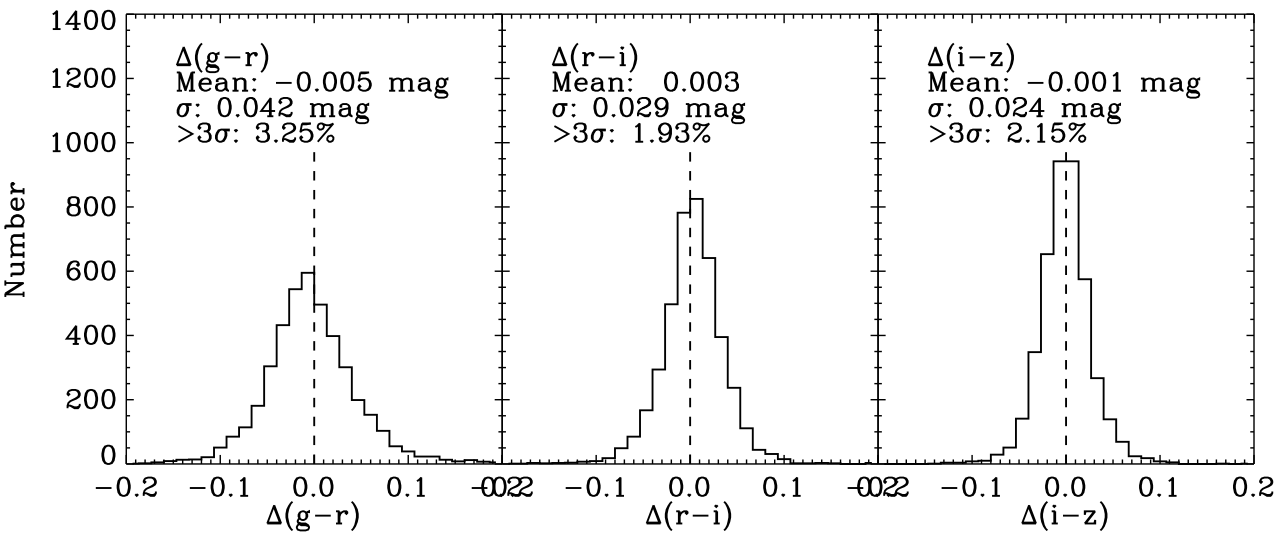}
\caption{Top: histogram showing the difference in the $g$-, $r$-, and $i$-bands between synthetic magnitudes from MaStar and observed magnitudes from PS1 for our per-visit spectra. This shows our absolute calibration is 0.051 mag (4.7\%) in the $r$-band. Bottom: histograms showing the difference in $g-r$, $r-i$, and $i-z$ between synthetic color and the observed color in PS1, also for per-visit spectra. This shows that our relative calibration is 0.42 mag (3.9\%) in $g-r$, 0.029 mag (2.7\%) in $r-i$, and 0.024 mag (2.2\%) in $i-z$. The fraction of outliers beyond 3$\sigma$ are also given in the legend, which are generally around 2-3\% for colors.}
\label{fig:colordiff_hist_ps1}
\end{center}
\end{figure*}

Figure~\ref{fig:colordiff_hist_ps1} shows the distribution of the differences between synthetic magnitudes from per-visit spectra in MaStar and observed magnitudes from PS1, and the distribution of the differences in color. This shows that the absolute calibration is 0.051 mag (or 4.7\%) in the $r$-band, with 4.29\% beyond $\pm3\sigma$, and the relative calibration between broadbands are accurate to 0.042 mag (3.9\%) in $g-r$, with 3.25\% beyond $\pm3\sigma$, 0.029 mag (2.7\%) in $r-i$, with 1.93\% beyond 3$\sigma$, and 0.024 mag (2.2\%) in $i-z$, with 2.15\% beyond 3$\sigma$. Of course, this scatter also includes the uncertainty contribution from PS1, which is about 1\%. Removing that would reduce the numbers slightly. These numbers indicate that both our absolute and relative calibrations are much better than the calibration of SDSS-III/BOSS, which used the same fiber size as we do but single fibers. Compared to SDSS-I and -II which used larger fiber sizes than we do (see their numbers in Section~\ref{sec:aperturecorrection}), we are slightly better in terms of the standard deviations, but are significantly better in terms of the fraction of outliers. Note that one could choose to combine multiple per-visit spectra of each star to reduce the calibration error further and to improve the S/N. 

Secondly, we can use repeated observations to evaluate the stability in flux calibration. For many stars, we have multiple visits (nights of observations). We compare the spectra among the multiple visits by constructing pairs of the spectra for the same star (identified by plate-IFU) observed on different nights. Among all of the good spectra, we have 4508 pairs of these repeated observations. First, we smooth the spectra using a window of 50 pixels so that the difference between the repeats are not significantly contributed by random noise in the spectra. Second, we normalize them by the flux at 5450\AA, and then take ratios between the two spectra in each pair. Third, for each pixel, we take the RMS of all the ratios among pairs in which both spectra have a smoothed S/N greater than 50 at that pixel, and then divide by $\sqrt{2}$ to get the typical uncertainty associated with one spectrum at each wavelength.  

\begin{figure}
\begin{center}
\includegraphics[width=0.49\textwidth]{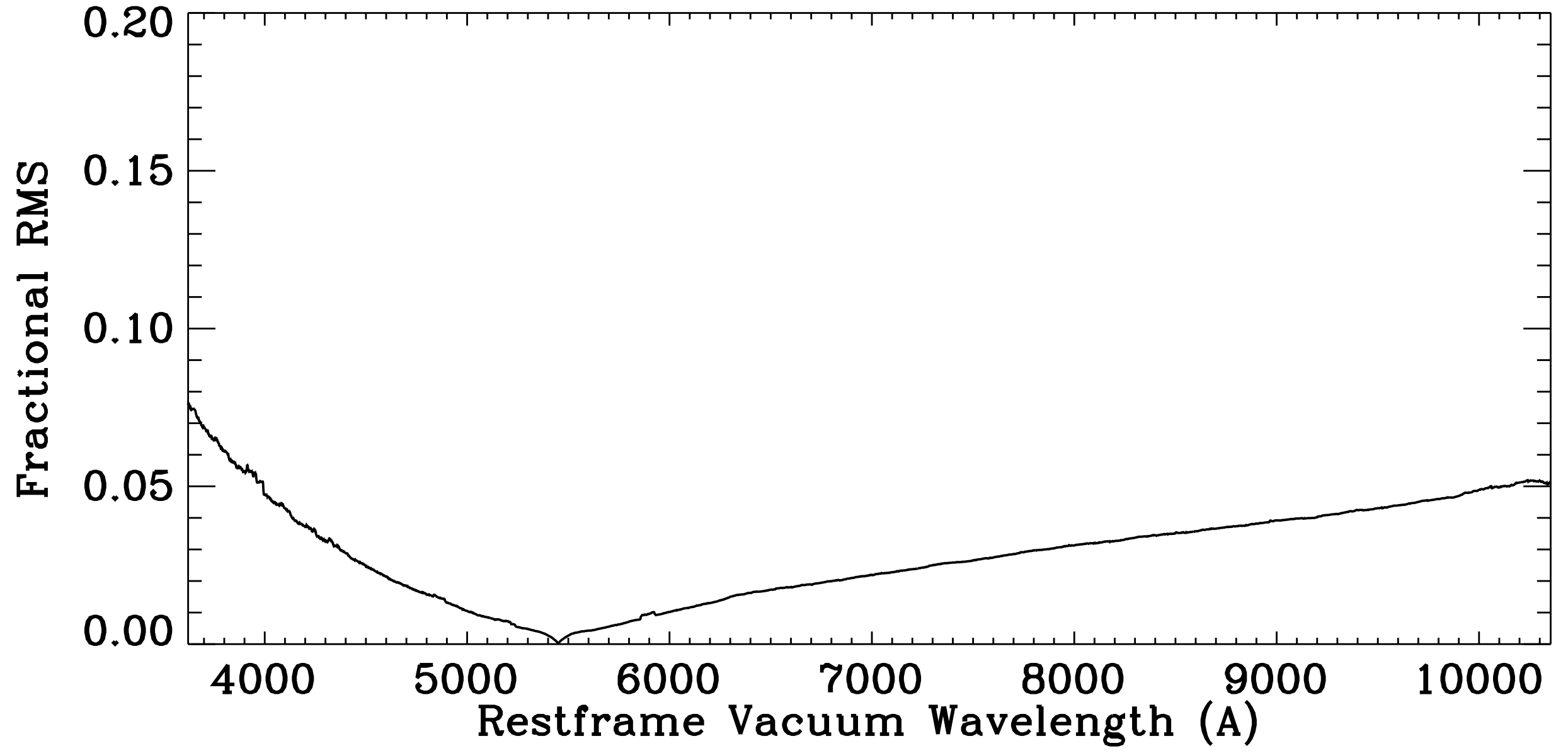}
\caption{Fractional uncertainty in the spectra due to flux-calibration systematics, relative to 5450\AA. This is derived using repeated observations of the same star. See the text for how it is produced.} 
\label{fig:mastarcalrms}
\end{center}
\end{figure}

Figure~\ref{fig:mastarcalrms} shows the resulting fractional uncertainty. This is dominated by systematics due to flux calibration errors. Due to the requirement on the smoothed signal-to-noise ratios, random noise contributes less than 2\% to the fractional uncertainty. The bumps in the spectra (e.g. between 3900\AA\ and 4000\AA) are due to the slightly reduced S/N at the wavelengths of absorption features (such the Ca~II H \& K lines). With respect to 5450\AA, our relative calibrations are better than 5\% between 3993\AA\ and 10,139\AA. The worst calibration is at the extremely blue end and it has a sigma of about 7.6\% relative to 5450\AA. 



\subsection{Radial Velocity Correction Stability and Accuracy}

The radial velocities for the spectra are determined using the BOSZ theoretical templates. In the GOODVISITS table of the {\it mastarall} summary catalog and the {\it mastar-goodspec} file, we provide the heliocentric velocity measured for each plate-IFU combination, the 1-$\sigma$ error of the velocity, and an error code (``V\_ERRCODE''). 

\begin{figure}
\begin{center}
\includegraphics[width=0.49\textwidth]{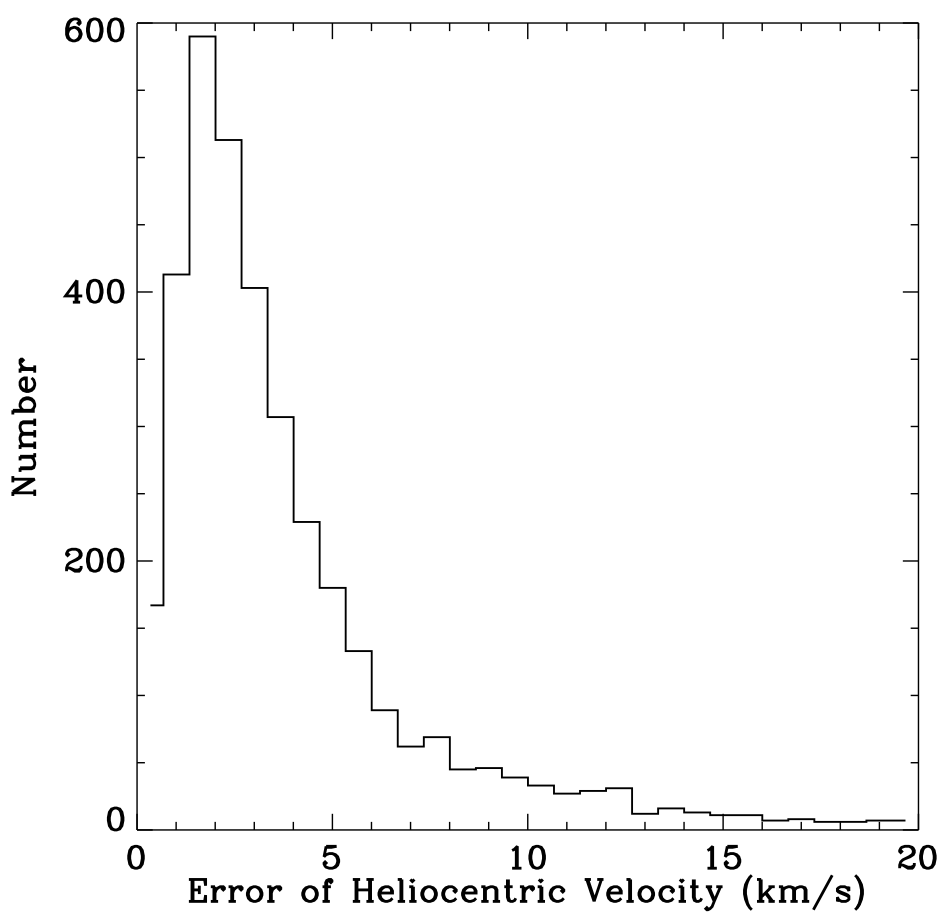}
\caption{Distribution of uncertainties in the measured heliocentric velocities.}
\label{fig:heliovrms}
\end{center}
\end{figure}

Figure~\ref{fig:heliovrms} shows the distribution of the radial velocity uncertainties for all the unique plate-IFUs among the good per-visit spectra. This includes only those that have more than one good exposure for a given plate-IFU, as otherwise we would not be able to assess the velocity error. The median error of heliocentric velocity is 2.9~km/s. 

\begin{figure}
\begin{center}
\includegraphics[width=0.49\textwidth]{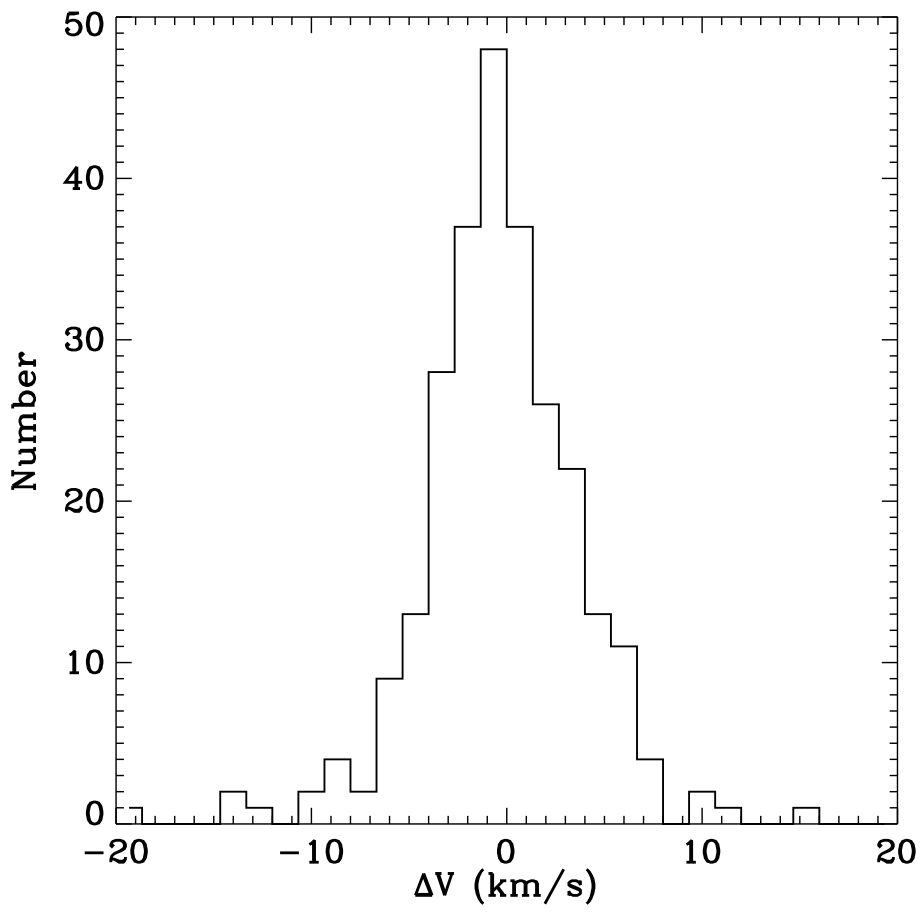}
\caption{Distribution of velocity differences between repeated observations of the same star on different plates.}
\label{fig:heliov_repeats}
\end{center}
\end{figure}

We check the stability of the heliocentric velocities by comparing repeated observations of the same star observed on different plates. We have 271 pairs of repeated observations of the same star on different plates, where both plates yielded a heliocentric velocity with V\_ERROCODE$=0$ and VERR$< 999.0$~km/s. Figure~\ref{fig:heliov_repeats} shows the distribution of the difference in derived velocity between these repeated observations. The distribution has a sigma of 3.5 km/s, which is consistent with the expectation given by the error of the individual measurements. There are a small fraction of cases, 7 out of 271 (2.6\%),  with discrepancies larger than 20~km/s. Some of these could be stars with genuine radial velocity variations, such as members of a binary. We will investigate the cause for these cases in the future, and remove binaries from the final library. 

\begin{figure*}
\begin{center}
\includegraphics[width=0.95\textwidth]{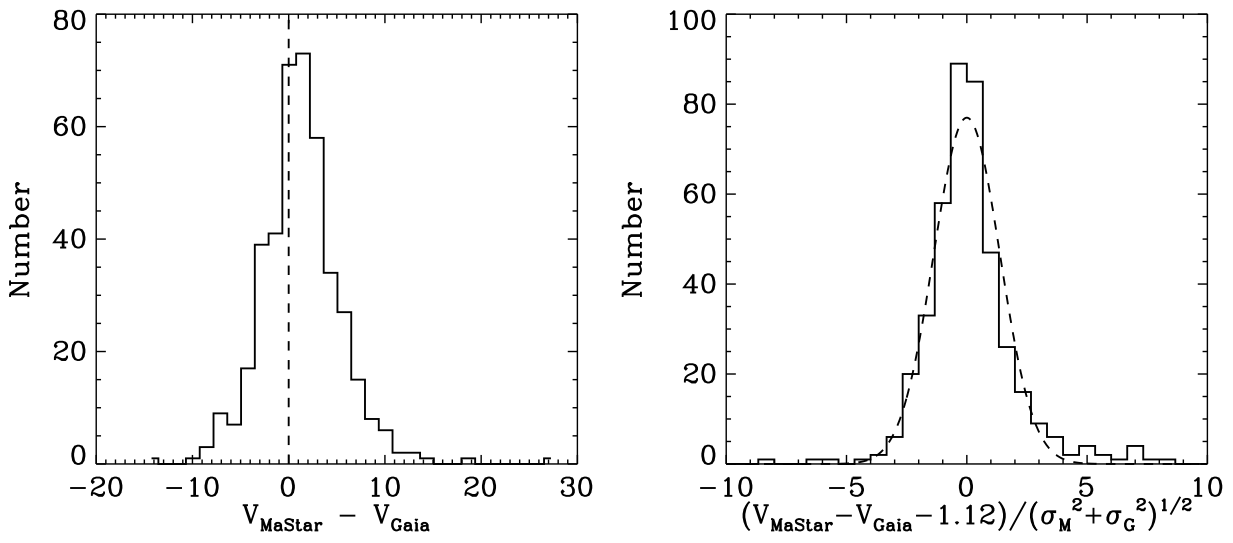}
\caption{Left: Distribution of heliocentric radial velocity differences between MaStar and {\it Gaia} measurements for 417 stars that have good velocities from both sources. There is a 1.12 km/s systematic offset. The standard deviation is 3.68 km/s. Right: Distribution of the velocity difference (after removing the systematic difference) divided by the reported uncertainty. The Gaussian curve has a $\sigma=1.375$. This shows our velocity errors are slightly underestimated. The outliers beyond 3$\sigma$ could be stars with genuine radial velocity variations.}
\label{fig:vcomp_gaia}
\end{center}
\end{figure*}

For a small fraction of our stars, {\it Gaia} DR2 provided radial velocity measurements. Here we compare our measurements against {\it Gaia} DR2. The crossmatching with {\it Gaia} DR2 source catalog is described below in Section~\ref{sec:parametercoverage}. Among the clean {\it Gaia} crossmatches, we have 417 plate-IFUs with {\it Gaia} radial velocity measurements, V\_ERRCODE$=0$, and VERR$<999.0$ km/s. The left panel of Figure~\ref{fig:vcomp_gaia} shows the distribution of the heliocentric velocity difference between our measurements and those from {\it Gaia}. The two sets of measurements are quite consistent. The mean systematic offset for $V_{\rm MaStar}- V_{Gaia}$ is 1.12 km/s. The standard deviation of the difference is 3.68 km/s, with only 1.4\% of the plate-IFUs distributed beyond 3$\sigma$ of the mean. 

This subsample that has {\it Gaia} radial velocity measurements are brighter than most of the MaStar targets. Their $r$-band magnitudes range from 11.5 to 14.8. Thus they tend to have smaller velocity errors than most of the other MaStar targets. We further check whether the difference between our velocity measurements and {\it Gaia}'s are consistent with the reported uncertainty. We combine the errors from both measurements quadratically and then divide the velocity difference by the combined error after removing the 1.12~km/s systematic offset. The resulting distribution has a standard deviation (around zero) of 1.375, as shown in the right panel of Figure~\ref{fig:vcomp_gaia}. This means our velocity error may be slightly underestimated. However, this subsample, especially those outliers (4.6\%), could also include stars with genuine radial velocity variations. From this comparison, we confirm that our radial velocity measurements are largely accurate, with a systematic error on the order of 1 km/s and the reported velocity uncertainty slightly underestimated by $\sim40\%$. If we assume the velocity uncertainty for other stars are underestimated by the same level, then the actual median velocity uncertainty for the whole sample should be around 4 km/s. 

\subsection{Distribution of the Line Spread Function}

The spectral resolution of the MaStar spectra can vary between stars and between different visits of the same star, depending on the exact focus of the spectrograph. It is difficult to maintain a steady focus for spectrographs that ride with the telescope and are exposed to the ambient environment, due to temperature variation and flexure of the instrument. Different fibers in the same spectrograph can also have different resolution, since the focal plane is not completely flat and the CCDs are usually not flat either. It is critical to accurately characterize the resolution as a function of wavelength for each individual spectrum. 

As mentioned in Section~\ref{sec:pipelineLSF}, we provide the detailed spectral resolution information for each spectrum released. Since it varies with wavelength, the resolution is given as a vector with the same dimension as the flux vector. We strongly recommend the users to take the varying resolution into account when using the spectra, using the instructions given in Section~\ref{sec:pipelineLSF}.

\begin{figure}
\begin{center}
\includegraphics[width=0.49\textwidth]{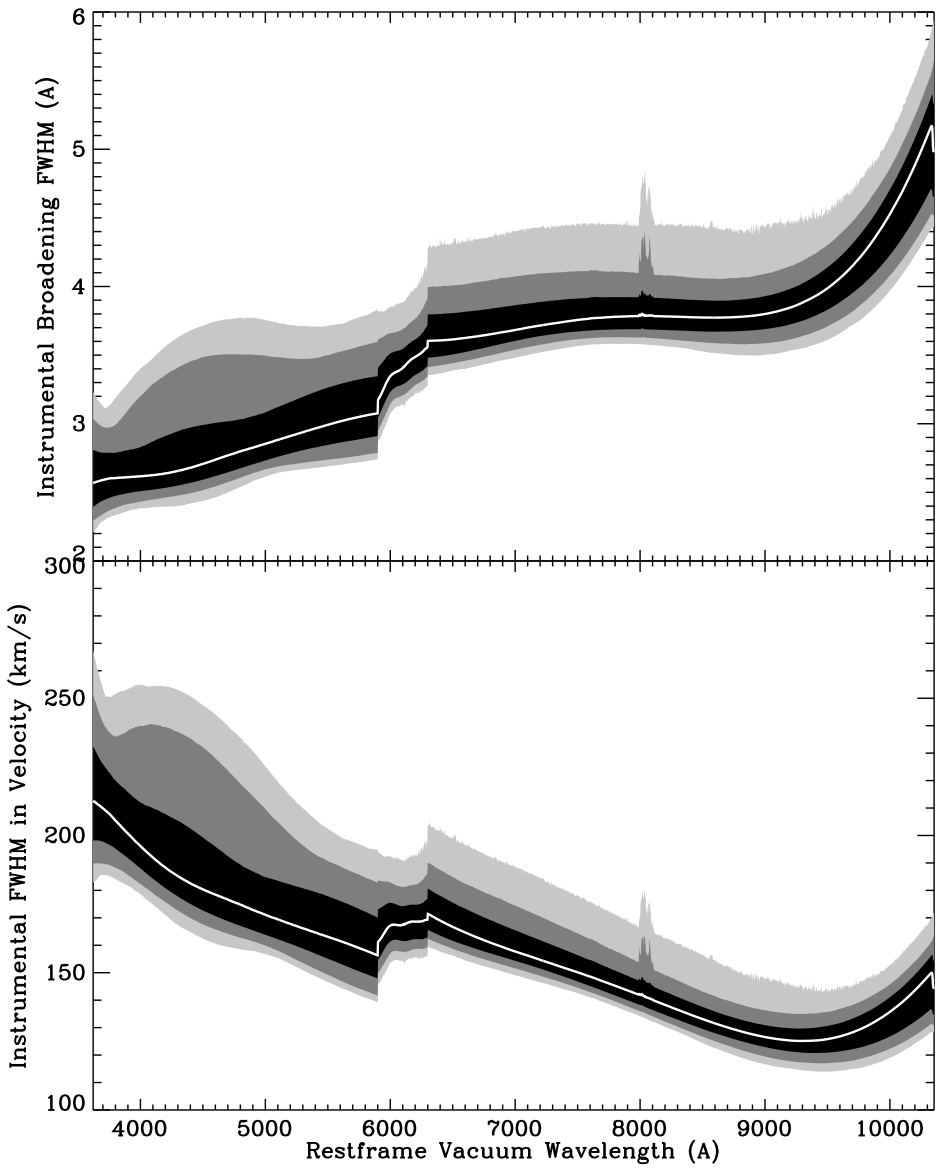}
\caption{Distribution of the spectral resolution as a function of wavelength among all good per-visit spectra, in units of \AA\ (top panel) or km/s (bottom panel). The white lines indicate the median resolution, while the three different gray scales, from black to light grey, indicate zones of 68.3-, 95-, and 99.7-percentiles around the median, respectively}
\label{fig:mastarresolution}
\end{center}
\end{figure}

Figure~\ref{fig:mastarresolution} shows the distribution of spectral resolution as a function of wavelength among all the good visits spectra. 

The variation in spectral resolution is the reason that we are not stacking multiple visits spectra together for the same star. In order to properly stack them, we would have to degrade the spectral resolution to the lowest resolution among the set. This is undesirable for some applications. 

In future SDSS data releases, we will provide versions of the library in which we make all spectra to have the same resolution vector to make  them more convenient to use. 

\section{Stellar-Parameter Coverage}\label{sec:parametercoverage}


In the current version of the library, we do not provide a catalog of associated stellar parameters for all of our stars. We only provide a catalog for a portion of the stars whose parameters are available from APOGEE/APOGEE-2, SEGUE, and LAMOST. These parameters are used as input in our target selection. They are not homogeneously derived, and should be treated with great caution. 

We are in the process of determining stellar parameters for all of the stars, which will primarily be based on our own MaStar spectra. The parameters will be presented in a future publication. 





The availability of {\it Gaia} parallax information makes it possible to provide a rough estimate of our stellar-parameter coverage using color-luminosity diagrams. We matched our library with {\it Gaia} DR2. For the 3321 unique stars (3326 unique MaNGAIDs) with good per-visit spectra, 3318 stars (3323 MaNGAIDs) have one or more matches with {\it Gaia} within 3\arcsec\ at the corresponding epochs for the input catalogs of MaStar. Among these, 3171 stars (3176 MaNGAIDs) have a single match, with the largest angular distance being 2.06\arcsec. For the remaining 147 stars, we check if the match with the shortest angular distance is also the dominant source,\ by computing the contamination fraction using the same algorithm and threshold as described in Section~\ref{sec:isolationconstraints}. This results in an additional list of 26 stars for which other nearby sources are too faint or too distant to matter for practical purposes. In total, we are able to cleanly match 3197 unique stars (3202 MaNGAIDs) with Gaia DR2. 


We adopt the distance estimates provided by \cite{Bailer-Jones18}, which are derived from Gaia parallax and error using Bayesian inference with a weak and purely geometric prior. For the 3197 stars, only 3160 stars (3165 MaNGAIDs) have valid distance estimates from \cite{Bailer-Jones18}.
We then select those stars that satisfy {\it any} of the following criteria so that we could correct their color and magnitudes for foreground extinction. 
\begin{enumerate}
\item Have an E(B-V) value less than 0.1 mag according to the \cite{SchlegelFD98} dust map. 
\item Be more than 300~pc above or below the midplane of the Milky Way, so that most of the dust measured by the \cite{SchlegelFD98} map should be in front of the star. 
\item Have a distance less than 100~pc.
\end{enumerate}

The motivation for the first criterion is that, if the extinction is small, it would not change the resulting color and absolute magnitude significantly even if we are over-correcting them. The motivation for the second criterion is that most of the dust should be concentrated around $\pm300$~pc around the plane of the Milky Way, thus using the SFD dust map values would be appropriate for stars far away from the midplane. The motivation for the third criterion is that very nearby stars are not significantly extincted, no matter which direction they are at. These assumptions are only roughly correct. A more accurate approach would be to combine the distance information with a 3D dust map. We leave that for future investigation. For now, we just need to evaluate the rough parameter coverage, and this approach should be sufficient. 

Among those stars with distance estimates, there are 2851 stars (2856 MaNGAIDs) that satisfy at least one of these criteria. For those that satisfy the 3rd criterion, we do not correct their magnitudes or colors for extinction, as we expect little dust lies between these stars and us. For all the other stars, we correct them for extinction using the E(B-V) values given by \cite{SchlegelFD98} dust map. 

Since our sources come from a variety of photometry catalogs, we convert all the photometry to the SDSS photometric system. For those stars with PS1 photometry, we converted to SDSS photometry using the relationship given by \cite{Finkbeiner16}. For those with APASS photometry, we assume they are already in the SDSS system. For stars without input photometry from either PS1, APASS, or SDSS. We convert their {\it Gaia} photometry to the SDSS system using the relationship given by \cite{Evans18}. 

For extinction correction in the SDSS photometric system, we use the extinction coefficients given by \citep{Schlafly11}, assuming $R_V = 3.1$. For the extinction correction in the Gaia photometric system, we use the prescription given by \cite{Danielski18}. 

\begin{figure*}
\begin{center}
\includegraphics[width=0.49\textwidth]{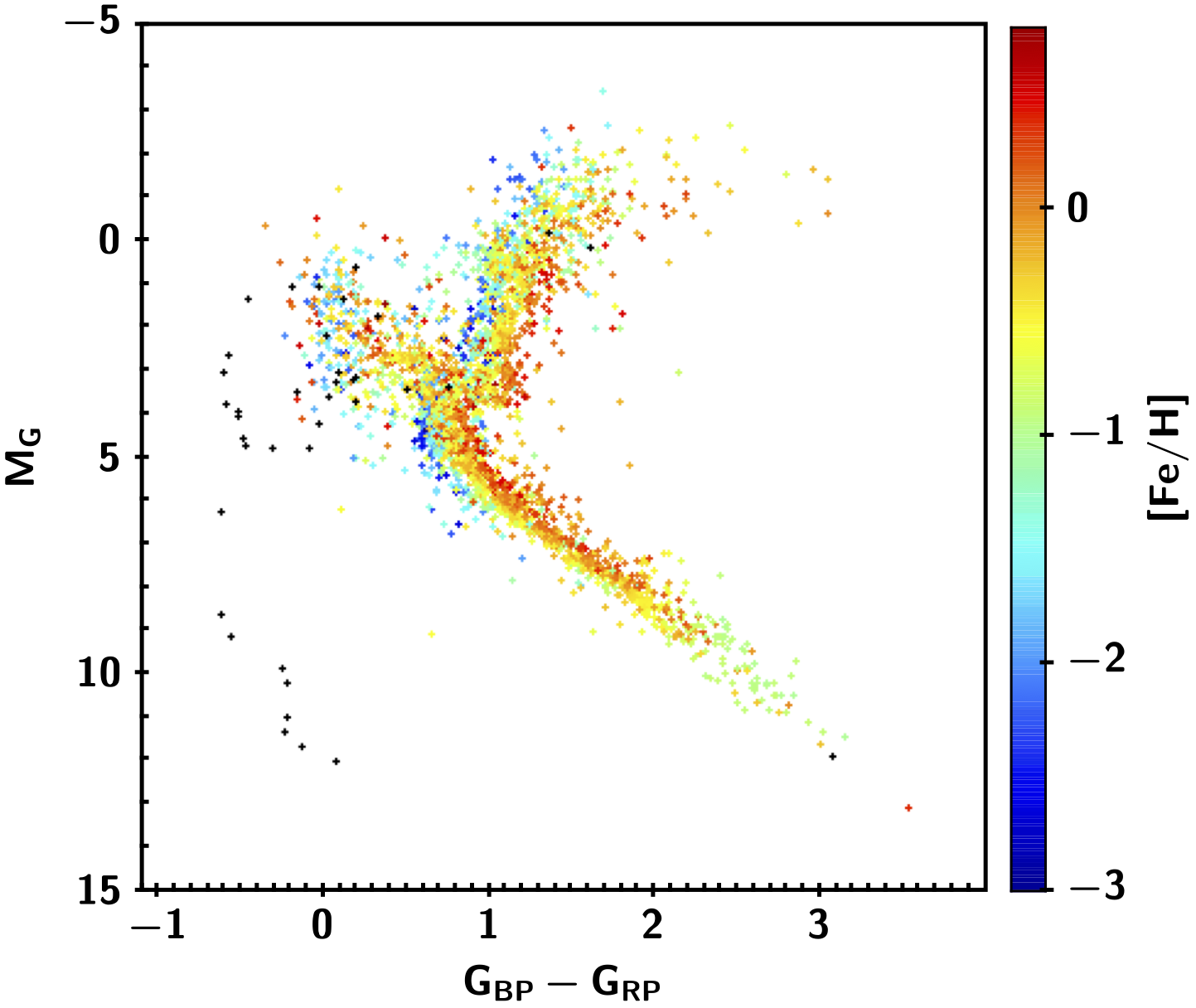}
\includegraphics[width=0.49\textwidth]{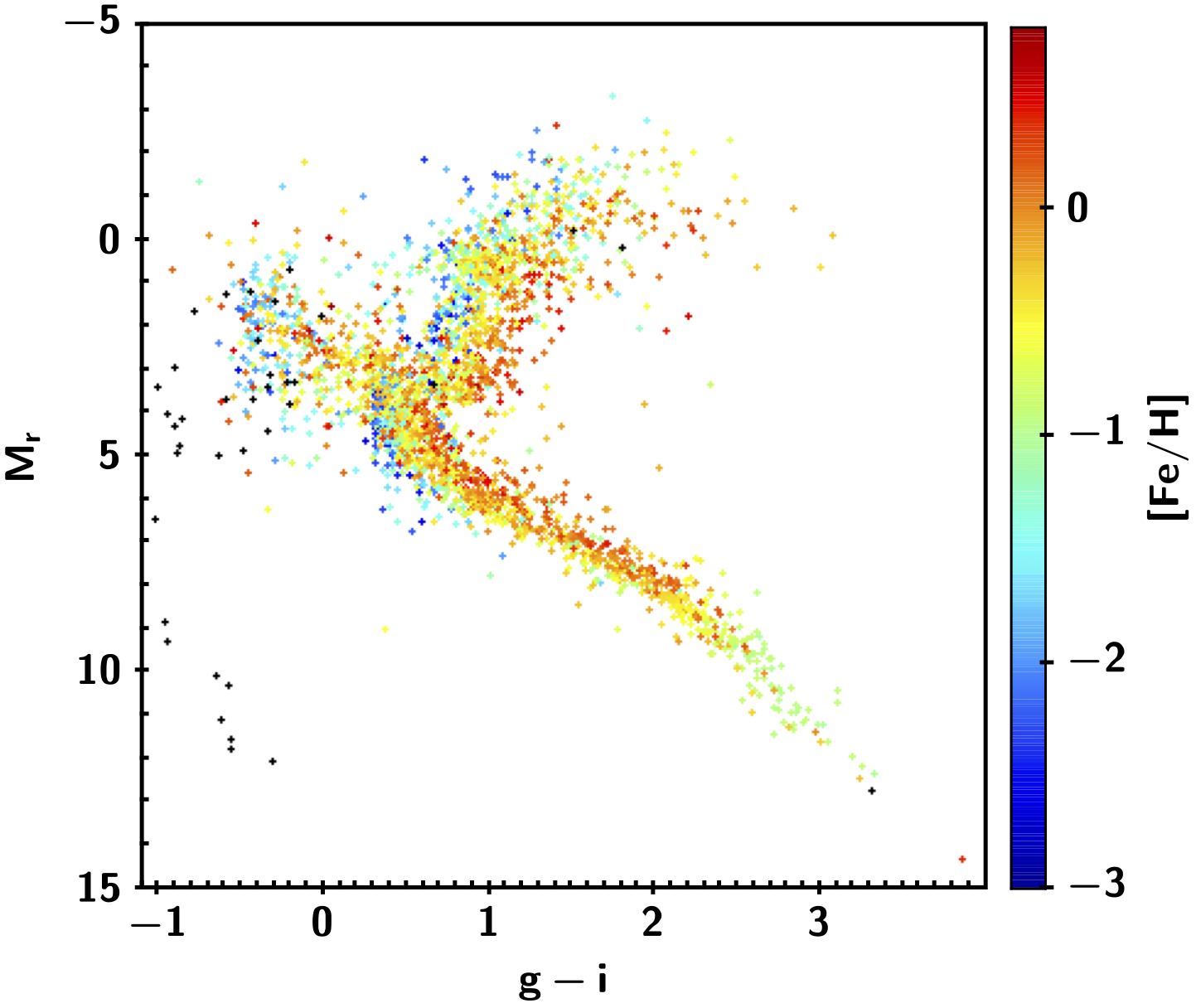}
\caption{Extinction-corrected color-luminosity diagrams for a subset of the good stars in our library in the {\it Gaia} bandpasses (left) and the SDSS bandpasses (right). The subset is selected to be those with unique matches in {\it Gaia} DR2, and with either moderate extinction, with significant height above or below the plane of the Milky Way, or with a small distance from the Sun. The color-coding of points indicate metallicity estimates. Those with unknown metallicities are plotted in black.}
\label{fig:gaiacmd}
\end{center}
\end{figure*}

With the distance estimates and extinction correction, we derive the color and absolute magnitude for our stars. Figure~\ref{fig:gaiacmd} shows the color-absolute magnitude diagram in both the Gaia photometric system (left) and the SDSS photometric system for these stars (right). We have color-coded the points with metallicities. When metallicity information is available from either APOGEE, SEGUE, or LAMOST, we employ it. Otherwise, we take the metallicity measurement from our preliminary measurements, based on our spectra using the ULySS pipeline \citep{Koleva09, Koleva11}, with MILES \citep{Sanchez-Blazquez06} as the training set. The reason we choose to mix the input parameters with those measured using ULySS is because there are significant discrepancies between the two for some parts of the parameter space for which we think the input parameters are more reliable.

These figures show that we have very good color-luminosity coverage over a wide range of metallicity. 

\begin{figure*}
\begin{center}
\includegraphics[width=0.95\textwidth]{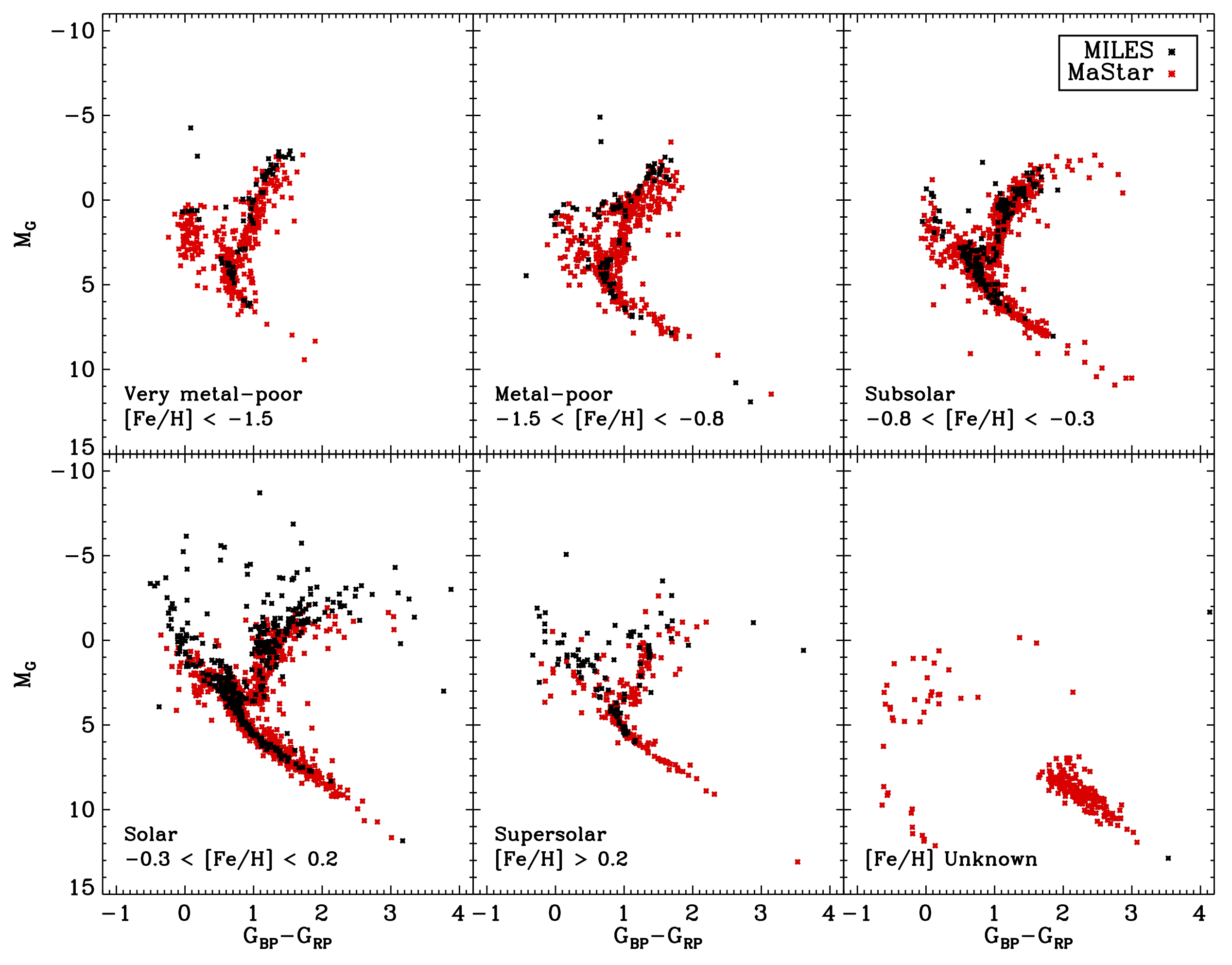}
\caption{Comparison of the coverage in extinction-corrected color-luminosity diagrams between MILES (black) and MaStar (red) in 5 metallicity bins. The last panel is for stars with no or unreliable metallicity measurements. MaStar has much more extensive and more contiguous coverage for cool dwarfs, red giants, and low metallicity stars. The coverages of the two libraries in the super-Solar metallicity bin are very complementary to each other.} 
\label{fig:gaiacmd_miles_feh}
\end{center}
\end{figure*}

Figure~\ref{fig:gaiacmd_miles_feh} compares our stellar-parameter coverage with MILES, grouped by metallicity. As discussed above, when we do not have input metallicity measurement for a star, we use the [Fe/H] measurement derived from ULySS . The results from ULySS is unreliable for dwarfs cooler than 4000K, perhaps due to the shortage of such cool dwarfs in the MILES training sample. The fitting method combined with the MILES training set appears to introduce an artificially tight correlation between $T_{\rm eff}$ and [Fe/H] for these cool dwarfs, in a way that the derived metallicity decreases as $T_{\rm eff}$ decreases. Therefore, we regard these metallicity measurements below [Fe/H] of $-0.3$ as unreliable, and show the HR diagram for those stars by themselves. The cool dwarfs in the other metallicity bins all have their metallicity given by an input spectroscopic catalog. 

For MILES, the [Fe/H] measurements come from \cite{Prugniel11}, who derived the atmospheric parameters for MILES stars using the ULySS code with the ELODIE library as a reference. 

To make a comparison with MILES, we cross-match the MILES library with Gaia DR2. Using the SIMBAD names provided by \cite{Cenarro07}, we found the coordinates and proper motion information for all the MILES library stars on the SIMBAD database. Then we compute their coordinates at the epoch of Gaia DR2 and find corresponding matches. In Appendix~\ref{sec:milescorrection}, we provide details and recommendations for others who may be interested in identifying the MILES stars for other purposes. 


Among the 985 stars in MILES, we were able to find a match in {\it Gaia} DR2 for 969 stars within 3\arcsec, and with similar magnitudes. Several stars have significantly different magnitudes in the V band. They are either due to variable stars or due to saturation in Gaia. Among the 16 missing stars, 13 are brighter than the bright limit of Gaia DR2 and are thus not included in Gaia DR2; the other three reside in clusters and they only have ambiguous matches beyond 3\arcsec. 
Among the 969 stars, we also select those ones that could be corrected for extinction using the set of three criteria described above. Here, in order to include more stars in the comparison, we relax the threshold to 0.2 mag in criterion 1 and to 200~pc in criterion 3. This yields 836 stars in the MILES library and 2989 unique stars in the MaStar library. Compared to MILES, the {\it current release} of the MaStar library has a much more extensive and more contiguous coverage in the cool dwarf regime in all metallicity bins, and a more contiguous coverage in the red giant branch, especially the lower part of it, in all metallicity bins. We also have much better coverage among blue main sequence and blue horizontal branch stars in the two metal-poor bins. However, in the Solar and super-Solar metallicity bins, our {\it current} coverages is not as good as MILES for the very hot part of the main sequence and the supergiants. We are working on improving the coverage for main sequence OB stars and the supergiants. These stars are too luminous to be found with large numbers within our regular magnitude range. We have to reduce the exposure time significantly to obtain them, which we are doing at the time of writing.  

In our final release we expect to triple the number of stars, and exceed the parameter coverage of MILES in all aspects. 

\begin{figure}
\begin{center}
\includegraphics[width=0.48\textwidth]{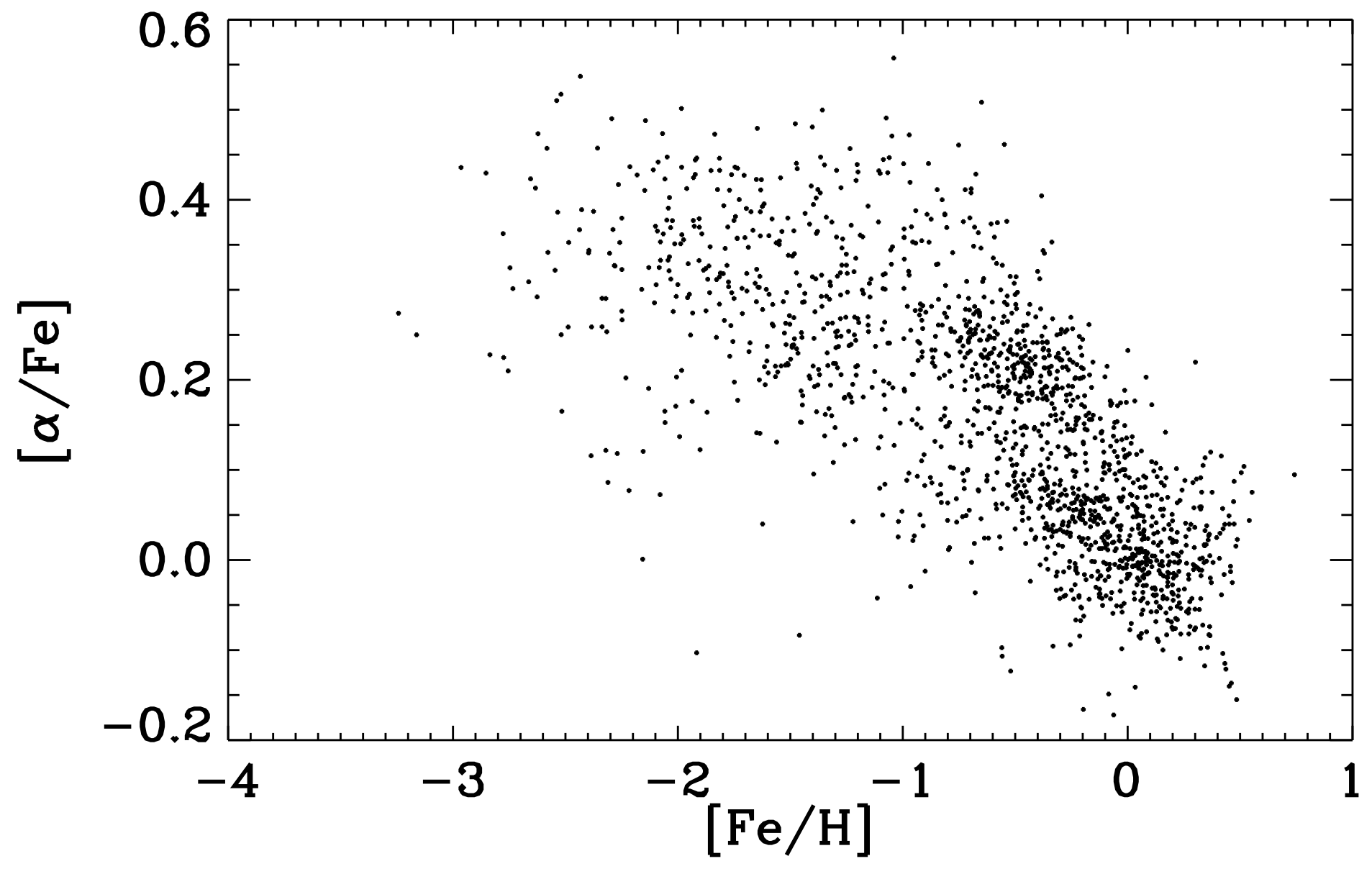}
\caption{Distribution of half of the MaStar sample in the [${\rm \alpha}$/Fe] vs. [Fe/H] space. This only shows those stars with known [${\rm \alpha}$/Fe] and [Fe/H] values from input catalogs. The distribution here is much more uniform than shown in Figure~\ref{fig:pre_alpha_distri}.}
\label{fig:mastar_alphafe}
\end{center}
\end{figure}

In Figure~\ref{fig:mastar_alphafe}, we show the current coverage of MaStar in the [${\rm \alpha}$/Fe] vs. [Fe/H] space. This plot only includes 1589 stars for which we have [${\rm \alpha}$/Fe] and [Fe/H] available from input stellar-parameter catalogs, which is less than half of the whole good-stars sample. This shows that our targeting strategy successfully achieved the sampling in both the high- and the low-sequences in this space. Comparing to the distribution in the input catalog as shown in Figure~\ref{fig:pre_alpha_distri}, the sampling of the two sequences is much more even. 

\subsection{Comparison with MILES spectra}

\begin{figure*}
    \centering
    \includegraphics[width=0.99\textwidth]{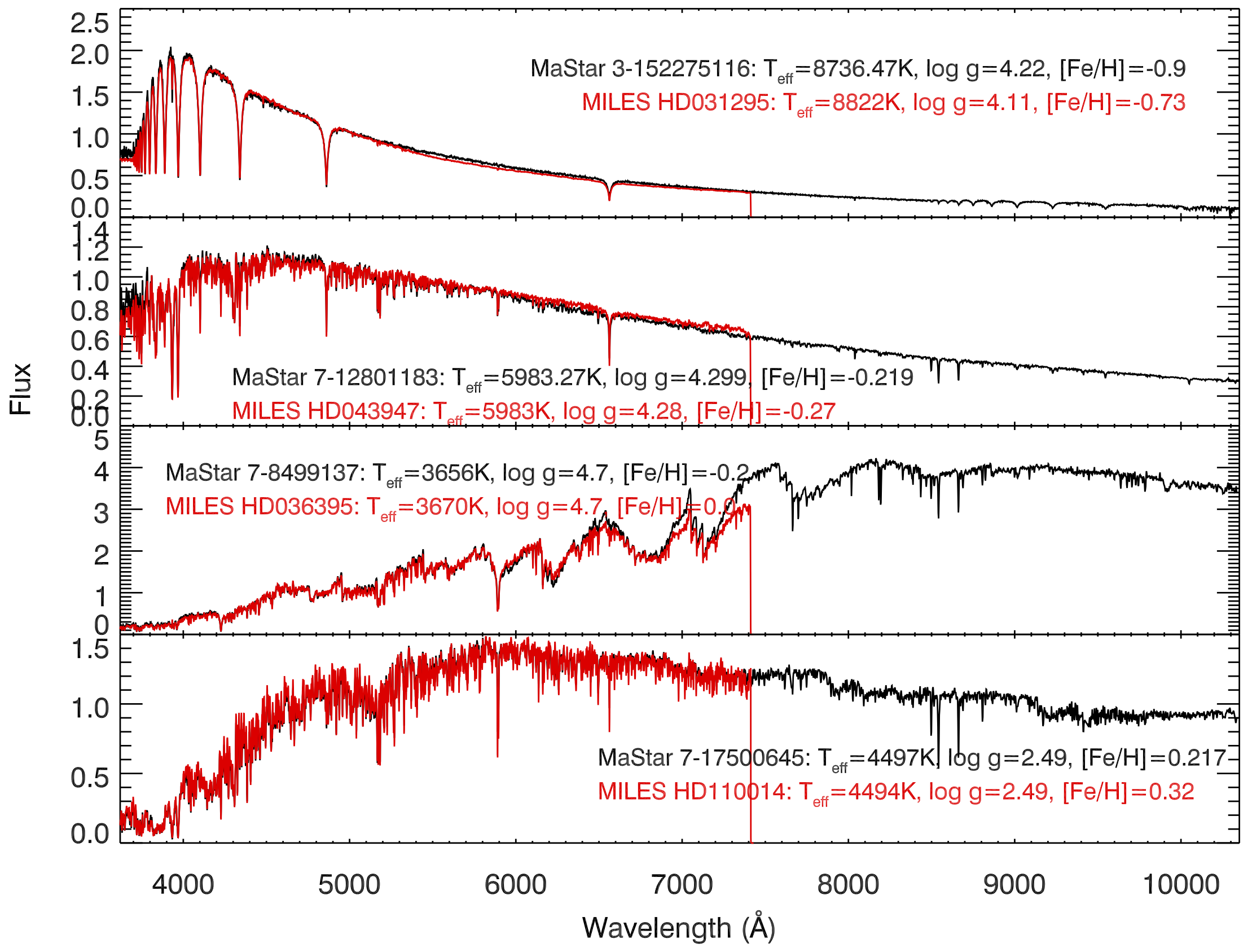}
    \caption{Comparison between MaStar spectra and MILES spectra for stars with similar parameters. From top to bottom, we show one hot main sequence star, one warm main sequence star, one cool main sequence star, and one red giant star. Both MaStar and MILES spectra are normalized by the median value between 5000A and 5050A. The wavelength are converted from vacuum to air to match MILES. The spectra agree well in general, with small differences in flux calibration.}
    \label{fig:spec_comp_miles}
\end{figure*}

We make a few comparisons with spectra in the MILES Library. Among the good stars in the current release, we do not have any stars in common with MILES. This is due to the large difference in the magnitude ranges covered by the two libraries. Therefore, for comparison, we picked stars with similar stellar parameters. The stellar parameters used for picking MaStar spectra come from the input catalogs. 
We show the comparison in Figure~\ref{fig:spec_comp_miles} for 3 main sequence stars with different temperatures, and one red giant branch star. 

We corrected the MaStar spectra for foreground dust extinction using the 3D dust maps provided by \cite{Green19} and the \cite{Fitzpatrick99} extinction law. 
 We normalized both MaStar and MILES spectra by the median flux between 5000A and 5050A to make direct comparison. The spectra from the two libraries agree very well in general. MaStar has much wider wavelength coverage. There are small differences in flux calibration at the level of a few percent, as expected from the flux calibration uncertainty of both MILES and MaStar. The difference is largest for the M dwarf comparison (the 3rd row in Figure~\ref{fig:spec_comp_miles}). In this case, the difference could also be due to differences in the intrinsic stellar parameters for the two stars, as deriving reliable parameters for these cool dwarfs is more challenging than for other parts of the parameter space. 

\section{Summary}\label{sec:summary}

We are assembling a large and comprehensive stellar spectra library with several thousand stars covering the wavelength from 3622~\AA\ to 10,354~\AA\ with a resolving power of $R\sim1800$. In this paper we describe the release of the first version, which consists of 3321 unique stars, with 8646 high-quality per-visit spectra in DR15 of SDSS-IV. The flux calibration is accurate to 3-4\%. Accurate line spread function measurements as a function of wavelength are provided for each spectrum. Compared to the MILES library, we have significantly expanded the coverage in the cool-dwarf regime and the low-metallicity regime, and provided more contiguous coverage in other parts of the parameter space. This library will form the basis for a new generation of stellar population synthesis models and be especially suited for the analysis of MaNGA galaxies. The observations are still ongoing and the final version of the library will be at least three times larger; we expect to expand the coverage in other parts of the parameter space as well. 

\acknowledgements
We thank all the volunteers who helped with the visual inspection of the stellar library spectra to assess their quality. The visual inspectors are (in alphabetical order) Maria Argudo, Rachael Beaton, Francesco Belfiore,
Dmitry Bizyaev, Yanping Chen, Lluís Galbany, Hsiang-Chih Hwang, Julie Imig, Xihan Ji, Amy Jones, David Law, Daniel Lazarz, Jianhui Lian, Karen Masters, Sofia Meneses-Gyotia, Preethi Nair, Zach Pace, Thomas Peterken, Sandro Rembold, José Sanchez-Gallego, Eddie Schlafly, Guy Stringfellow, Michael Talbot, Daniel Thomas, Christy Tremonti, Kyle Westfall, Ronald Wilhelm, Renbin Yan, Fangting Yuan, and Kai Zhang.

We thank Hans-Walter Rix for his very useful comments on the draft of the paper. We acknowledge the support by the NSF grant AST-1715898. Support for this work was also partially provided by NASA through Hubble Fellowship grant \#51386.01 awarded to R.L.B.by the Space Telescope Science Institute, which is operated by the Association of  Universities for Research in Astronomy, Inc., for NASA, under contract NAS 5-26555. J.~F-B. acknowledges support from grant AYA2016-77237-C3-1-P from the Spanish Ministry of Economy and Competitiveness (MINECO)
TCB acknowledges partial support for this work from grant PHY 14-30152:   Physics Frontier Center /JINA Center for the Evolution of the Elements (JINA-CEE), awarded by the US National Science Foundation.

Funding for the Sloan Digital Sky Survey IV has been provided by the Alfred P. Sloan Foundation, the U.S. Department of Energy Office of Science, and the Participating Institutions. SDSS acknowledges support and resources from the Center for High-Performance Computing at the University of Utah. The SDSS web site is www.sdss.org.

SDSS is managed by the Astrophysical Research Consortium for the Participating Institutions of the SDSS Collaboration including the Brazilian Participation Group, the Carnegie Institution for Science, Carnegie Mellon University, the Chilean Participation Group, the French Participation Group, Harvard-Smithsonian Center for Astrophysics, Instituto de Astrofísica de Canarias, The Johns Hopkins University, Kavli Institute for the Physics and Mathematics of the Universe (IPMU) / University of Tokyo, the Korean Participation Group, Lawrence Berkeley National Laboratory, Leibniz Institut für Astrophysik Potsdam (AIP), Max-Planck-Institut für Astronomie (MPIA Heidelberg), Max-Planck-Institut für Astrophysik (MPA Garching), Max-Planck-Institut für Extraterrestrische Physik (MPE), National Astronomical Observatories of China, New Mexico State University, New York University, University of Notre Dame, Observatório Nacional / MCTI, The Ohio State University, Pennsylvania State University, Shanghai Astronomical Observatory, United Kingdom Participation Group, Universidad Nacional Autónoma de México, University of Arizona, University of Colorado Boulder, University of Oxford, University of Portsmouth, University of Utah, University of Virginia, University of Washington, University of Wisconsin, Vanderbilt University, and Yale University.

This research was made possible through the use of the AAVSO Photometric All-Sky Survey (APASS), funded by the Robert Martin Ayers Sciences Fund.

LAMOST is operated and managed by National Astronomical Observatories, Chinese Academy of Sciences.

The Pan-STARRS1 Surveys (PS1) and the PS1 public science archive have been made possible through contributions by the Institute for Astronomy, the University of Hawaii, the Pan-STARRS Project Office, the Max-Planck Society and its participating institutes, the Max Planck Institute for Astronomy, Heidelberg and the Max Planck Institute for Extraterrestrial Physics, Garching, The Johns Hopkins University, Durham University, the University of Edinburgh, the Queen's University Belfast, the Harvard-Smithsonian Center for Astrophysics, the Las Cumbres Observatory Global Telescope Network Incorporated, the National Central University of Taiwan, the Space Telescope Science Institute, the National Aeronautics and Space Administration under Grant No. NNX08AR22G issued through the Planetary Science Division of the NASA Science Mission Directorate, the National Science Foundation Grant No. AST-1238877, the University of Maryland, Eotvos Lorand University (ELTE), the Los Alamos National Laboratory, and the Gordon and Betty Moore Foundation.

This work has made use of data from the European Space Agency (ESA) mission
{\it Gaia} (\url{https://www.cosmos.esa.int/gaia}), processed by the {\it Gaia}
Data Processing and Analysis Consortium (DPAC,
\url{https://www.cosmos.esa.int/web/gaia/dpac/consortium}). Funding for the DPAC
has been provided by national institutions, in particular the institutions
participating in the {\it Gaia} Multilateral Agreement.

\appendix

\section{MaStar Data Products}
Here we describe the main MaStar data products. We provide two summary files, mastarall and mastar-goodspec files, one contains the metadata and the other contains the spectra. They are described in detail in Section~\ref{sec:summarystat}. Here we provide basic info for the file structure, and links to the software and data model. More details of the data can be found on the SDSS DR15 website: https://www.sdss.org/dr15/mastar/

\subsection{Metadata Summary File (mastarall)}\label{apdx:mastarall}

The mastarall file collates metadata of all target stars and all observational visits. The information is listed in four binary FITS tables. The details are listed in Table~\ref{tab:mastarall}. 

{\bf Written by:} https://svn.sdss.org/public/repo/manga/mastar/mastarproc/tags/v1\_0\_2/pro/mastarall.pro

{\bf Data Model:} https://data.sdss.org/datamodel/files/MANGA\_SPECTRO\_MASTAR/DRPVER/MPROCVER/mastarall-DRPVER-MPROCVER.html

\begin{table}
\caption{Data Model for mastarall-(DRPVER)-(MPROCVER).fits}
\begin{tabular}{lll}
\hline\hline
HDU & Extension Name & Description \\
\hline
0   &   ...  & Empty \\
1   &  GOODSTARS  & Summary table for stars that have at least one good visit spectrum \\
2   &  GOODVISITS  & Summary table for all of the good visits of the good stars \\
3   &  ALLSTARS  & Summary table for all stars that have been observed at least once, regardless of quality \\
4   &  ALLVISITS & Summary table for all of the visits of all of the stars\\
\hline
\end{tabular}
\label{tab:mastarall}
\end{table}

\begin{deluxetable}{cll}
\tablecaption{Data model for the GOODSTARS and ALLSTARS table}
\tablehead{
\colhead{Column number} & \colhead{Label} & \colhead{Description}
}
\startdata
1 & DRPVER & Version of mangadrp used for reduction \\
2 & MPROCVER & Version of mastarproc used to produce this file \\
3 & MANGAID & The XX-XXXXXX format MaNGA ID\\
4 & MINMJD & Minimum MJD for all observations of this star \\
5 & MAXMJD & Maximum MJD for all observations of this star \\
6 & NVISITS & Number of visits for this star (including good and bad observations)\\
7 & NPLATES & Number of different plates this star is involved \\
8 & OBJRA & RA for the target (J2000, but not necessarily at epoch 2000) \\
9 & OBJDEC & Dec for the target (J2000, but not necessarily at epoch 2000) \\
10 & CATALOGRA & RA of this object in the photometry catalog specified by PHOTOCAT\\
11 & CATALOGDEC & Dec of this object in the photometry catalog specified by PHOTOCAT\\
12 & CAT\_EPOCH & Epoch of the astrometry (approximate epoch for PS1 and APASS)\\
13 & PSFMAG\_1 & PSF magnitude in passband\_1. The passband depends on PHOTOCAT \\
14 & PSFMAG\_2 & PSF magnitude in passband\_2. The passband depends on PHOTOCAT \\
15 & PSFMAG\_3 & PSF magnitude in passband\_3. The passband depends on PHOTOCAT \\
16 & PSFMAG\_4 & PSF magnitude in passband\_4. The passband depends on PHOTOCAT \\
17 & PSFMAG\_5 & PSF magnitude in passband\_5. The passband depends on PHOTOCAT \\
18 & MNGTARG2 & Bitmask giving information about target selection and source of photometry\\
19 & INPUT\_LOGG & Surface gravity ($\log g$) in the input stellar parameter catalog\\
20 & INPUT\_TEFF & Effective temperature ($T_{\rm eff}$) in the input stellar parameter catalog\\
21 & INPUT\_FE\_H & [Fe/H] in the input stellar parameter catalog\\
22 & INPUT\_ALPHA\_M & [alpha/M] in the input stellar parameter catalog\\
23 & INPUT\_SOURCE & Source of the input stellar parameters\\
24 & PHOTOCAT & Photometry catalog (also specified by MNGTARG2)
\enddata
\tablecomments{The GOODSTARS table is published in its entirety in the electronic edition of the {\it Astrophysical Journal}.  A portion is shown here for guidance regarding its form and content.}
\label{tab:goodstars}
\end{deluxetable}

The format of the GOODSTARS and ALLSTARS tables are shown in Table~\ref{tab:goodstars}. The GOODSTARS table is also available in the electronic edition of the {\it Astrophysical Journal}. 

\begin{deluxetable}{cll}
\tablecaption{Data model for the GOODVISITS and ALLVISITS table}
\tablehead{
\colhead{Column number} & \colhead{Label} & \colhead{Description}
}
\startdata
1 & DRPVER & Version of mangadrp used for reduction \\
2 & MPROCVER & Version of mastarproc used to produce this file \\
3 & MANGAID & The XX-XXXXXX format MaNGA ID\\
4 & PLATE & Plate number\\
5 & IFUDESIGN & IFU bundle number \\
6 & MJD & Modified Julian Date for the observation\\
7 & IFURA & RA for the center of the IFU (could be offset from star)\\
8 & IFUDEC & Dec for the center of the IFU (could be offset from star)\\
9 & OBJRA & RA for the target (J2000, but not necessarily at epoch 2000)\\
10 & OBJDEC & Dec for the target (J2000, but not necessarily at epoch 2000)\\
11 & PSFMAG\_1 & PSF magnitude in passband\_1. \\
12 & PSFMAG\_2 & PSF magnitude in passband\_2. \\
13 & PSFMAG\_3 & PSF magnitude in passband\_3. \\
14 & PSFMAG\_4 & PSF magnitude in passband\_4. \\
15 & PSFMAG\_5 & PSF magnitude in passband\_5. \\
16 & MNGTARG2 & Bitmask giving information about target selection and source of photometry\\
17 & NEXP & Total number of exposures during this visit\\
18 & HELIOV & Heliocentric velocity used to shift the spectra to the rest frame. \\
& & If HELIOV=0, then the spectra are still in the observed frame.\\
19 & VERR & 1-sigma error on the heliocentric velocity\\
20 & V\_ERRCODE & Error code for radial velocity search (0 is good, nonzero is bad)\\
21 & MJDQUAL & Bitmask for spectral quality flags
\enddata
\tablecomments{The GOODVISITS table is also published in its entirety in the electronic edition of the {\it Astrophysical Journal}.  A portion is shown here for guidance regarding its form and content.}
\label{tab:goodvisits}
\end{deluxetable}

The format of the GOODVISITS and ALLVISITS tables are shown in Table~\ref{tab:goodvisits}. The GOODVISITS table is also available in the electronic edition of the {\it Astrophysical Journal}. 

\subsection{Spectra Summary File (mastar-goodspec)} 

The mastar-goodspec file contains the per-visit spectra and metadata for all good visits. Its entries are in the same order as the entries in the GOODVISITS table of the mastarall file. The wavelength array for all spectra is logarithmically spaced with $\Delta \log \lambda = 10^{-4}$, and are given in vacuum. 

{\bf Written by:} https://svn.sdss.org/public/repo/manga/mastar/mastarproc/tags/v1\_0\_2/pro/collectstellar.pro

{\bf Data Model:} https://data.sdss.org/datamodel/files/MANGA\_SPECTRO\_MASTAR/DRPVER/MPROCVER/mastar-goodspec-DRPVER-MPROCVER.html

\subsection{Spectra of Individual Stars}
Besides being collected in the spectra summary file (`mastar-goodspec'), the spectra for each individual star can also be found under MaNGA DRP reduction directory, specifically under the subdirectory called `mastar' under the subdirectory for each plate. The files called `mastar-LOG-[PLATE]-[IFUDESIGN].fits.gz' contains the final per-exposure and per-visit spectra for each star. One can also find those spectra that are considered to be of poor quality in these directories. Table~\ref{tab:mastar-log} lists the basic content of these files. The wavelength array for all spectra is logarithmically spaced with $\Delta \log \lambda = 10^{-4}$, and are given in vacuum. 

{\bf Written by:} https://svn.sdss.org/public/repo/manga/mangadrp/tags/v2\_4\_3/pro/spec3d/mdrp\_mastar.pro

{\bf Data model:} https://data.sdss.org/datamodel/files/MANGA\_SPECTRO\_REDUX/DRPVER/PLATE4/mastar/mastar-LOG-PLATE-IFU.html

\begin{table}
\caption{Data Model for mastar-LOG-[PLATE]-[IFUDESIGN].fits.gz}
\label{tab:mastar-log}
\begin{tabular}{lll}
\hline\hline
HDU & Extension Name & Description \\ \hline
0 & ... & Empty except for global header \\
1 & MASTAR & Binary table providing all of the per-visit spectra for a star and associated metadata\\
2 & OBSINFO & Binary table of auxillary information of the observations, one row for each exposure\\
3 & FITDETAIL & Binary table of per-exposure spectra and the associated metadata for all exposures taken on all visits\\
\hline
\end{tabular}
\end{table}

\section{Meanings of bitmasks used in MaStar data products}

\subsection{Pixel-level Mask for Spectra}
The pixel-level mask gives indication of quality issues related with each pixel in a spectrum. The mask associated with each spectrum can be found as the column 'MASK' in both the spectra summary file and the individual spectra file. The meaning of this pixel-level bitmask is identical to that used by MaNGA, which is listed in Table 13 of \cite{Law16}. 

\subsection{Quality Bitmask for Spectra}
The MASTAR\_QUAL bitmask provides indications of the overall spectral quality. This bitmask is applied in EXPQUAL, MJDQUAL, and MSTRQUAL keywords in various files. In the FITDETAIL extension of the mastar-LOG-[PLATE]-[IFUDESIGN] files, the column EXPQUAL provides the quality indication for each per-exposure spectrum. In the summary spectra file (`mastar-goodspec') and the MASTAR extension of the mastar-LOG-[PLATE]-[IFUDESIGN] files, the column MJDQUAL provides the quality indication for each per-visit spectrum. In the global header of the mastar-LOG-[PLATE]-[IFUDESIGN] file, the MSTRQUAL provides an overall quality indication for that plate-IFU. Table ~\ref{tab:mastarqual} lists the meaning of each bit. These are explained in more detail in Section~\ref{sec:qualityflagging}

\begin{table}
\caption{MASTAR\_QUAL Data-Quality Bitmask (Applied to MJDQUAL, EXPQUAL, and MSTRQUAL).}
\label{tab:mastarqual}
\begin{tabular}{lrll}
\hline\hline
Bit & Value &Label & Description \\ \hline
     0 &           1 & NODATA &  No Data\\
       1 &           2 & SKYSUBBAD &  Bad sky subtraction in one or more frames\\
       2 &           4 & HIGHSCAT &  High scattered light in one or more frames\\
       3 &           8 & BADFLUX &  Bad flux calibration\\
       4 &          16 & LOWCOV &  PSF-covering fraction by fiber is too small ($<10$\%)\\
       5 &          32 & POORCAL &  Poor throughput\\
       6 &          64 & BADHELIORV &  High variance between stellar RVs\\
       7 &         128 & MANUAL &  Flagged as problematic by visual inspection\\
       8 &         256 & EMLINE &  Spectrum contain emission lines\\
       9 &         512 & LOWSN &  Per-MJD Spectrum has median S/N $\leq 15$\\
      30 &  1,073,741,824 & CRITICAL &  Critical failure in one or more frames\\\hline
\label{tab:mastar_qual}
\end{tabular}
\end{table}

\subsection{Targeting Bitmask}

The MANGA\_TARGET2 bitmask provides targeting information for MaStar targets. It is given as the `MNGTARG2' column in all extensions of the metadata summary file (`mastarall') and the spectra summary file (`mastar-goodspec'). Table~\ref{tab:mangatarget2} lists the meaning of each bit. In particular, bits 7,8,11,12,13, 15, and 16 are useful for knowing which photometric system the PSFMAG for each star is based on. The corresponding description of the PSFMAG is given in Table~\ref{tab:psfmag}. 

\begin{table}
\caption{MANGA\_TARGET2 Bitmask (Abbreviated as MNGTARG2)}
\label{tab:mangatarget2}
\begin{tabular}{lrll}
\hline\hline
Bit & Value &Label & Description \\ \hline
       0 &            1 & NONE & (not used) \\
       1 &            2 & SKY &  sky fibers \\
       2 &            4 & STELLIB\_SDSS\_COM & Commissioning selection using SDSS photometry\\
       3 &            8 & STELLIB\_2MASS\_COM & Commissioning selection using 2MASS photometry\\
       4 &           16 & STELLIB\_KNOWN\_COM & Commissioning selection of known parameter stars\\
       5 &           32 & STELLIB\_COM\_mar2015 & Commissioning selection in March 2015\\
       6 &           64 & STELLIB\_COM\_jun2015 & Commissioning selection in June 2015\\
       7 &          128 & STELLIB\_PS1 & Library stars using PS1 photometry\\
       8 &          256 & STELLIB\_APASS & Library stars using APASS photometry\\
       9 &          512 & STELLIB\_PHOTO\_COM & Commissioning selection using photo-derived parameters\\
      10 &         1,024 & STELLIB\_aug2015 & Global Selection since Aug 2015\\
      11 &         2,048 & STELLIB\_SDSS & Library stars using SDSS photometry\\
      12 &         4,096 & STELLIB\_GAIA & Library stars using Gaia DR1 photometry, $G$ band only\\
      13 &         8,192 & STELLIB\_TYCHO2 & Library stars using TYCHO2 photometry ($B$ and $V$ in place of $u$ and $r$)\\
      14 &        16,384 & STELLIB\_BRIGHT & bright stars observed with short exposures\\
      15 &        32,768 & STELLIB\_UNRELIABLE & Library stars with unreliable photometry\\
      16 &        65,536 & STELLIB\_GAIADR2 & Library stars using Gaia DR2 photometry, $G$, $G_{\rm BP}$, $G_{\rm RP}$\\
      20 &      1,048,576 & STD\_FSTAR\_COM & MaNGA commissioning selection of F type flux standards \\
      21 &      2,097,152 & STD\_WD\_COM & MaNGA commissioning of white dwarf flux standards \\
      22 &      4,194,304 & STD\_STD\_COM & Other standards for MaNGA commissioning \\
      23 &      8,388,608 & STD\_FSTAR & MaNGA selection of F type flux standards (based on SDSS photometry) \\
      24 &     16,777,216 & STD\_WD & White dwarf standards\\
      25 &     33,554,432 & STD\_APASS\_COM & Commissioning selection of stds using APASS photometry\\
      26 &     67,108,864 & STD\_PS1\_COM & Commissioning selection of stds using PS1 photometry\\
      27 &    134,217,728 & STD\_BRIGHT & standards on bright star plates\\
\hline
\end{tabular}
\end{table}

\begin{table}
\caption{Filter Bands of PSFMAG for Different MANGA\_TARGET2 bits}
\label{tab:psfmag}
\begin{tabular}{lrl}
\hline\hline
PHOTOCAT & MNGTARG2 bit & Filter bands for PSFMAG [0]-[4] \\
\hline
PS1 &  7 & [None, $g$, $r$, $i$, $z$]\\
APASS & 8 & [None, $g$, $r$, $i$, None] \\
SDSS & 11 & [$u$, $g$, $r$, $i$, $z$]\\
{\it Gaia} DR1 & 12 & [None, $G$, None, None, None]\\
TYCHO2 & 13 & [$B$, None, $V$, None, None]\\
{\it Gaia} DR2 & 16 & [None, $G$, $G_{\rm BP}$, $G_{\rm RP}$, None]\\
\hline
\end{tabular}
\end{table}

\section{Matching the MILES Library to Gaia DR2}\label{sec:milescorrection}

We found there are some errors in the coordinates of the MILES stars as given on the MILES website (http://www.iac.es/proyecto/miles/pages/stellar-libraries/the-catalogue.php). Using these coordinates, for a large fraction of the stars, we cannot find a match for them to sources on SIMBAD or Gaia DR2. 

Using the SIMBAD names for the stars provided by \cite{Cenarro07}, we were able to match all 985 stars to sources on SIMBAD. Most of these already have been crossmatched to Gaia DR2 and have updated astrometry and proper motion information available. We make use of these coordinates and proper motion to compute the coordinates for Equinox J2000.0 at Epoch J2015.5 which is the epoch for Gaia DR2.  We then crossmatched them with Gaia DR2 sources using the online query tool offered by Gaia Archive. We found matches for 969 stars within 3\arcsec\ and with similar magnitudes. For the remaining 16 stars, 13 of them (HD029139, HD039801, HD054605, HD057061, HD060179, HD081797, HD085235, HD089484, HD095735, HD124897, HD146051, HD164058, HD020902) are too bright to be found in Gaia, and 3 of them (M71 KC-147, M71 KC-263, NGC 288 77) are in clusters and only have ambiguous matches beyond 3\arcsec. 


For others who may be interested in identifying the MILES sources, we recommend using the SIMBAD names provided by \cite{Cenarro07} to obtain astrometry from SIMBAD. We advise against using the coordinates given on the MILES website. For four of the stars(BD+090352, HD000249, HD151217, HD152601), the coordinates provided are in error. For 23 of the 28 stars in the cluster M71, the coordiantes are given in Equinox B1950 rather than J2000. And due to the lack of epoch information, many others stars can have coordinate offsets up to 70 arcsec.

\section{Details of the Targets Selection}
\label{sec:app_target}

\subsection{Magnitude Limits}

We select stars that are brighter than 17.5\ in either the $g$-band or $i$-band. This ensures we have a S/N of more than 50\ with 8 15-minute exposures.

We set the bright limit of our target selection to 12.7\ in both the $g$- and $i$-bands. Given the throughput of the instrument\citep{Yan16}, assuming the star is perfectly centered in the central fiber and observed under 1\arcsec\ seeing, the count in the brightest pixel in the blue or red camera would be about 25,000 for a star of magnitude 12.7 in the $g$- or $i$-bands, respectively. A brighter star would risk being saturated. For one of the amplifiers of the red camera in Spectrograph 1 (r1), the detector exhibits slight non-linearity when the raw count is above 33,000. 

Starting from Plate 9800, we adopted a bright limit of 11.7 in both the $g$- and $i$-bands in order to include more intrinsically luminous stars. For stars with either $g$-band or $i$-band magnitudes between 11.7 and 12.7, we intentionally offset the fiber bundles by 1.443\arcsec\ to the north of the target so that the center of the star falls in the gaps between fibers to avoid saturating the detector. We are still able to recover the correct flux for the star using the aperture-correction technique, as described in Section~\ref{sec:aperturecorrection}. 


\subsection{Homogenization of Various Catalogs}\label{sec:homogenization}


A large fraction of our targets are selected from stellar parameter catalogs, including the APOGEE ASPCAP catalog \citep{Holtzman15, GarciaPerez16, Holtzman18}, SEGUE SSPP catalog \citep{Lee08a, Lee08b, AllendePrieto08, AllendePrieto14}, and the LAMOST LEGUE catalog \citep{LAMOSTDR1}. Here, we describe in detail the selection and homogenization of their stellar parameters. 

In the ASPCAP catalog, we first remove stars with velocity variations detected at more than 5$\sigma$, using an velocity uncertainty floor of 0.2~km/s, to avoid targeting binary stars which could have contaminated spectra. We also remove those stars with bit BAD\_PIXELS, COMMISSIONING, VERY\_BRIGHT\_NEIGHBOR, or LOW\_SNR set in the APOGEE\_STARFLAG bitmask, and those with bits STAR\_BAD, SN\_BAD, or NO\_ASPCAP\_RESULT set in the APOGEE\_ASPCAPFLAG bitmask. In the SEGUE catalog, we remove those with a non-zero ZWARNING column. 

As the stellar parameters from different catalogs could have systematic offsets from each other, we applied shifts to the parameters from these catalogs to make them roughly consistent with each other before selecting targets. By comparing the parameters for the stars in common, we found that SEGUE and LAMOST parameters are very consistent with each other, thus no shift is necessary. Using stars in common between APOGEE and LAMOST, we found small systematic differences in stellar parameters between them. 
We shift APOGEE to be consistent with LAMOST, because LAMOST has the greatest number of stars, and overlaps with both APOGEE and SEGUE in their magnitude range. It does not matter for our purposes which catalog has a closer-to-truth parameters. What matters is that they are consistent with each other. 

The corrections applied to the APOGEE stellar parameters are different for giants and dwarfs. ASPCAP provides the uncalibrated parameters (`fparam') and the calibrated parameters (`param'). The latter is only available for giants. For giants, which are defined as those with an uncalibrated $\log g <3.8$, we adjust the calibrated ASPCAP parameters from APOGEE (`param') as:
\begin{equation}
\log g = \log g_{\rm aspcap} +0.144 .
\end{equation}

For dwarfs, defined to have an uncalibrated $\log g > 3.8$, we apply a temperature-dependent log g correction based on comparison with LAMOST:
\begin{equation}
\log g = (-5.3407\times10^{-5} T_{\rm aspcap} + 1.320) \log g_{\rm aspcap} ,
\end{equation}
where both $\log g_{\rm aspcap}$ and $T_{\rm aspcap}$ are from the uncalibrated parameters.

For temperature, there is no difference between the calibrated and uncalibrated parameters in the ASPCAP catalog. Relative to LAMOST, the median shift is 54K: 
\begin{equation}
T_{\rm eff} = T_{\rm aspcap}+54~{\rm K} .
\end{equation}

For metallicity, ASPCAP provides metallicity in terms of [M/H]. Here, ``M'' does not strictly represents the total metallicity (total number of nuclei for elements heavier than helium). Rather, [M/H] represents the scaling factor applied to all elements other than H, He, C, and ${\rm \alpha}$-elements, using the solar-abundance pattern for these elements. This is the abundance setting in the atmosphere grids generated by \citep{Meszaros12}, which are used by ASPCAP to produce template spectra to fit the data. Therefore, the [M/H] quoted by ASPCAP is the solar-scaled metal abundance. The atmosphere grids produced by \cite{Meszaros12} allow independent variations of C and $\alpha$-elements relative to other metals. Thus they can deviate from the solar abundance pattern. As a result, the [M/H] derived by ASPCAP follows [Fe/H] closely. We adopt the [M/H] and [${\rm \alpha}$/M] values provided by ASPCAP, but treat and quote them as [Fe/H] and [${\rm \alpha}$/Fe].

The difference is negligible between LAMOST and APOGEE. For stars with calibrated [Fe/H] available, we adopt the calibrated metallicity and made no correction. For those stars without calibrated [Fe/H] available, we adopt the uncalibrated metallicity, and made a small correction to match the calibrated parameters:
\begin{equation}
[{\rm Fe/H}] = [{\rm Fe/H}]_{\rm aspcap} - 0.02.
\end{equation}

For $[{\rm \alpha/Fe}]$, for stars with calibrated $[{\rm \alpha/Fe}]$ available, we adopt it without any correction. Otherwise,we adopt the uncalibrated $[{\rm \alpha/Fe}]$ and made a small correction to match the calibrated parameters:
\begin{equation}
[{\rm \alpha/Fe}] = [{\rm \alpha/Fe}]_{\rm aspcap}-0.0413. 
\end{equation}

\subsection{Selection of Known-Parameter Stars}
In practice, there are several constraints. First, we do not dictate which fields we observe as we piggyback on APOGEE-2N observations. Secondly, we have a fixed number of designs in each field, and cannot observe more than $17\times n_{\rm design}$ stars in each field. Thirdly, not all fields have known-parameter stars. Therefore, we have to design an effective algorithm in order to observe rare stars where they are available. The problems are how to define ``rareness'' and how to decide the priority among different types of rare stars.

Our method to solve this problem under these practical constraints is to select targets globally among all of the fields. Because APOGEE-2N provides us with a list of all the fields they will observe and the number of designs for each field, we can predict the number of opportunities each star can be selected. Collecting all the stellar-parameter catalogs within these fields also allows us to build a density distribution in the multi-dimensional parameter space. We can then pick the stars with appropriate weight so that the resulting distribution is flat in the parameter space. 

In practice, we first run the selection in a three-dimensional parameter space ($T_{\rm eff}$, $\log g$, and [Fe/H]). We treat $[{\rm \alpha/Fe}]$ separately because only some of the catalogs have this information. We find all the APOGEE, SEGUE, and LAMOST stars in all of the APOGEE-2 fields, select only stars with magnitudes falling between the limits defined above. We place all these stars  in the three-dimensional space, and then compute the local density of each star by counting its neighbors within a box of dimension 0.05 in $\Theta$ (defined as 5040K/$T_{\rm eff}$), 0.3 dex in $\log g$, and 0.3 dex in [Fe/H] centered on that star. The size of the box is chosen to be comparable to the uncertainty in these parameters. When computing the local density, we do not simply add the number of stars in that box. We sum up $n_{\rm design}$ of all the stars in that box, because this is the number of opportunities that each star could be selected. This procedure gives us the local density around each star. 

We assign an initial weight to each star that is proportional to the inverse of the local density. With some further adjustments to these weights, which are described below, we eventually select stars randomly with probabilities proportional to the final weights. This results in a list of stars roughly, but not strictly, sorted by their local density. The reason we do not do the strict sorting is to avoid rare stars taking up all available plate real estate, and leaving no place for the more common stars. 

To express the above in mathematical terms, the local density around each star is: 
\begin{equation}
\rho_i =  \sum_{\rm box} n_{\rm design}.
\end{equation}
The probability assigned to each star is:
\begin{equation}
P_i = {1\over \rho_i} = {1 \over \sum_{\rm box} n_{\rm design}}.
\end{equation}
We normalize $P_i$ so that the sum of all $P_i$ is 1.0. 

Assuming all the stars in that box have roughly the same $\rho_i$ and the same $P_i$, the total probability of targeting stars in the box around star $i$ is: 
\begin{equation}
\sum P_i n_{\rm design,i} = 1.
\end{equation}

Therefore, this arrangement will provide roughly the same number of stars in each box populated with sufficient number of stars. 

There are three further adjustments done to the weights before we run the probabilistic random selection. (a) First, we reduce the weights for some stars if the part of the parameter space they cover has already been sampled sufficiently by previously-designed plates. (b) Secondly, we adjust the weights to prioritize some source catalogs over others. (c) Thirdly, we adjust them to flatten the sampling in $[{\rm \alpha/Fe}]$ space. 

a) The reduction of weights for the already-sampled parameter space only takes effect after we have designed some plates. Each month, when we design new plates to be observed, we rerun the target-selection code to reassess the sampling of the parameter space. This gives us the opportunity to reduce the weight for those stars whose parameter space has already been sufficiently sampled. We set the threshold for sufficient sampling to the ratio between the number of total targets we expect to observe and the rough number of bins in the parameter space. We reduce the weight in proportion to the ratio between the number of stars already-observed around each star and this threshold. If we have observed more stars than this threshold, then we do not take any more targets from that bin by assigning the given star a very low weight. 

The exact number for the threshold of sufficient sampling changes with time, and is set separately for stars with $[{\rm \alpha/Fe}]$ information and the stars without this information. For stars without $[{\rm \alpha/Fe}]$ available, the number of already-observed stars are counted in the 3D neighborhood space ($\pm0.05$ in $\Theta$, $\pm0.3$ in $\log g$, and $\pm0.3$ in [Fe/H]). The threshold is about 18. For stars with $[{\rm \alpha/Fe}]$ available, the number of already-observed stars are counted in the 4D neighborhood space ($\pm0.05$ in $\Theta$, $\pm0.3$ in $\log g$, $\pm0.3$ in [Fe/H], and $\pm0.05$ in $[{\rm \alpha/Fe}]$) and the threshold for sufficient sampling is about 6. 

b) The three catalogs we adopt have different accuracies on their stellar parameters. Because APOGEE uses relatively high-resolution spectroscopy, its parameters are much more reliable than those from SEGUE and LAMOST. This will help our final stellar-parameter measurements considerably. Therefore, whenever the same types of stars are available in more than one catalogs, we prefer in order, from highest to lowest, APOGEE, SEGUE, and LAMOST. We prefer SEGUE over LAMOST because SEGUE has better flux calibration than LAMOST.

Therefore, when we have stars from all three surveys in a given bin of the stellar-parameter space, we would like the chance of selecting APOGEE stars to be 10 times higher than the chance of selecting SEGUE stars, and the latter to be ten times higher than the chance of selecting LAMOST stars. To achieve this, we adjust the weights assigned to targets from different sources. The weights have to be adjusted according to the number of stars from each source, so that the collective probability for picking stars from that source differ by a factor of 10. At the same time, we maintain the total sum of weights for each bin, so that the total probability for drawing stars from that bin does not change. 

\begin{figure}
\begin{center}
\includegraphics[width=0.45\textwidth]{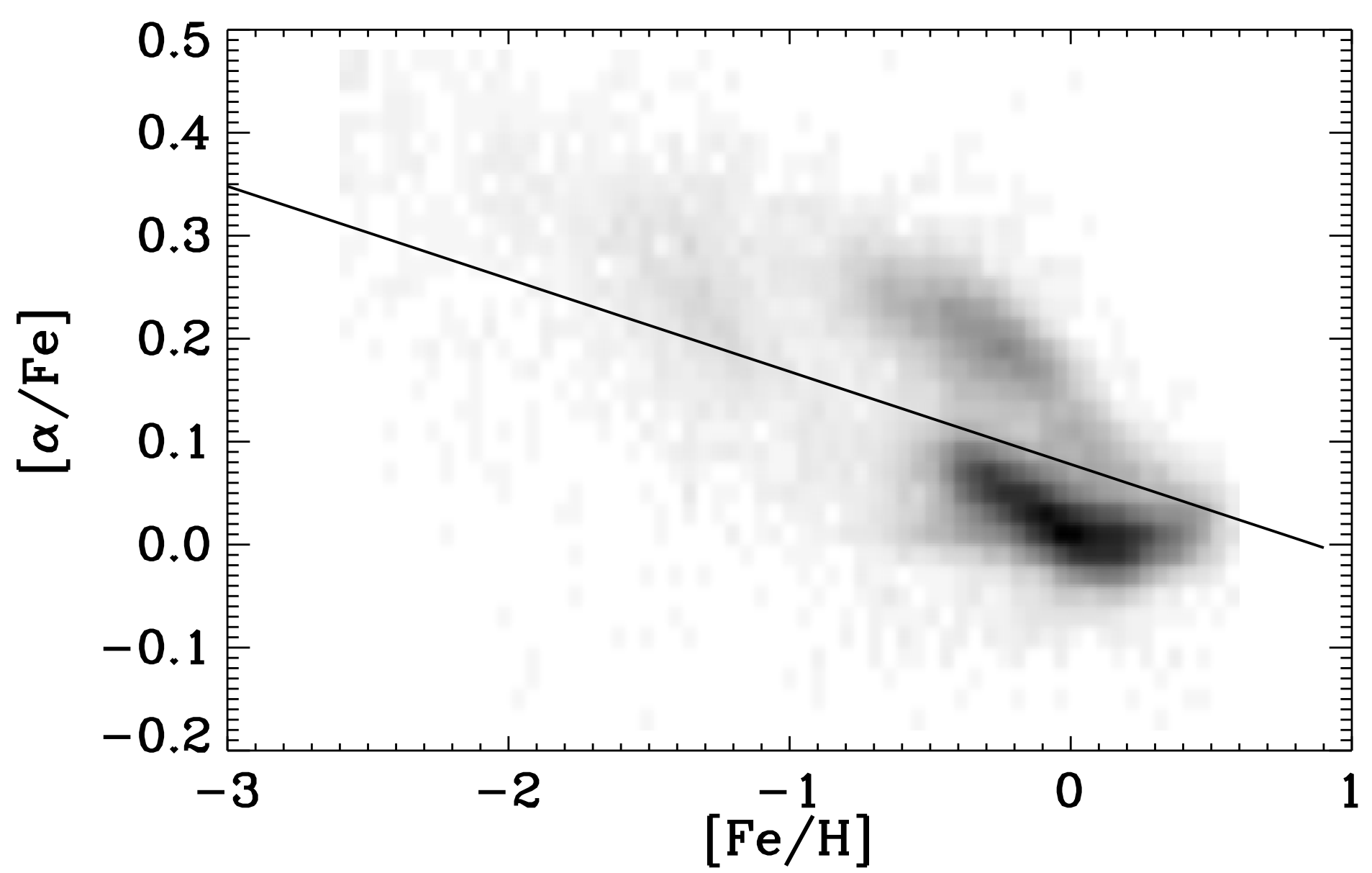}
\caption{Distribution of APOGEE stars in [${\rm \alpha}$/Fe] vs. [Fe/H] space. One can see two sequences which merge at high [Fe/H]. These parameters have been adjusted according to the prescription in Section~\ref{sec:homogenization}. The solid line indicates our fiducial demarcation. We try to balance the sampling of stars above and below this demarcation.}
\label{fig:pre_alpha_distri}
\end{center}
\end{figure}
c) We also adjust the weights to flatten the distribution of stars in the [${\rm \alpha}$/Fe] space. This is done only for stars in APOGEE and SEGUE which have $[{\rm \alpha/Fe}]$ measurements available, and for some LAMOST stars that can be found in the LAMOST-Cannon catalog \citep{Ho.A17} which provided $[{\rm \alpha/Fe}]$ for 450,000 giant stars in LAMOST DR2. 
Figure~\ref{fig:pre_alpha_distri} shows the distribution of APOGEE stars in [${\rm \alpha}$/Fe] vs. [Fe/H]  space. At the low-[Fe/H] end, there are two sequences in this plot with an offset in [${\rm \alpha}$/Fe] between them, corresponding roughly to the thin-disk and thick-disk populations. The two sequences merge at high [Fe/H]. We define a fiducial demarcation between the two sequences, which can be described by the following equation: 
\begin{equation}
[\alpha/Fe] = -0.09 [Fe/H] + 0.078 .
\end{equation}
We aim to select the same number of stars above and below the demarcation at each [Fe/H]. This is done by binning stars in [Fe/H] into several bins, with boundaries set at [Fe/H]=[$-4.5, -0.9, -0.5, -0.2$, 0.1, 0.4, 0.9], and then adjusting the weight for stars above and below the demarcation, so that they have the same total probability to be targeted. 


We then have a second step to pick the stars for each field and for each design. When designing each plate, we pick out the stars in the global list that belong to the field being designed, while keeping their order in the global list. We then select stars from top to bottom so that we first select the rarer stars on that plate. 

In our plate design, we need to coordinate with the APOGEE-2 targeting for infrared spectroscopy. To allow fibers to be pluggable, we have to reject MaStar targets that are closer than 115.74\arcsec\ to any APOGEE target or another MaStar target. This requirement rejects some of the stars, and we have to keep going down the list to find a sufficient number of stars. 

Whenever APOGEE-2 changes their field plan, we rerun our algorithm to include stars in any new fields and/or modify their $n_{\rm design}$. 

Due to the tight monthly schedule for plate design and the need to get targets on plates, this selection algorithm described above was gradually implemented and improved over the years. As a result, one should {\it never} use this sample for any statistical studies, and the sample is not meant to be unbiased at all. The strength of the sample is the wide range of parameter space coverage, with significant oversampling for rare combinations of stellar parameters.


\subsection{Early Commissioning Selection}
The MaStar program was conceived relatively late in the planning of SDSS-IV. The first rounds of target selection were therefore performed using a preliminary pipeline before the full pipeline was developed. 
In June of 2015, we switched over to the selection described above for the majority of MaStar plates. As a user of the library, one only needs to be aware that the target selection for those early plates (with plate numbers smaller than 8500) was different. There were also mistakes that caused some very bright stars to be observed; these are flagged and rejected in the final summary catalog and summary spectra files. 

\subsection{Selection of Stars Based on Photometry}

In many APOGEE-2 fields, we have no or very few stars with known stellar parameters. In these fields, we use a photometric selection to select preferentially very hot and very cool stars. 

This selection is done by SED fitting to optical and infrared photometry.  We generated PARSEC theoretical SEDs \citep{Bressan12} using the PARSEC online service\footnote{http://stev.oapd.inaf.it/cgi-bin/cmd\_2.7} (version 1.1) in Pan-STARRS1 grizy bands, 2MASS J,H,K bands, and WISE W1 and W2 bands, for a grid of ages, metallicities, and extinction values. The input parameters are set as following: $t$ = 0.1, 0.2, 0.5, 1, 2, 5, 10 Gyr; $Z=0.0002$ to 0.05 in steps of $\Delta Z$=0.0002; and $A_v$=0, 0.05, 0.1, 0.15 mag. Other parameters are set according to the default. Since the PARSEC online service does not have APASS filter information, we convert the APASS magnitudes into PS1 magnitudes using an empirical relation derived using a sample of stars with both PS1 and APASS measurements. The best-fit SED model is determined according to the minimum $\chi^2$ method. For each star, from the best-fit model, we obtain $T_{\rm eff}$, $\log g$, [Fe/H], age, and total extinction $A_v$, where [Fe/H] is defined to be $\log (Z/Z_\odot)$ with $Z_\odot=0.019$.

By comparing to a sample of stars with stellar parameters available from LAMOST, we found that the $T_{\rm eff}$ are well-determined for the great majority of stars. However, $\log g$ and [Fe/H] are not well-constrained. Therefore, for the photometry-based selection, we only use temperature in our selection. 

For the cool end, we select only stars cooler than 3981K. For the hot end, we select only stars hotter than 8000K. We bin this sample in $\Theta$ ($=5040K/T_{\rm eff}$) space with a binsize of $0.05$. We assign weight to the stars according to the inverse of the number of stars in each bin. This increases the weight for the more extreme stars, and gives them higher priority. 

We have also used an OB star catalog \citep{Liu15}, a Carbon star catalog \citep{Ji16}, an M giant star catalog \citep{Zhong15}, and an M dwarf star catalog \citep{Guo15} provided by the LAMOST team. We consider these to have higher fidelity than SED fitting, and prioritize them before the other photometrically-selected stars. 

\subsection{Optimization of Stellar-Parameter Coverage}

Because MaStar is a piggyback program, we have to yield priority to APOGEE-2 targets in plate design. This means many of our high priority targets do not get a fiber bundle assigned in the end due to target conflicts or fiber collisions. To alleviate the impact of this, we rerun our target selection algorithm after each run of plate design to reoptimize the weight distribution, taking into account those targets that are already allocated and the number of remaining designs in the remaining fields. We also took the opportunity to fix bugs and improve the selection. Again, the goal of the program is to cover as wide a parameter space as possible. Thus, continued reoptimization is beneficial.

\subsection{Ancillary Programs to Patch Parameter Space}

There have been ancillary observing opportunities in both MaNGA and APOGEE-2N programs, which have allowed us to widen our stellar-parameter space. 

MaNGA had a small shortage of fields to observe during Local Sidereal Times of 4.25-5.25 hr. This provides an opportunity to go outside the planned APOGEE-2 footprint. We selected two fields to observe during this time. The fields are chosen to contain rare metal-poor stars with relatively more high-weight stars around them. 

SDSS-North also had a call for proposals to use some extra bright time as a result of faster-than-expected survey speed. We were granted 74 hours of MaStar-led observing. We planned these fields to focus on  OB stars in star-forming regions,  supergiants, and metal-poor M dwarfs.


Some parts of the parameter space are only accessible at brighter magnitudes. These regions include hot main sequence stars, blue supergiants, very luminous red giant branch (RGB) stars, and asymptotic giant branch (AGB) stars. For example, for stars with an absolute magnitude brighter than $-5.7$, which is roughly that of late O stars, they have to be more than 30 kpc away from us to fall within our nominal magnitude limits (fainter than 11.7 in both the $g$- and $i$-bands). In addition, most of these young hot stars are in the Galactic disk. Thus, we cannot afford to observe distant stars, as they would be behind a significant amount of foreground dust. Therefore, in order to have these very luminous stars in our library, we have to find them at brighter magnitudes, and use shorter exposure times to observe them. 

Starting from this year (2018), we are designing some plates for which we adopt a much brighter magnitude limit, and observe them with shorter exposure times. With 30 second exposures, we can observe stars as bright as 8.0 mag in the $g$- or $i$-band. This would allow us to build a bigger sample of more luminous stars to widen our stellar-parameter coverage. DR15 will not include these stars yet, but future releases will. 

The release of parallax information for 1.3 billion sources in  {\it Gaia} DR2 \citep{GaiaMission, GaiaDR2} makes it much easier to select stars from extreme parts of the HR diagram. Taking advantage of {\it Gaia}, we are patching our library with hot blue stars, supergiants, tip-of-the-RGB stars and AGB stars, and very cool M dwarfs. This will also improve the parameter coverage of the final sample. 

\subsection{Isolation Constraints}\label{sec:isolationconstraints}

Flux calibration is critical for an empirical stellar library. Therefore, we need to exclude stars that could be contaminated by neighboring stars in projection. Because we are using fiber bundles to observe our targets, we actually make use of the flux ratios between the central fiber and the six surrounding fibers to constrain the exact centering of the star relative to the fiber aperture. We use this to infer the wavelength-dependent flux loss to calibrate the final spectra (see Section~\ref{sec:aperturecorrection}). Therefore, in order to ensure accurate aperture correction, we require that there is no significant contamination to the six surrounding fibers by other stars. 

More precisely, we require the flux contribution by neighboring stars to any of the six surrounding fibers used in the aperture correction to be less than 10\% of the flux contribution by the target star in that surrounding fiber. Assuming a very poor seeing (2.5\arcsec) and unlucky placement of the target and contaminating stars relative to the fibers, the above requirement translates to the following: The flux contributed by a contaminating star in a fiber placed in the direction of the contaminating star, at the distance of the contaminating star or 3\arcsec, whichever is smaller, needs to be less than 0.28\% of the total flux of the target star. For stars with more than one neighbor, we require the sum of all contaminating flux computed this way to be less than 0.28\% of the total flux of the target star, ignoring potentially different directions of the stars as if they were all in the same direction, which is a very conservative assumption. For a star to be a candidate target, this isolation requirement needs to be satisfied in every filter available. 



The isolation constraint requires our photometry catalogs to be quite complete at both the bright and faint ends. The faint limit needs to be 4.8 mag deeper than the faintest target to be absolutely complete. 
For targets that can be found in the PS1 catalog, the contamination is computed using the full PS1 catalog, which is deep enough for the magnitude limit of 17.5 for our targets. For targets found in the APASS catalog, but not found in PS1, the contamination is computed using our combined PS1/APASS catalog which is limited to 17.5 in the $g$- or $i$-band. This is not deep enough, although 85\% of these are brighter than 15.5, and they only make up 5\% of all the stars to be considered. 

The incompleteness at the bright end also matters for this isolation constraint. Neither the PS1 or the APASS catalogs are sufficiently complete for that. However, bright stars are too rare to cause issues most of the time. 

With {\it Gaia} DR2 providing photometry for nearly all stars between magnitude $\sim3$ and 21 in the $G$-band, we are now switching to use both Gaia DR2 and PS1 for contamination checking, which should pick up almost all contaminating sources. 

\subsection{Selection of Standard Stars}\label{sec:app_fluxcal}


Our standard stars are not the traditional Oke standards \citep{Oke90}, but are F stars with reasonably smooth spectra. These stars can be sufficiently well-modeled by theoretical spectra such as those based on Kurucz model atmospheres. This is the basis of our spectrophotometry for both MaNGA and MaStar. 

In MaNGA, we select F subdwarfs as standards using a set of color criteria on the SDSS photometry as described by \cite{Yan16b}. This method only works for a small fraction of fields observed by MaStar. For most of the fields, which are outside the SDSS footprint, we use PS1 and APASS for selecting the standards. 

We convert the PS1 magnitudes to SDSS filters using the relations provided by \cite{Finkbeiner16}. We measure the extinction using the Rayleigh-Jeans Color Excess method (RJCE; \citealt{Majewski11}), based on 2MASS $H$-band and WISE $W2$ magnitudes, then cap it at the value given by the \cite{SchlegelFD98} (SFD) dust map. We correct the optical magnitudes for extinction using this extinction value and the coefficients from \cite{Schlafly11} for Rv=3.1. We then define $m_{\rm dist}$ as follows, and select those stars with $m_{\rm dist} < 0.12$ as our standard star candidates: 

\begin{eqnarray}
m_{\rm dist} = [&(g-r - 0.3)^2 + (r-i - 0.09)^2 \nonumber\\
               & + (i-z-0.02)^2]^{1/2} .
\end{eqnarray}

For APASS, we only use the $g-r$ and $r-i$ colors to select standards. We define $m_{\rm dist}$ as follows, and select only those stars with $m_{\rm dist}< 0.08$ as our standard star candidates. 

\begin{equation}
m_{\rm dist} = [(g-r - 0.3)^2 + (r-i-0.09)^2]^{1/2} .
\end{equation}

Although we are using the RJCE method for extinction estimation, we recognize that the errors in infrared magnitudes could become significantly amplified in the optical, leading to large errors in our selection. Therefore, among stars that satisfy the color criteria, we preferentially picked those with the least extinction as our final standards. 

We select F stars for spectrophotometry because they are warm enough to have a relatively smooth spectrum and are still cool enough to be abundant in the field and not enter the regime where Balmer lines become less sensitive to temperature. In order to select the correct stars, MaNGA uses 5-band photometry that includes the $u$-band.  The lack of $u$-band photometry in PS1 and APASS, and the larger uncertainty in APASS magnitudes, often lead to warmer stars being selected as standards. We expanded the template set used by the pipeline to deal with this issue.

\bibliographystyle{aasjournal}
\bibliography{astro_refs}

\begin{thebibliography}{}
\expandafter\ifx\csname natexlab\endcsname\relax\def\natexlab#1{#1}\fi
\providecommand{\url}[1]{\href{#1}{#1}}
\providecommand{\dodoi}[1]{doi:~\href{http://doi.org/#1}{\nolinkurl{#1}}}
\providecommand{\doeprint}[1]{\href{http://ascl.net/#1}{\nolinkurl{http://ascl.net/#1}}}
\providecommand{\doarXiv}[1]{\href{https://arxiv.org/abs/#1}{\nolinkurl{https://arxiv.org/abs/#1}}}

\bibitem[{{Abolfathi} {et~al.}(2018){Abolfathi}, {Aguado}, {Aguilar}, {Allende
  Prieto}, {Almeida}, {Ananna}, {Anders}, {Anderson}, {Andrews}, {Anguiano}, \&
  et~al.}]{SDSSDR14}
{Abolfathi}, B., {Aguado}, D.~S., {Aguilar}, G., {et~al.} 2018, \apjs, 235, 42,
  \dodoi{10.3847/1538-4365/aa9e8a}

\bibitem[{{Adelman-McCarthy} {et~al.}(2008)}]{SDSSDR6}
{Adelman-McCarthy}, J.~K., {et~al.} 2008, \apjs, 175, 297,
  \dodoi{10.1086/524984}

\bibitem[{{Aguado} {et~al.}(2019){Aguado}, {Ahumada}, {Almeida}, {Anderson},
  {Andrews}, {Anguiano}, {Aquino Ort{\'\i}z}, {Arag{\'o}n-Salamanca},
  {Argudo-Fern{\'a}ndez}, {Aubert}, \& et~al.}]{SDSSDR15}
{Aguado}, D.~S., {Ahumada}, R., {Almeida}, A., {et~al.} 2019, \apjs, 240, 23,
  \dodoi{10.3847/1538-4365/aaf651}

\bibitem[{{Alam} {et~al.}(2015){Alam}, {Albareti}, {Allende Prieto}, {Anders},
  {Anderson}, {Anderton}, {Andrews}, {Armengaud}, {Aubourg}, {Bailey}, \&
  et~al.}]{SDSSDR12}
{Alam}, S., {Albareti}, F.~D., {Allende Prieto}, C., {et~al.} 2015, \apjs, 219,
  12, \dodoi{10.1088/0067-0049/219/1/12}

\bibitem[{{Albareti} {et~al.}(2017){Albareti}, {Allende Prieto}, {Almeida},
  {Anders}, {Anderson}, {Andrews}, {Arag{\'o}n-Salamanca},
  {Argudo-Fern{\'a}ndez}, {Armengaud}, {Aubourg}, \& et~al.}]{SDSSDR13}
{Albareti}, F.~D., {Allende Prieto}, C., {Almeida}, A., {et~al.} 2017, \apjs,
  233, 25, \dodoi{10.3847/1538-4365/aa8992}

\bibitem[{{Allende Prieto} {et~al.}(2008){Allende Prieto}, {Sivarani}, {Beers},
  {Lee}, {Koesterke}, {Shetrone}, {Sneden}, {Lambert}, {Wilhelm}, {Rockosi},
  {Lai}, {Yanny}, {Ivans}, {Johnson}, {Aoki}, {Bailer-Jones}, \& {Re
  Fiorentin}}]{AllendePrieto08}
{Allende Prieto}, C., {Sivarani}, T., {Beers}, T.~C., {et~al.} 2008, \aj, 136,
  2070, \dodoi{10.1088/0004-6256/136/5/2070}

\bibitem[{{Allende Prieto} {et~al.}(2014){Allende Prieto},
  {Fern{\'a}ndez-Alvar}, {Schlesinger}, {Lee}, {Morrison}, {Schneider},
  {Beers}, {Bizyaev}, {Ebelke}, {Malanushenko}, {Malanushenko}, {Oravetz},
  {Pan}, {Simmons}, {Simmerer}, {Sobeck}, \& {Robin}}]{AllendePrieto14}
{Allende Prieto}, C., {Fern{\'a}ndez-Alvar}, E., {Schlesinger}, K.~J., {et~al.}
  2014, \aap, 568, A7, \dodoi{10.1051/0004-6361/201424053}

\bibitem[{{Asplund} {et~al.}(2005){Asplund}, {Grevesse}, \&
  {Sauval}}]{Asplund05}
{Asplund}, M., {Grevesse}, N., \& {Sauval}, A.~J. 2005, in Astronomical Society
  of the Pacific Conference Series, Vol. 336, Cosmic Abundances as Records of
  Stellar Evolution and Nucleosynthesis, ed. T.~G. {Barnes}, III \& F.~N.
  {Bash}, 25

\bibitem[{{Bailer-Jones} {et~al.}(2018){Bailer-Jones}, {Rybizki}, {Fouesneau},
  {Mantelet}, \& {Andrae}}]{Bailer-Jones18}
{Bailer-Jones}, C.~A.~L., {Rybizki}, J., {Fouesneau}, M., {Mantelet}, G., \&
  {Andrae}, R. 2018, \aj, 156, 58, \dodoi{10.3847/1538-3881/aacb21}

\bibitem[{{Barbuy} {et~al.}(2003){Barbuy}, {Perrin}, {Katz}, {Coelho},
  {Cayrel}, {Spite}, \& {Van't Veer-Menneret}}]{Barbuy03}
{Barbuy}, B., {Perrin}, M.-N., {Katz}, D., {et~al.} 2003, \aap, 404, 661,
  \dodoi{10.1051/0004-6361:20030496}

\bibitem[{{Blanton} {et~al.}(2017){Blanton}, {Bershady}, {Abolfathi},
  {Albareti}, {Allende Prieto}, {Almeida}, {Alonso-Garc{\'{\i}}a}, {Anders},
  {Anderson}, {Andrews}, \& et~al.}]{Blanton17}
{Blanton}, M.~R., {Bershady}, M.~A., {Abolfathi}, B., {et~al.} 2017, \aj, 154,
  28, \dodoi{10.3847/1538-3881/aa7567}

\bibitem[{{Bohlin} {et~al.}(2017){Bohlin}, {M{\'e}sz{\'a}ros}, {Fleming},
  {Gordon}, {Koekemoer}, \& {Kov{\'a}cs}}]{BohlinM17}
{Bohlin}, R.~C., {M{\'e}sz{\'a}ros}, S., {Fleming}, S.~W., {et~al.} 2017, \aj,
  153, 234, \dodoi{10.3847/1538-3881/aa6ba9}

\bibitem[{{Bressan} {et~al.}(2012){Bressan}, {Marigo}, {Girardi}, {Salasnich},
  {Dal Cero}, {Rubele}, \& {Nanni}}]{Bressan12}
{Bressan}, A., {Marigo}, P., {Girardi}, L., {et~al.} 2012, \mnras, 427, 127,
  \dodoi{10.1111/j.1365-2966.2012.21948.x}

\bibitem[{{Bruzual} \& {Charlot}(2003)}]{BC03}
{Bruzual}, G., \& {Charlot}, S. 2003, \mnras, 344, 1000,
  \dodoi{10.1046/j.1365-8711.2003.06897.x}

\bibitem[{{Bundy} {et~al.}(2015){Bundy}, {Bershady}, {Law}, {Yan}, {Drory},
  {MacDonald}, {Wake}, {Cherinka}, {S{\'a}nchez-Gallego}, {Weijmans}, {Thomas},
  {Tremonti}, {Masters}, {Coccato}, {Diamond-Stanic}, {Arag{\'o}n-Salamanca},
  {Avila-Reese}, {Badenes}, {Falc{\'o}n-Barroso}, {Belfiore}, {Bizyaev},
  {Blanc}, {Bland-Hawthorn}, {Blanton}, {Brownstein}, {Byler}, {Cappellari},
  {Conroy}, {Dutton}, {Emsellem}, {Etherington}, {Frinchaboy}, {Fu}, {Gunn},
  {Harding}, {Johnston}, {Kauffmann}, {Kinemuchi}, {Klaene}, {Knapen},
  {Leauthaud}, {Li}, {Lin}, {Maiolino}, {Malanushenko}, {Malanushenko}, {Mao},
  {Maraston}, {McDermid}, {Merrifield}, {Nichol}, {Oravetz}, {Pan}, {Parejko},
  {Sanchez}, {Schlegel}, {Simmons}, {Steele}, {Steinmetz}, {Thanjavur},
  {Thompson}, {Tinker}, {van den Bosch}, {Westfall}, {Wilkinson}, {Wright},
  {Xiao}, \& {Zhang}}]{Bundy15}
{Bundy}, K., {Bershady}, M.~A., {Law}, D.~R., {et~al.} 2015, \apj, 798, 7,
  \dodoi{10.1088/0004-637X/798/1/7}

\bibitem[{{Cenarro} {et~al.}(2001){Cenarro}, {Cardiel}, {Gorgas}, {Peletier},
  {Vazdekis}, \& {Prada}}]{Cenarro01}
{Cenarro}, A.~J., {Cardiel}, N., {Gorgas}, J., {et~al.} 2001, \mnras, 326, 959,
  \dodoi{10.1046/j.1365-8711.2001.04688.x}

\bibitem[{{Cenarro} {et~al.}(2007){Cenarro}, {Peletier},
  {S{\'a}nchez-Bl{\'a}zquez}, {Selam}, {Toloba}, {Cardiel},
  {Falc{\'o}n-Barroso}, {Gorgas}, {Jim{\'e}nez-Vicente}, \&
  {Vazdekis}}]{Cenarro07}
{Cenarro}, A.~J., {Peletier}, R.~F., {S{\'a}nchez-Bl{\'a}zquez}, P., {et~al.}
  2007, \mnras, 374, 664, \dodoi{10.1111/j.1365-2966.2006.11196.x}

\bibitem[{{Chambers} {et~al.}(2016){Chambers}, {Magnier}, {Metcalfe},
  {Flewelling}, {Huber}, {Waters}, {Denneau}, {Draper}, {Farrow}, {Finkbeiner},
  {Holmberg}, {Koppenhoefer}, {Price}, {Saglia}, {Schlafly}, {Smartt},
  {Sweeney}, {Wainscoat}, {Burgett}, {Grav}, {Heasley}, {Hodapp}, {Jedicke},
  {Kaiser}, {Kudritzki}, {Luppino}, {Lupton}, {Monet}, {Morgan}, {Onaka},
  {Stubbs}, {Tonry}, {Banados}, {Bell}, {Bender}, {Bernard}, {Botticella},
  {Casertano}, {Chastel}, {Chen}, {Chen}, {Cole}, {Deacon}, {Frenk},
  {Fitzsimmons}, {Gezari}, {Goessl}, {Goggia}, {Goldman}, {Grebel}, {Hambly},
  {Hasinger}, {Heavens}, {Heckman}, {Henderson}, {Henning}, {Holman}, {Hopp},
  {Ip}, {Isani}, {Keyes}, {Koekemoer}, {Kotak}, {Long}, {Lucey}, {Liu},
  {Martin}, {McLean}, {Morganson}, {Murphy}, {Nieto-Santisteban}, {Norberg},
  {Peacock}, {Pier}, {Postman}, {Primak}, {Rae}, {Rest}, {Riess}, {Riffeser},
  {Rix}, {Roser}, {Schilbach}, {Schultz}, {Scolnic}, {Szalay}, {Seitz},
  {Shiao}, {Small}, {Smith}, {Soderblom}, {Taylor}, {Thakar}, {Thiel},
  {Thilker}, {Urata}, {Valenti}, {Walter}, {Watters}, {Werner}, {White},
  {Wood-Vasey}, \& {Wyse}}]{Chambers16}
{Chambers}, K.~C., {Magnier}, E.~A., {Metcalfe}, N., {et~al.} 2016, ArXiv
  e-prints.
\newblock \doarXiv{1612.05560}

\bibitem[{{Chen} {et~al.}(2014){Chen}, {Trager}, {Peletier}, {Lan{\c c}on},
  {Vazdekis}, {Prugniel}, {Silva}, \& {Gonneau}}]{Chen14}
{Chen}, Y.-P., {Trager}, S.~C., {Peletier}, R.~F., {et~al.} 2014, \aap, 565,
  A117, \dodoi{10.1051/0004-6361/201322505}

\bibitem[{{Coelho} {et~al.}(2005){Coelho}, {Barbuy}, {Mel{\'e}ndez},
  {Schiavon}, \& {Castilho}}]{Coelho05}
{Coelho}, P., {Barbuy}, B., {Mel{\'e}ndez}, J., {Schiavon}, R.~P., \&
  {Castilho}, B.~V. 2005, \aap, 443, 735, \dodoi{10.1051/0004-6361:20053511}

\bibitem[{{Coelho} {et~al.}(2007){Coelho}, {Bruzual}, {Charlot}, {Weiss},
  {Barbuy}, \& {Ferguson}}]{Coelho07}
{Coelho}, P., {Bruzual}, G., {Charlot}, S., {et~al.} 2007, \mnras, 382, 498,
  \dodoi{10.1111/j.1365-2966.2007.12364.x}

\bibitem[{{Coelho}(2014)}]{Coelho14}
{Coelho}, P.~R.~T. 2014, \mnras, 440, 1027, \dodoi{10.1093/mnras/stu365}

\bibitem[{{Conroy}(2013)}]{Conroy13}
{Conroy}, C. 2013, \araa, 51, 393, \dodoi{10.1146/annurev-astro-082812-141017}

\bibitem[{{Conroy} {et~al.}(2009){Conroy}, {Gunn}, \& {White}}]{Conroy09}
{Conroy}, C., {Gunn}, J.~E., \& {White}, M. 2009, \apj, 699, 486,
  \dodoi{10.1088/0004-637X/699/1/486}

\bibitem[{{Cui} {et~al.}(2012){Cui}, {Zhao}, {Chu}, {Li}, {Li}, {Zhang}, {Su},
  {Yao}, {Wang}, {Xing}, {Li}, {Zhu}, {Wang}, {Gu}, {Luo}, {Xu}, {Zhang},
  {Liu}, {Zhang}, {Yang}, {Cao}, {Chen}, {Chen}, {Chen}, {Chen}, {Chu}, {Feng},
  {Gong}, {Hou}, {Hu}, {Hu}, {Hu}, {Jia}, {Jiang}, {Jiang}, {Jiang}, {Jin},
  {Li}, {Li}, {Li}, {Liu}, {Liu}, {Lu}, {Mao}, {Men}, {Qi}, {Qi}, {Shi},
  {Tang}, {Tao}, {Wang}, {Wang}, {Wang}, {Wang}, {Wang}, {Wang}, {Wang},
  {Wang}, {Wang}, {Wang}, {Wang}, {Wang}, {Xu}, {Xu}, {Yang}, {Yu}, {Yuan},
  {Yuan}, {Zhai}, {Zhang}, {Zhang}, {Zhang}, {Zhao}, {Zhou}, {Zhou}, {Zhu}, \&
  {Zou}}]{Cui2012}
{Cui}, X.-Q., {Zhao}, Y.-H., {Chu}, Y.-Q., {et~al.} 2012, Research in Astronomy
  and Astrophysics, 12, 1197, \dodoi{10.1088/1674-4527/12/9/003}

\bibitem[{{Danielski} {et~al.}(2018){Danielski}, {Babusiaux}, {Ruiz-Dern},
  {Sartoretti}, \& {Arenou}}]{Danielski18}
{Danielski}, C., {Babusiaux}, C., {Ruiz-Dern}, L., {Sartoretti}, P., \&
  {Arenou}, F. 2018, \aap, 614, A19, \dodoi{10.1051/0004-6361/201732327}

\bibitem[{{Dawson} {et~al.}(2013){Dawson}, {Schlegel}, {Ahn}, {Anderson},
  {Aubourg}, {Bailey}, {Barkhouser}, {Bautista}, {Beifiori}, {Berlind},
  {Bhardwaj}, {Bizyaev}, {Blake}, {Blanton}, {Blomqvist}, {Bolton}, {Borde},
  {Bovy}, {Brandt}, {Brewington}, {Brinkmann}, {Brown}, {Brownstein}, {Bundy},
  {Busca}, {Carithers}, {Carnero}, {Carr}, {Chen}, {Comparat}, {Connolly},
  {Cope}, {Croft}, {Cuesta}, {da Costa}, {Davenport}, {Delubac}, {de Putter},
  {Dhital}, {Ealet}, {Ebelke}, {Eisenstein}, {Escoffier}, {Fan}, {Filiz Ak},
  {Finley}, {Font-Ribera}, {G{\'e}nova-Santos}, {Gunn}, {Guo}, {Haggard},
  {Hall}, {Hamilton}, {Harris}, {Harris}, {Ho}, {Hogg}, {Holder}, {Honscheid},
  {Huehnerhoff}, {Jordan}, {Jordan}, {Kauffmann}, {Kazin}, {Kirkby}, {Klaene},
  {Kneib}, {Le Goff}, {Lee}, {Long}, {Loomis}, {Lundgren}, {Lupton}, {Maia},
  {Makler}, {Malanushenko}, {Malanushenko}, {Mandelbaum}, {Manera}, {Maraston},
  {Margala}, {Masters}, {McBride}, {McDonald}, {McGreer}, {McMahon}, {Mena},
  {Miralda-Escud{\'e}}, {Montero-Dorta}, {Montesano}, {Muna}, {Myers},
  {Naugle}, {Nichol}, {Noterdaeme}, {Nuza}, {Olmstead}, {Oravetz}, {Oravetz},
  {Owen}, {Padmanabhan}, {Palanque-Delabrouille}, {Pan}, {Parejko},
  {P{\^a}ris}, {Percival}, {P{\'e}rez-Fournon}, {P{\'e}rez-R{\`a}fols},
  {Petitjean}, {Pfaffenberger}, {Pforr}, {Pieri}, {Prada}, {Price-Whelan},
  {Raddick}, {Rebolo}, {Rich}, {Richards}, {Rockosi}, {Roe}, {Ross}, {Ross},
  {Rossi}, {Rubi{\~n}o-Martin}, {Samushia}, {S{\'a}nchez}, {Sayres}, {Schmidt},
  {Schneider}, {Sc{\'o}ccola}, {Seo}, {Shelden}, {Sheldon}, {Shen}, {Shu},
  {Slosar}, {Smee}, {Snedden}, {Stauffer}, {Steele}, {Strauss}, {Streblyanska},
  {Suzuki}, {Swanson}, {Tal}, {Tanaka}, {Thomas}, {Tinker}, {Tojeiro},
  {Tremonti}, {Vargas Maga{\~n}a}, {Verde}, {Viel}, {Wake}, {Watson}, {Weaver},
  {Weinberg}, {Weiner}, {West}, {White}, {Wood-Vasey}, {Yeche}, {Zehavi},
  {Zhao}, \& {Zheng}}]{Dawson13}
{Dawson}, K.~S., {Schlegel}, D.~J., {Ahn}, C.~P., {et~al.} 2013, \aj, 145, 10,
  \dodoi{10.1088/0004-6256/145/1/10}

\bibitem[{{Dawson} {et~al.}(2016){Dawson}, {Kneib}, {Percival}, {Alam},
  {Albareti}, {Anderson}, {Armengaud}, {Aubourg}, {Bailey}, {Bautista},
  {Berlind}, {Bershady}, {Beutler}, {Bizyaev}, {Blanton}, {Blomqvist},
  {Bolton}, {Bovy}, {Brandt}, {Brinkmann}, {Brownstein}, {Burtin}, {Busca},
  {Cai}, {Chuang}, {Clerc}, {Comparat}, {Cope}, {Croft}, {Cruz-Gonzalez}, {da
  Costa}, {Cousinou}, {Darling}, {de la Macorra}, {de la Torre}, {Delubac}, {du
  Mas des Bourboux}, {Dwelly}, {Ealet}, {Eisenstein}, {Eracleous}, {Escoffier},
  {Fan}, {Finoguenov}, {Font-Ribera}, {Frinchaboy}, {Gaulme}, {Georgakakis},
  {Green}, {Guo}, {Guy}, {Ho}, {Holder}, {Huehnerhoff}, {Hutchinson}, {Jing},
  {Jullo}, {Kamble}, {Kinemuchi}, {Kirkby}, {Kitaura}, {Klaene}, {Laher},
  {Lang}, {Laurent}, {Le Goff}, {Li}, {Liang}, {Lima}, {Lin}, {Lin}, {Lin},
  {Long}, {Lundgren}, {MacDonald}, {Geimba Maia}, {Malanushenko},
  {Malanushenko}, {Mariappan}, {McBride}, {McGreer}, {M{\'e}nard}, {Merloni},
  {Meza}, {Montero-Dorta}, {Muna}, {Myers}, {Nandra}, {Naugle}, {Newman},
  {Noterdaeme}, {Nugent}, {Ogando}, {Olmstead}, {Oravetz}, {Oravetz},
  {Padmanabhan}, {Palanque-Delabrouille}, {Pan}, {Parejko}, {P{\^a}ris},
  {Peacock}, {Petitjean}, {Pieri}, {Pisani}, {Prada}, {Prakash}, {Raichoor},
  {Reid}, {Rich}, {Ridl}, {Rodriguez-Torres}, {Carnero Rosell}, {Ross},
  {Rossi}, {Ruan}, {Salvato}, {Sayres}, {Schneider}, {Schlegel}, {Seljak},
  {Seo}, {Sesar}, {Shandera}, {Shu}, {Slosar}, {Sobreira}, {Streblyanska},
  {Suzuki}, {Taylor}, {Tao}, {Tinker}, {Tojeiro}, {Vargas-Maga{\~n}a}, {Wang},
  {Weaver}, {Weinberg}, {White}, {Wood-Vasey}, {Yeche}, {Zhai}, {Zhao}, {Zhao},
  {Zheng}, {Ben Zhu}, \& {Zou}}]{Dawson16}
{Dawson}, K.~S., {Kneib}, J.-P., {Percival}, W.~J., {et~al.} 2016, \aj, 151,
  44, \dodoi{10.3847/0004-6256/151/2/44}

\bibitem[{{de Laverny} {et~al.}(2012){de Laverny}, {Recio-Blanco}, {Worley}, \&
  {Plez}}]{deLaverny12}
{de Laverny}, P., {Recio-Blanco}, A., {Worley}, C.~C., \& {Plez}, B. 2012,
  \aap, 544, A126, \dodoi{10.1051/0004-6361/201219330}

\bibitem[{{Deng} {et~al.}(2012){Deng}, {Newberg}, {Liu}, {Carlin}, {Beers},
  {Chen}, {Chen}, {Christlieb}, {Grillmair}, {Guhathakurta}, {Han}, {Hou},
  {Lee}, {L{\'e}pine}, {Li}, {Liu}, {Pan}, {Sellwood}, {Wang}, {Wang}, {Yang},
  {Yanny}, {Zhang}, {Zhang}, {Zheng}, \& {Zhu}}]{Deng12}
{Deng}, L.-C., {Newberg}, H.~J., {Liu}, C., {et~al.} 2012, Research in
  Astronomy and Astrophysics, 12, 735, \dodoi{10.1088/1674-4527/12/7/003}

\bibitem[{{Diaz} {et~al.}(1989){Diaz}, {Terlevich}, \& {Terlevich}}]{Diaz89}
{Diaz}, A.~I., {Terlevich}, E., \& {Terlevich}, R. 1989, \mnras, 239, 325,
  \dodoi{10.1093/mnras/239.2.325}

\bibitem[{{Drory} {et~al.}(2015){Drory}, {MacDonald}, {Bershady}, {Bundy},
  {Gunn}, {Law}, {Smith}, {Stoll}, {Tremonti}, {Wake}, {Yan}, {Weijmans},
  {Byler}, {Cherinka}, {Cope}, {Eigenbrot}, {Harding}, {Holder}, {Huehnerhoff},
  {Jaehnig}, {Jansen}, {Klaene}, {Paat}, {Percival}, \& {Sayres}}]{Drory15}
{Drory}, N., {MacDonald}, N., {Bershady}, M.~A., {et~al.} 2015, \aj, 149, 77,
  \dodoi{10.1088/0004-6256/149/2/77}

\bibitem[{{Evans} {et~al.}(2018){Evans}, {Riello}, {De Angeli}, {Carrasco},
  {Montegriffo}, {Fabricius}, {Jordi}, {Palaversa}, {Diener}, {Busso},
  {Cacciari}, {van Leeuwen}, {Burgess}, {Davidson}, {Harrison}, {Hodgkin},
  {Pancino}, {Richards}, {Altavilla}, {Balaguer-N{\'u}{\~n}ez}, {Barstow},
  {Bellazzini}, {Brown}, {Castellani}, {Cocozza}, {De Luise}, {Delgado},
  {Ducourant}, {Galleti}, {Gilmore}, {Giuffrida}, {Holl}, {Kewley}, {Koposov},
  {Marinoni}, {Marrese}, {Osborne}, {Piersimoni}, {Portell}, {Pulone},
  {Ragaini}, {Sanna}, {Terrett}, {Walton}, {Wevers}, \&
  {Wyrzykowski}}]{Evans18}
{Evans}, D.~W., {Riello}, M., {De Angeli}, F., {et~al.} 2018, \aap, 616, A4,
  \dodoi{10.1051/0004-6361/201832756}

\bibitem[{{Falc{\'o}n-Barroso} {et~al.}(2011){Falc{\'o}n-Barroso},
  {S{\'a}nchez-Bl{\'a}zquez}, {Vazdekis}, {Ricciardelli}, {Cardiel}, {Cenarro},
  {Gorgas}, \& {Peletier}}]{FalconBarroso11}
{Falc{\'o}n-Barroso}, J., {S{\'a}nchez-Bl{\'a}zquez}, P., {Vazdekis}, A.,
  {et~al.} 2011, \aap, 532, A95, \dodoi{10.1051/0004-6361/201116842}

\bibitem[{{Finkbeiner} {et~al.}(2016){Finkbeiner}, {Schlafly}, {Schlegel},
  {Padmanabhan}, {Juri{\'c}}, {Burgett}, {Chambers}, {Denneau}, {Draper},
  {Flewelling}, {Hodapp}, {Kaiser}, {Magnier}, {Metcalfe}, {Morgan}, {Price},
  {Stubbs}, \& {Tonry}}]{Finkbeiner16}
{Finkbeiner}, D.~P., {Schlafly}, E.~F., {Schlegel}, D.~J., {et~al.} 2016, \apj,
  822, 66, \dodoi{10.3847/0004-637X/822/2/66}

\bibitem[{{Fitzpatrick}(1999)}]{Fitzpatrick99}
{Fitzpatrick}, E.~L. 1999, \pasp, 111, 63, \dodoi{10.1086/316293}

\bibitem[{{Fr{\'e}maux} {et~al.}(2006){Fr{\'e}maux}, {Kupka}, {Boisson},
  {Joly}, \& {Tsymbal}}]{Fremaux06}
{Fr{\'e}maux}, J., {Kupka}, F., {Boisson}, C., {Joly}, M., \& {Tsymbal}, V.
  2006, \aap, 449, 109, \dodoi{10.1051/0004-6361:20053699}

\bibitem[{{Gaia Collaboration} {et~al.}(2016){Gaia Collaboration}, {Prusti},
  {de Bruijne}, {Brown}, {Vallenari}, {Babusiaux}, {Bailer-Jones}, {Bastian},
  {Biermann}, {Evans}, \& et~al.}]{GaiaMission}
{Gaia Collaboration}, {Prusti}, T., {de Bruijne}, J.~H.~J., {et~al.} 2016,
  \aap, 595, A1, \dodoi{10.1051/0004-6361/201629272}

\bibitem[{{Gaia Collaboration} {et~al.}(2018){Gaia Collaboration}, {Brown},
  {Vallenari}, {Prusti}, {de Bruijne}, {Babusiaux}, {Bailer-Jones}, {Biermann},
  {Evans}, {Eyer}, \& et~al.}]{GaiaDR2}
{Gaia Collaboration}, {Brown}, A.~G.~A., {Vallenari}, A., {et~al.} 2018, \aap,
  616, A1, \dodoi{10.1051/0004-6361/201833051}

\bibitem[{{Garc{\'{\i}}a P{\'e}rez} {et~al.}(2016){Garc{\'{\i}}a P{\'e}rez},
  {Allende Prieto}, {Holtzman}, {Shetrone}, {M{\'e}sz{\'a}ros}, {Bizyaev},
  {Carrera}, {Cunha}, {Garc{\'{\i}}a-Hern{\'a}ndez}, {Johnson}, {Majewski},
  {Nidever}, {Schiavon}, {Shane}, {Smith}, {Sobeck}, {Troup}, {Zamora},
  {Weinberg}, {Bovy}, {Eisenstein}, {Feuillet}, {Frinchaboy}, {Hayden},
  {Hearty}, {Nguyen}, {O'Connell}, {Pinsonneault}, {Wilson}, \&
  {Zasowski}}]{GarciaPerez16}
{Garc{\'{\i}}a P{\'e}rez}, A.~E., {Allende Prieto}, C., {Holtzman}, J.~A.,
  {et~al.} 2016, \aj, 151, 144, \dodoi{10.3847/0004-6256/151/6/144}

\bibitem[{{Gray} \& {Corbally}(1994)}]{GrayO94}
{Gray}, R.~O., \& {Corbally}, C.~J. 1994, \aj, 107, 742, \dodoi{10.1086/116893}

\bibitem[{{Green} {et~al.}(2019){Green}, {Schlafly}, {Zucker}, {Speagle}, \&
  {Finkbeiner}}]{Green19}
{Green}, G.~M., {Schlafly}, E.~F., {Zucker}, C., {Speagle}, J.~S., \&
  {Finkbeiner}, D.~P. 2019, arXiv e-prints, arXiv:1905.02734.
\newblock \doarXiv{1905.02734}

\bibitem[{{Gregg} {et~al.}(2006){Gregg}, {Silva}, {Rayner}, {Worthey},
  {Valdes}, {Pickles}, {Rose}, {Carney}, \& {Vacca}}]{Gregg06}
{Gregg}, M.~D., {Silva}, D., {Rayner}, J., {et~al.} 2006, in The 2005 HST
  Calibration Workshop: Hubble After the Transition to Two-Gyro Mode, ed. A.~M.
  {Koekemoer}, P.~{Goudfrooij}, \& L.~L. {Dressel}, 209

\bibitem[{{Gunn} \& {Stryker}(1983)}]{GunnS83}
{Gunn}, J.~E., \& {Stryker}, L.~L. 1983, \apjs, 52, 121, \dodoi{10.1086/190861}

\bibitem[{{Gunn} {et~al.}(2006){Gunn}, {Siegmund}, {Mannery}, {Owen}, {Hull},
  {Leger}, {Carey}, {Knapp}, {York}, {Boroski}, {Kent}, {Lupton}, {Rockosi},
  {Evans}, {Waddell}, {Anderson}, {Annis}, {Barentine}, {Bartoszek}, {Bastian},
  {Bracker}, {Brewington}, {Briegel}, {Brinkmann}, {Brown}, {Carr},
  {Czarapata}, {Drennan}, {Dombeck}, {Federwitz}, {Gillespie}, {Gonzales},
  {Hansen}, {Harvanek}, {Hayes}, {Jordan}, {Kinney}, {Klaene}, {Kleinman},
  {Kron}, {Kresinski}, {Lee}, {Limmongkol}, {Lindenmeyer}, {Long}, {Loomis},
  {McGehee}, {Mantsch}, {Neilsen}, {Neswold}, {Newman}, {Nitta}, {Peoples},
  {Pier}, {Prieto}, {Prosapio}, {Rivetta}, {Schneider}, {Snedden}, \&
  {Wang}}]{Gunn06}
{Gunn}, J.~E., {Siegmund}, W.~A., {Mannery}, E.~J., {et~al.} 2006, \aj, 131,
  2332, \dodoi{10.1086/500975}

\bibitem[{{Guo} {et~al.}(2015){Guo}, {Yi}, {Luo}, {Wang}, {Bai}, {Yang},
  {Song}, {Chen}, {Chen}, {Zuo}, {Du}, {Zhang}, {Li}, {Kong}, {Wang}, {Wu},
  {Wu}, {Zhao}, {Zhang}, {Hou}, {Wang}, \& {Yang}}]{Guo15}
{Guo}, Y.-X., {Yi}, Z.-P., {Luo}, A.-L., {et~al.} 2015, Research in Astronomy
  and Astrophysics, 15, 1182, \dodoi{10.1088/1674-4527/15/8/007}

\bibitem[{{Ho} {et~al.}(2017){Ho}, {Ness}, {Hogg}, {Rix}, {Liu}, {Yang},
  {Zhang}, {Hou}, \& {Wang}}]{Ho.A17}
{Ho}, A.~Y.~Q., {Ness}, M.~K., {Hogg}, D.~W., {et~al.} 2017, \apj, 836, 5,
  \dodoi{10.3847/1538-4357/836/1/5}

\bibitem[{{Holtzman} {et~al.}(2015){Holtzman}, {Shetrone}, {Johnson}, {Allende
  Prieto}, {Anders}, {Andrews}, {Beers}, {Bizyaev}, {Blanton}, {Bovy},
  {Carrera}, {Chojnowski}, {Cunha}, {Eisenstein}, {Feuillet}, {Frinchaboy},
  {Galbraith-Frew}, {Garc{\'{\i}}a P{\'e}rez}, {Garc{\'{\i}}a-Hern{\'a}ndez},
  {Hasselquist}, {Hayden}, {Hearty}, {Ivans}, {Majewski}, {Martell},
  {Meszaros}, {Muna}, {Nidever}, {Nguyen}, {O'Connell}, {Pan}, {Pinsonneault},
  {Robin}, {Schiavon}, {Shane}, {Sobeck}, {Smith}, {Troup}, {Weinberg},
  {Wilson}, {Wood-Vasey}, {Zamora}, \& {Zasowski}}]{Holtzman15}
{Holtzman}, J.~A., {Shetrone}, M., {Johnson}, J.~A., {et~al.} 2015, \aj, 150,
  148, \dodoi{10.1088/0004-6256/150/5/148}

\bibitem[{{Holtzman} {et~al.}(2018){Holtzman}, {Hasselquist}, {Shetrone},
  {Cunha}, {Allende Prieto}, {Anguiano}, {Bizyaev}, {Bovy}, {Casey},
  {Edvardsson}, {Johnson}, {J{\"o}nsson}, {Meszaros}, {Smith}, {Sobeck},
  {Zamora}, {Chojnowski}, {Fernandez-Trincado}, {Garcia-Hernandez}, {Majewski},
  {Pinsonneault}, {Souto}, {Stringfellow}, {Tayar}, {Troup}, \&
  {Zasowski}}]{Holtzman18}
{Holtzman}, J.~A., {Hasselquist}, S., {Shetrone}, M., {et~al.} 2018, \aj, 156,
  125, \dodoi{10.3847/1538-3881/aad4f9}

\bibitem[{{Ji} {et~al.}(2016){Ji}, {Cui}, {Liu}, {Luo}, {Zhao}, \&
  {Zhang}}]{Ji16}
{Ji}, W., {Cui}, W., {Liu}, C., {et~al.} 2016, \apjs, 226, 1,
  \dodoi{10.3847/0067-0049/226/1/1}

\bibitem[{{Kirby}(2011)}]{Kirby11}
{Kirby}, E.~N. 2011, \pasp, 123, 531, \dodoi{10.1086/660019}

\bibitem[{{Koleva} {et~al.}(2009){Koleva}, {Prugniel}, {Bouchard}, \&
  {Wu}}]{Koleva09}
{Koleva}, M., {Prugniel}, P., {Bouchard}, A., \& {Wu}, Y. 2009, \aap, 501,
  1269, \dodoi{10.1051/0004-6361/200811467}

\bibitem[{{Koleva} {et~al.}(2011){Koleva}, {Prugniel}, {de Rijcke}, \&
  {Zeilinger}}]{Koleva11}
{Koleva}, M., {Prugniel}, P., {de Rijcke}, S., \& {Zeilinger}, W.~W. 2011,
  \mnras, 417, 1643, \dodoi{10.1111/j.1365-2966.2011.19057.x}

\bibitem[{{Kurucz}(1979)}]{Kurucz79}
{Kurucz}, R.~L. 1979, \apjs, 40, 1, \dodoi{10.1086/190589}

\bibitem[{{Kurucz}(2011)}]{Kurucz11}
---. 2011, Canadian Journal of Physics, 89, 417, \dodoi{10.1139/p10-104}

\bibitem[{{Kurucz} \& {Avrett}(1981)}]{KuruczA81}
{Kurucz}, R.~L., \& {Avrett}, E.~H. 1981, SAO Special Report, 391

\bibitem[{{Lan{\c c}on} \& {Wood}(2000)}]{LanconW00}
{Lan{\c c}on}, A., \& {Wood}, P.~R. 2000, \aaps, 146, 217,
  \dodoi{10.1051/aas:2000269}

\bibitem[{{Law} {et~al.}(2015){Law}, {Yan}, {Bershady}, {Bundy}, {Cherinka},
  {Drory}, {MacDonald}, {S{\'a}nchez-Gallego}, {Wake}, {Weijmans}, {Blanton},
  {Klaene}, {Moran}, {Sanchez}, \& {Zhang}}]{Law15}
{Law}, D.~R., {Yan}, R., {Bershady}, M.~A., {et~al.} 2015, \aj, 150, 19,
  \dodoi{10.1088/0004-6256/150/1/19}

\bibitem[{{Law} {et~al.}(2016){Law}, {Cherinka}, {Yan}, {Andrews}, {Bershady},
  {Bizyaev}, {Blanc}, {Blanton}, {Bolton}, {Brownstein}, {Bundy}, {Chen},
  {Drory}, {D'Souza}, {Fu}, {Jones}, {Kauffmann}, {MacDonald}, {Masters},
  {Newman}, {Parejko}, {S{\'a}nchez-Gallego}, {S{\'a}nchez}, {Schlegel},
  {Thomas}, {Wake}, {Weijmans}, {Westfall}, \& {Zhang}}]{Law16}
{Law}, D.~R., {Cherinka}, B., {Yan}, R., {et~al.} 2016, \aj, 152, 83,
  \dodoi{10.3847/0004-6256/152/4/83}

\bibitem[{{Le Borgne} {et~al.}(2003){Le Borgne}, {Bruzual}, {Pell{\'o}},
  {Lan{\c{c}}on}, {Rocca-Volmerange}, {Sanahuja}, {Schaerer}, {Soubiran}, \&
  {V{\'\i}lchez-G{\'o}mez}}]{LeBorgne03}
{Le Borgne}, J.~F., {Bruzual}, G., {Pell{\'o}}, R., {et~al.} 2003, \aap, 402,
  433, \dodoi{10.1051/0004-6361:20030243}

\bibitem[{{Lee} {et~al.}(2008{\natexlab{a}}){Lee}, {Beers}, {Sivarani},
  {Allende Prieto}, {Koesterke}, {Wilhelm}, {Re Fiorentin}, {Bailer-Jones},
  {Norris}, {Rockosi}, {Yanny}, {Newberg}, {Covey}, {Zhang}, \& {Luo}}]{Lee08a}
{Lee}, Y.~S., {Beers}, T.~C., {Sivarani}, T., {et~al.} 2008{\natexlab{a}}, \aj,
  136, 2022, \dodoi{10.1088/0004-6256/136/5/2022}

\bibitem[{{Lee} {et~al.}(2008{\natexlab{b}}){Lee}, {Beers}, {Sivarani},
  {Johnson}, {An}, {Wilhelm}, {Allende Prieto}, {Koesterke}, {Re Fiorentin},
  {Bailer-Jones}, {Norris}, {Yanny}, {Rockosi}, {Newberg}, {Cudworth}, \&
  {Pan}}]{Lee08b}
---. 2008{\natexlab{b}}, \aj, 136, 2050, \dodoi{10.1088/0004-6256/136/5/2050}

\bibitem[{{Leitherer} {et~al.}(2010){Leitherer}, {Ortiz Ot{\'a}lvaro},
  {Bresolin}, {Kudritzki}, {Lo Faro}, {Pauldrach}, {Pettini}, \&
  {Rix}}]{Leitherer10}
{Leitherer}, C., {Ortiz Ot{\'a}lvaro}, P.~A., {Bresolin}, F., {et~al.} 2010,
  \apjs, 189, 309, \dodoi{10.1088/0067-0049/189/2/309}

\bibitem[{{Leitherer} {et~al.}(1999){Leitherer}, {Schaerer}, {Goldader},
  {Delgado}, {Robert}, {Kune}, {de Mello}, {Devost}, \&
  {Heckman}}]{Leitherer99}
{Leitherer}, C., {Schaerer}, D., {Goldader}, J.~D., {et~al.} 1999, \apjs, 123,
  3, \dodoi{10.1086/313233}

\bibitem[{{Lejeune} {et~al.}(1997){Lejeune}, {Cuisinier}, \&
  {Buser}}]{Lejeune97}
{Lejeune}, T., {Cuisinier}, F., \& {Buser}, R. 1997, \aaps, 125, 229,
  \dodoi{10.1051/aas:1997373}

\bibitem[{{Lejeune} {et~al.}(1998){Lejeune}, {Cuisinier}, \&
  {Buser}}]{Lejeune98}
---. 1998, \aaps, 130, 65, \dodoi{10.1051/aas:1998405}

\bibitem[{{Liu} {et~al.}(2015){Liu}, {Cui}, {Zhang}, {Wan}, {Deng}, {Hou},
  {Wang}, {Yang}, \& {Zhang}}]{Liu15}
{Liu}, C., {Cui}, W.-Y., {Zhang}, B., {et~al.} 2015, Research in Astronomy and
  Astrophysics, 15, 1137, \dodoi{10.1088/1674-4527/15/8/004}

\bibitem[{{Luo} {et~al.}(2015){Luo}, {Zhao}, {Zhao}, {Deng}, {Liu}, {Jing},
  {Wang}, {Zhang}, {Shi}, {Cui}, {Chu}, {Li}, {Bai}, {Wu}, {Cai}, {Cao}, {Cao},
  {Carlin}, {Chen}, {Chen}, {Chen}, {Chen}, {Chen}, {Chen}, {Chen},
  {Christlieb}, {Chu}, {Cui}, {Dong}, {Du}, {Fan}, {Feng}, {Fu}, {Gao}, {Gong},
  {Gu}, {Guo}, {Han}, {He}, {Hou}, {Hou}, {Hou}, {Hu}, {Hu}, {Hu}, {Huo},
  {Jia}, {Jiang}, {Jiang}, {Jiang}, {Jin}, {Kong}, {Kong}, {Lei}, {Li}, {Li},
  {Li}, {Li}, {Li}, {Li}, {Li}, {Li}, {Li}, {Li}, {Li}, {Li}, {Liang}, {Lin},
  {Liu}, {Liu}, {Liu}, {Liu}, {Lu}, {Luo}, {Mao}, {Newberg}, {Ni}, {Qi}, {Qi},
  {Shen}, {Shi}, {Song}, {Song}, {Su}, {Su}, {Tang}, {Tao}, {Tian}, {Wang},
  {Wang}, {Wang}, {Wang}, {Wang}, {Wang}, {Wang}, {Wang}, {Wang}, {Wang},
  {Wang}, {Wang}, {Wang}, {Wang}, {Wang}, {Wang}, {Wang}, {Wang}, {Wang},
  {Wang}, {Wei}, {Wei}, {Wu}, {Wu}, {Wu}, {Wu}, {Xing}, {Xu}, {Xu}, {Xu},
  {Yan}, {Yang}, {Yang}, {Yang}, {Yang}, {Yao}, {Yu}, {Yuan}, {Yuan}, {Yuan},
  {Yuan}, {Zhai}, {Zhang}, {Zhang}, {Zhang}, {Zhang}, {Zhang}, {Zhang},
  {Zhang}, {Zhang}, {Zhao}, {Zhou}, {Zhou}, {Zhu}, {Zhu}, {Zou}, \&
  {Zuo}}]{LAMOSTDR1}
{Luo}, A.-L., {Zhao}, Y.-H., {Zhao}, G., {et~al.} 2015, Research in Astronomy
  and Astrophysics, 15, 1095, \dodoi{10.1088/1674-4527/15/8/002}

\bibitem[{{Majewski} {et~al.}(2016){Majewski}, {APOGEE Team}, \& {APOGEE-2
  Team}}]{Majewski16}
{Majewski}, S.~R., {APOGEE Team}, \& {APOGEE-2 Team}. 2016, Astronomische
  Nachrichten, 337, 863, \dodoi{10.1002/asna.201612387}

\bibitem[{{Majewski} {et~al.}(2011){Majewski}, {Zasowski}, \&
  {Nidever}}]{Majewski11}
{Majewski}, S.~R., {Zasowski}, G., \& {Nidever}, D.~L. 2011, \apj, 739, 25,
  \dodoi{10.1088/0004-637X/739/1/25}

\bibitem[{{Maraston}(2005)}]{Maraston05}
{Maraston}, C. 2005, \mnras, 362, 799, \dodoi{10.1111/j.1365-2966.2005.09270.x}

\bibitem[{{Maraston} \& {Str{\"o}mb{\"a}ck}(2011)}]{MarastonS11}
{Maraston}, C., \& {Str{\"o}mb{\"a}ck}, G. 2011, \mnras, 418, 2785,
  \dodoi{10.1111/j.1365-2966.2011.19738.x}

\bibitem[{{Martins} {et~al.}(2005){Martins}, {Gonz{\'a}lez Delgado},
  {Leitherer}, {Cervi{\~n}o}, \& {Hauschildt}}]{Martins05}
{Martins}, L.~P., {Gonz{\'a}lez Delgado}, R.~M., {Leitherer}, C.,
  {Cervi{\~n}o}, M., \& {Hauschildt}, P. 2005, \mnras, 358, 49,
  \dodoi{10.1111/j.1365-2966.2005.08703.x}

\bibitem[{{M{\'e}sz{\'a}ros} {et~al.}(2012){M{\'e}sz{\'a}ros}, {Allende
  Prieto}, {Edvardsson}, {Castelli}, {Garc{\'{\i}}a P{\'e}rez}, {Gustafsson},
  {Majewski}, {Plez}, {Schiavon}, {Shetrone}, \& {de Vicente}}]{Meszaros12}
{M{\'e}sz{\'a}ros}, S., {Allende Prieto}, C., {Edvardsson}, B., {et~al.} 2012,
  \aj, 144, 120, \dodoi{10.1088/0004-6256/144/4/120}

\bibitem[{{Munari} {et~al.}(2005){Munari}, {Sordo}, {Castelli}, \&
  {Zwitter}}]{Munari05}
{Munari}, U., {Sordo}, R., {Castelli}, F., \& {Zwitter}, T. 2005, \aap, 442,
  1127, \dodoi{10.1051/0004-6361:20042490}

\bibitem[{{Murphy} \& {Meiksin}(2004)}]{MurphyM04}
{Murphy}, T., \& {Meiksin}, A. 2004, \mnras, 351, 1430,
  \dodoi{10.1111/j.1365-2966.2004.07895.x}

\bibitem[{{Oke}(1990)}]{Oke90}
{Oke}, J.~B. 1990, \aj, 99, 1621, \dodoi{10.1086/115444}

\bibitem[{{Palacios} {et~al.}(2010){Palacios}, {Gebran}, {Josselin}, {Martins},
  {Plez}, {Belmas}, \& {L{\`e}bre}}]{Palacios10}
{Palacios}, A., {Gebran}, M., {Josselin}, E., {et~al.} 2010, \aap, 516, A13,
  \dodoi{10.1051/0004-6361/200913932}

\bibitem[{{Pickles}(1985)}]{Pickles85}
{Pickles}, A.~J. 1985, \apjs, 59, 33, \dodoi{10.1086/191061}

\bibitem[{{Pickles}(1998)}]{Pickles98}
---. 1998, \pasp, 110, 863, \dodoi{10.1086/316197}

\bibitem[{{Prugniel} \& {Soubiran}(2001)}]{PrugnielS01}
{Prugniel}, P., \& {Soubiran}, C. 2001, \aap, 369, 1048,
  \dodoi{10.1051/0004-6361:20010163}

\bibitem[{{Prugniel} \& {Soubiran}(2004)}]{PrugnielS04}
---. 2004, ArXiv Astrophysics e-prints

\bibitem[{{Prugniel} {et~al.}(2007){Prugniel}, {Soubiran}, {Koleva}, \& {Le
  Borgne}}]{Prugniel07}
{Prugniel}, P., {Soubiran}, C., {Koleva}, M., \& {Le Borgne}, D. 2007, VizieR
  Online Data Catalog, 3251

\bibitem[{{Prugniel} {et~al.}(2011){Prugniel}, {Vauglin}, \&
  {Koleva}}]{Prugniel11}
{Prugniel}, P., {Vauglin}, I., \& {Koleva}, M. 2011, \aap, 531, A165,
  \dodoi{10.1051/0004-6361/201116769}

\bibitem[{{Rayner} {et~al.}(2009){Rayner}, {Cushing}, \& {Vacca}}]{Rayner09}
{Rayner}, J.~T., {Cushing}, M.~C., \& {Vacca}, W.~D. 2009, \apjs, 185, 289,
  \dodoi{10.1088/0067-0049/185/2/289}

\bibitem[{{R{\"o}ck} {et~al.}(2016){R{\"o}ck}, {Vazdekis}, {Ricciardelli},
  {Peletier}, {Knapen}, \& {Falc{\'o}n-Barroso}}]{Rock16}
{R{\"o}ck}, B., {Vazdekis}, A., {Ricciardelli}, E., {et~al.} 2016, \aap, 589,
  A73, \dodoi{10.1051/0004-6361/201527570}

\bibitem[{{Rodr{\'{\i}}guez-Merino} {et~al.}(2005){Rodr{\'{\i}}guez-Merino},
  {Chavez}, {Bertone}, \& {Buzzoni}}]{Rodriguez-Merino05}
{Rodr{\'{\i}}guez-Merino}, L.~H., {Chavez}, M., {Bertone}, E., \& {Buzzoni}, A.
  2005, \apj, 626, 411, \dodoi{10.1086/429858}

\bibitem[{{S{\'a}nchez-Bl{\'a}zquez} {et~al.}(2006){S{\'a}nchez-Bl{\'a}zquez},
  {Peletier}, {Jim{\'e}nez-Vicente}, {Cardiel}, {Cenarro},
  {Falc{\'o}n-Barroso}, {Gorgas}, {Selam}, \& {Vazdekis}}]{Sanchez-Blazquez06}
{S{\'a}nchez-Bl{\'a}zquez}, P., {Peletier}, R.~F., {Jim{\'e}nez-Vicente}, J.,
  {et~al.} 2006, \mnras, 371, 703, \dodoi{10.1111/j.1365-2966.2006.10699.x}

\bibitem[{{Schlafly} \& {Finkbeiner}(2011)}]{Schlafly11}
{Schlafly}, E.~F., \& {Finkbeiner}, D.~P. 2011, \apj, 737, 103,
  \dodoi{10.1088/0004-637X/737/2/103}

\bibitem[{{Schlegel} {et~al.}(1998){Schlegel}, {Finkbeiner}, \&
  {Davis}}]{SchlegelFD98}
{Schlegel}, D.~J., {Finkbeiner}, D.~P., \& {Davis}, M. 1998, \apj, 500, 525,
  \dodoi{10.1086/305772}

\bibitem[{{Silva} \& {Cornell}(1992)}]{SilvaC92}
{Silva}, D.~R., \& {Cornell}, M.~E. 1992, \apjs, 81, 865,
  \dodoi{10.1086/191706}

\bibitem[{{Smee} {et~al.}(2013){Smee}, {Gunn}, {Uomoto}, {Roe}, {Schlegel},
  {Rockosi}, {Carr}, {Leger}, {Dawson}, {Olmstead}, {Brinkmann}, {Owen},
  {Barkhouser}, {Honscheid}, {Harding}, {Long}, {Lupton}, {Loomis}, {Anderson},
  {Annis}, {Bernardi}, {Bhardwaj}, {Bizyaev}, {Bolton}, {Brewington}, {Briggs},
  {Burles}, {Burns}, {Castander}, {Connolly}, {Davenport}, {Ebelke}, {Epps},
  {Feldman}, {Friedman}, {Frieman}, {Heckman}, {Hull}, {Knapp}, {Lawrence},
  {Loveday}, {Mannery}, {Malanushenko}, {Malanushenko}, {Merrelli}, {Muna},
  {Newman}, {Nichol}, {Oravetz}, {Pan}, {Pope}, {Ricketts}, {Shelden},
  {Sandford}, {Siegmund}, {Simmons}, {Smith}, {Snedden}, {Schneider},
  {SubbaRao}, {Tremonti}, {Waddell}, \& {York}}]{Smee13}
{Smee}, S.~A., {Gunn}, J.~E., {Uomoto}, A., {et~al.} 2013, \aj, 146, 32,
  \dodoi{10.1088/0004-6256/146/2/32}

\bibitem[{{Sordo} {et~al.}(2010){Sordo}, {Vallenari}, {Tantalo}, {Allard},
  {Blomme}, {Bouret}, {Brott}, {Fremat}, {Martayan}, {Damerdji}, {Edvardsson},
  {Josselin}, {Plez}, {Kochukhov}, {Kontizas}, {Munari}, {Saguner}, {Zorec},
  {Schweitzer}, \& {Tsalmantza}}]{Sordo10}
{Sordo}, R., {Vallenari}, A., {Tantalo}, R., {et~al.} 2010, \apss, 328, 331,
  \dodoi{10.1007/s10509-010-0272-7}

\bibitem[{{Soubiran} {et~al.}(1998){Soubiran}, {Katz}, \&
  {Cayrel}}]{Soubiran98}
{Soubiran}, C., {Katz}, D., \& {Cayrel}, R. 1998, \aaps, 133, 221,
  \dodoi{10.1051/aas:1998456}

\bibitem[{{Valdes} {et~al.}(2004){Valdes}, {Gupta}, {Rose}, {Singh}, \&
  {Bell}}]{Valdes04}
{Valdes}, F., {Gupta}, R., {Rose}, J.~A., {Singh}, H.~P., \& {Bell}, D.~J.
  2004, \apjs, 152, 251, \dodoi{10.1086/386343}

\bibitem[{{Vazdekis} {et~al.}(2012){Vazdekis}, {Ricciardelli}, {Cenarro},
  {Rivero-Gonz{\'a}lez}, {D{\'{\i}}az-Garc{\'{\i}}a}, \&
  {Falc{\'o}n-Barroso}}]{Vazdekis12}
{Vazdekis}, A., {Ricciardelli}, E., {Cenarro}, A.~J., {et~al.} 2012, \mnras,
  424, 157, \dodoi{10.1111/j.1365-2966.2012.21179.x}

\bibitem[{{Vazdekis} {et~al.}(2010){Vazdekis}, {S{\'a}nchez-Bl{\'a}zquez},
  {Falc{\'o}n-Barroso}, {Cenarro}, {Beasley}, {Cardiel}, {Gorgas}, \&
  {Peletier}}]{Vazdekis10}
{Vazdekis}, A., {S{\'a}nchez-Bl{\'a}zquez}, P., {Falc{\'o}n-Barroso}, J.,
  {et~al.} 2010, \mnras, 404, 1639, \dodoi{10.1111/j.1365-2966.2010.16407.x}

\bibitem[{{Villaume} {et~al.}(2017){Villaume}, {Conroy}, {Johnson}, {Rayner},
  {Mann}, \& {van Dokkum}}]{Villaume17}
{Villaume}, A., {Conroy}, C., {Johnson}, B., {et~al.} 2017, \apjs, 230, 23,
  \dodoi{10.3847/1538-4365/aa72ed}

\bibitem[{{Wake} {et~al.}(2017){Wake}, {Bundy}, {Diamond-Stanic}, {Yan},
  {Blanton}, {Bershady}, {S{\'a}nchez-Gallego}, {Drory}, {Jones}, {Kauffmann},
  {Law}, {Li}, {MacDonald}, {Masters}, {Thomas}, {Tinker}, {Weijmans}, \&
  {Brownstein}}]{Wake17}
{Wake}, D.~A., {Bundy}, K., {Diamond-Stanic}, A.~M., {et~al.} 2017, \aj, 154,
  86, \dodoi{10.3847/1538-3881/aa7ecc}

\bibitem[{{Westera} {et~al.}(2002){Westera}, {Lejeune}, {Buser}, {Cuisinier},
  \& {Bruzual}}]{Westera02}
{Westera}, P., {Lejeune}, T., {Buser}, R., {Cuisinier}, F., \& {Bruzual}, G.
  2002, \aap, 381, 524, \dodoi{10.1051/0004-6361:20011493}

\bibitem[{{Wilson} {et~al.}(2012){Wilson}, {Hearty}, {Skrutskie}, {Majewski},
  {Schiavon}, {Eisenstein}, {Gunn}, {Holtzman}, {Nidever}, {Gillespie},
  {Weinberg}, {Blank}, {Henderson}, {Smee}, {Barkhouser}, {Harding}, {Hope},
  {Fitzgerald}, {Stolberg}, {Arns}, {Nelson}, {Brunner}, {Burton}, {Walker},
  {Lam}, {Maseman}, {Barr}, {Leger}, {Carey}, {MacDonald}, {Ebelke}, {Beland},
  {Horne}, {Young}, {Rieke}, {Rieke}, {O'Brien}, {Crane}, {Carr}, {Harrison},
  {Stoll}, {Vernieri}, {Shetrone}, {Allende-Prieto}, {Johnson}, {Frinchaboy},
  {Zasowski}, {Garcia Perez}, {Bizyaev}, {Cunha}, {Smith}, {Meszaros}, {Zhao},
  {Hayden}, {Chojnowski}, {Andrews}, {Loomis}, {Owen}, {Klaene}, {Brinkmann},
  {Stauffer}, {Long}, {Jordan}, {Holder}, {Cope}, {Naugle}, {Pfaffenberger},
  {Schlegel}, {Blanton}, {Muna}, {Weaver}, {Snedden}, {Pan}, {Brewington},
  {Malanushenko}, {Malanushenko}, {Simmons}, {Oravetz}, {Mahadevan}, \&
  {Halverson}}]{Wilson12}
{Wilson}, J.~C., {Hearty}, F., {Skrutskie}, M.~F., {et~al.} 2012, in \procspie,
  Vol. 8446, Ground-based and Airborne Instrumentation for Astronomy IV, 84460H

\bibitem[{{Worthey} {et~al.}(1994){Worthey}, {Faber}, {Gonzalez}, \&
  {Burstein}}]{Worthey94}
{Worthey}, G., {Faber}, S.~M., {Gonzalez}, J.~J., \& {Burstein}, D. 1994,
  \apjs, 94, 687, \dodoi{10.1086/192087}

\bibitem[{{Yan} {et~al.}(2016{\natexlab{a}}){Yan}, {Bundy}, {Law}, {Bershady},
  {Andrews}, {Cherinka}, {Diamond-Stanic}, {Drory}, {MacDonald},
  {S{\'a}nchez-Gallego}, {Thomas}, {Wake}, {Weijmans}, {Westfall}, {Zhang},
  {Arag{\'o}n-Salamanca}, {Belfiore}, {Bizyaev}, {Blanc}, {Blanton},
  {Brownstein}, {Cappellari}, {D'Souza}, {Emsellem}, {Fu}, {Gaulme}, {Graham},
  {Goddard}, {Gunn}, {Harding}, {Jones}, {Kinemuchi}, {Li}, {Li}, {Maiolino},
  {Mao}, {Maraston}, {Masters}, {Merrifield}, {Oravetz}, {Pan}, {Parejko},
  {Sanchez}, {Schlegel}, {Simmons}, {Thanjavur}, {Tinker}, {Tremonti}, {van den
  Bosch}, \& {Zheng}}]{Yan16b}
{Yan}, R., {Bundy}, K., {Law}, D.~R., {et~al.} 2016{\natexlab{a}}, \aj, 152,
  197, \dodoi{10.3847/0004-6256/152/6/197}

\bibitem[{{Yan} {et~al.}(2016{\natexlab{b}}){Yan}, {Tremonti}, {Bershady},
  {Law}, {Schlegel}, {Bundy}, {Drory}, {MacDonald}, {Bizyaev}, {Blanc},
  {Blanton}, {Cherinka}, {Eigenbrot}, {Gunn}, {Harding}, {Hogg},
  {S{\'a}nchez-Gallego}, {S{\'a}nchez}, {Wake}, {Weijmans}, {Xiao}, \&
  {Zhang}}]{Yan16}
{Yan}, R., {Tremonti}, C., {Bershady}, M.~A., {et~al.} 2016{\natexlab{b}}, \aj,
  151, 8, \dodoi{10.3847/0004-6256/151/1/8}

\bibitem[{{Yanny} {et~al.}(2009){Yanny}, {Rockosi}, {Newberg}, {Knapp},
  {Adelman-McCarthy}, {Alcorn}, {Allam}, {Allende Prieto}, {An}, {Anderson},
  {Anderson}, {Bailer-Jones}, {Bastian}, {Beers}, {Bell}, {Belokurov},
  {Bizyaev}, {Blythe}, {Bochanski}, {Boroski}, {Brinchmann}, {Brinkmann},
  {Brewington}, {Carey}, {Cudworth}, {Evans}, {Evans}, {Gates}, {G{\"a}nsicke},
  {Gillespie}, {Gilmore}, {Nebot Gomez-Moran}, {Grebel}, {Greenwell}, {Gunn},
  {Jordan}, {Jordan}, {Harding}, {Harris}, {Hendry}, {Holder}, {Ivans},
  {Ivezi{\v c}}, {Jester}, {Johnson}, {Kent}, {Kleinman}, {Kniazev},
  {Krzesinski}, {Kron}, {Kuropatkin}, {Lebedeva}, {Lee}, {French Leger},
  {L{\'e}pine}, {Levine}, {Lin}, {Long}, {Loomis}, {Lupton}, {Malanushenko},
  {Malanushenko}, {Margon}, {Martinez-Delgado}, {McGehee}, {Monet}, {Morrison},
  {Munn}, {Neilsen}, {Nitta}, {Norris}, {Oravetz}, {Owen}, {Padmanabhan},
  {Pan}, {Peterson}, {Pier}, {Platson}, {Re Fiorentin}, {Richards}, {Rix},
  {Schlegel}, {Schneider}, {Schreiber}, {Schwope}, {Sibley}, {Simmons},
  {Snedden}, {Allyn Smith}, {Stark}, {Stauffer}, {Steinmetz}, {Stoughton},
  {SubbaRao}, {Szalay}, {Szkody}, {Thakar}, {Sivarani}, {Tucker}, {Uomoto},
  {Vanden Berk}, {Vidrih}, {Wadadekar}, {Watters}, {Wilhelm}, {Wyse}, {Yarger},
  \& {Zucker}}]{Yanny09}
{Yanny}, B., {Rockosi}, C., {Newberg}, H.~J., {et~al.} 2009, \aj, 137, 4377,
  \dodoi{10.1088/0004-6256/137/5/4377}

\bibitem[{{Zasowski} {et~al.}(2013){Zasowski}, {Johnson}, {Frinchaboy},
  {Majewski}, {Nidever}, {Rocha Pinto}, {Girardi}, {Andrews}, {Chojnowski},
  {Cudworth}, {Jackson}, {Munn}, {Skrutskie}, {Beaton}, {Blake}, {Covey},
  {Deshpande}, {Epstein}, {Fabbian}, {Fleming}, {Garcia Hernandez}, {Herrero},
  {Mahadevan}, {M{\'e}sz{\'a}ros}, {Schultheis}, {Sellgren}, {Terrien}, {van
  Saders}, {Allende Prieto}, {Bizyaev}, {Burton}, {Cunha}, {da Costa},
  {Hasselquist}, {Hearty}, {Holtzman}, {Garc{\'{\i}}a P{\'e}rez}, {Maia},
  {O'Connell}, {O'Donnell}, {Pinsonneault}, {Santiago}, {Schiavon}, {Shetrone},
  {Smith}, \& {Wilson}}]{Zasowski13}
{Zasowski}, G., {Johnson}, J.~A., {Frinchaboy}, P.~M., {et~al.} 2013, \aj, 146,
  81, \dodoi{10.1088/0004-6256/146/4/81}

\bibitem[{{Zhao} {et~al.}(2012){Zhao}, {Zhao}, {Chu}, {Jing}, \&
  {Deng}}]{Zhao12}
{Zhao}, G., {Zhao}, Y.-H., {Chu}, Y.-Q., {Jing}, Y.-P., \& {Deng}, L.-C. 2012,
  Research in Astronomy and Astrophysics, 12, 723,
  \dodoi{10.1088/1674-4527/12/7/002}

\bibitem[{{Zhong} {et~al.}(2015){Zhong}, {L{\'e}pine}, {Li}, {Chen}, {Hou},
  {Yang}, {Li}, {Zhang}, \& {Hou}}]{Zhong15}
{Zhong}, J., {L{\'e}pine}, S., {Li}, J., {et~al.} 2015, Research in Astronomy
  and Astrophysics, 15, 1154, \dodoi{10.1088/1674-4527/15/8/005}

\bibitem[{{Zwitter} {et~al.}(2004){Zwitter}, {Castelli}, \&
  {Munari}}]{Zwitter04}
{Zwitter}, T., {Castelli}, F., \& {Munari}, U. 2004, \aap, 417, 1055,
  \dodoi{10.1051/0004-6361:20034324}

\end{thebibliography}

\end{document}